\documentclass[useAMS,usenatbib]{mn2e}
\usepackage{times}
\usepackage{graphicx}
\usepackage{subfigure}
\usepackage{makecell}
\usepackage[table, x11names]{xcolor}
\usepackage{tabu}
\usepackage{bm}
\newcommand{\ds}{\displaystyle}

\def\M{{\rmn M}}
\def\S{{\rmn S}}

\usepackage{mathtools}

\usepackage{stmaryrd}

\usepackage{titlesec}
\titleformat{\section}{\large\bfseries\uppercase}{\thesection}{1em}{}
\titleformat{\subsection}{\Large\bfseries}{\thesubsection}{1em}{}
\titleformat{\subsubsection}{\Large\itshape}{\thesubsubsection}{1em}{}
\onecolumn
\usepackage{caption}
\title[Galileo disposal strategy]{Galileo disposal strategy: stability, chaos and predictability\thanks{Some results of this paper were presented at the 25th International Symposium on Space Flight Dynamics (ISSFD), 2015, Munich, Germany.}} 

\author[A. J. Rosengren et al.]
{Aaron J. Rosengren,$^{1}$\thanks{E-mail: a.rosengren@ifac.cnr.it} 
J\'{e}r\^{o}me Daquin,$^{2,3}$ 
Elisa Maria Alessi,$^{1}$ 
Florent Deleflie,$^{3}$ \and
Alessandro Rossi$^{1}$ and Giovanni B. Valsecchi$^{1,4}$\\
$^{1}$IFAC-CNR, Via Madonna del Piano 10, 50019 Sesto Fiorentino (FI), Italy\\
$^{2}$Thales Services, 3 Impasse de l'Europe, 31400 Toulouse, France\\
$^{3}$IMCCE/Observatoire de Paris, Universit\'{e} Lille1, 1 Impasse de l'Observatoire, 59000 Lille, France\\
$^{4}$IAPS-INAF, Via Fosso del Cavaliere 100, 00133 Roma, Italy}
\begin{document}

\date{}

\pagerange{\pageref{firstpage}--\pageref{lastpage}} \pubyear{2015}

\maketitle

\label{firstpage}

\begin{abstract}
\Large 
Recent studies have shown that the medium-Earth orbit (MEO) region of the Global Navigation Satellite Systems is permeated by a devious network of lunisolar secular resonances, which can interact to produce chaotic and diffusive motions. The precarious state of the four navigation constellations, perched on the threshold of instability, makes it understandable why all past efforts to define stable graveyard orbits, especially in the case of Galileo, were bound to fail; the region is far too complex to allow of an adoption of the simple geosynchronous disposal strategy. We retrace one such recent attempt, funded by ESA's General Studies Programme in the frame of the GreenOPS initiative, that uses a systematic parametric approach and the straightforward maximum-eccentricity method to identify long-term stable regions, suitable for graveyards, as well as large-scale excursions in eccentricity, which can be used for post-mission deorbiting of constellation satellites. We then apply our new results on the stunningly rich dynamical structure of the MEO region toward the analysis of these disposal strategies for Galileo, and discuss the practical implications of resonances and chaos in this regime. We outline how the identification of the hyperbolic and elliptic fixed points of the resonances near Galileo can lead to explicit criteria for defining optimal disposal strategies. 
\end{abstract}

\begin{keywords}
celestial mechanics -- chaos -- methods: analytical -- methods: numerical --  planets and satellites: dynamical evolution and stability --- planets and satellites: general.
\end{keywords}

\Large

% -------------------------------------------------------------------------------------------------------------------------------------- 
%     INTRODUCTION
% -------------------------------------------------------------------------------------------------------------------------------------- 
\section{Introduction}
\label{sec:intro}

The application of the mathematical tools and techniques of nonlinear dynamics has provided astronomers with a deeper understanding of the dynamical processes that have helped to shape the Solar System \citep{aM02}. Resonant phenomena connected with the commensurability of frequencies of interacting motions abound in celestial mechanics and have both dynamical and theoretical importance. A succession of remarkable features in the asteroid belt, known as the Kirkwood gaps, vividly illustrates the physical significance of resonances and chaos in real systems. Considerable impetus was imparted over the past three decades to the study and understanding of this type of chaotic unpredictability and its manifestation in other astronomical problems.

With chaotic motions being a natural consequence of even the most simplest of systems, it may no longer be sensible to investigate the ``exact'' trajectory of a celestial body (natural or artificial) in a given time interval \citep[q.v.][and references therein]{rZ15}. Far beyond the Lyapunov time, the characteristic time over which an orbit is said to remain predictable, it is not possible to reproduce the same time evolution if the system is chaotic, due to the exponential growth of uncertainties (in the initial state, mis-modelling effects, numerical errors, etc.). The irregular and haphazard character of the chaotic path of a celestial body reflects a similar irregularity in the trajectories of stochastic systems, as if the former were influenced by a random perturbation even though, in fact, the motion is governed by purely deterministic dynamical equations. There is, however, an essential difference: ``classical (that is, non-quantum mechanical) chaotic systems are not in any sense intrinsically random or unpredictable,'' as John Barrow puts it, ``they merely possess extreme sensitivity to ignorance \citep{jB10}.'' Despite the unpredictability of the path of a particular orbit, chaotic systems can exhibit statistical regularities, and have stable, predictable, long-term, average behaviours \citep{aLmL92,jM92}. The lesson is that the time evolution of a chaotic system can only be described in statistical terms; one must study the statistical properties of ensembles of stochastic orbits \citep[e.g.,][]{jL09,rZ15}.

Our knowledge about the stability of the orbits of artificial Earth satellites is still incomplete. Despite over fifty years of space activities, we know amazingly little about the dynamical environment occupied by artificial satellites and space debris. Strange as it may seem, we understand the structure and evolution of the, mostly invisible, trans-Neptunian belts of small bodies \citep[q.v.][]{aM02} far better than we understand that of the artificial bodies that orbit our terrestrial abode. Before these remnants of Solar-System formation diverted the interests and energies of space-age astronomers, such astrodynamical problems stood in the foremost rank of astronomical research work \citep{dB59}. The kind of Newtonian determinism brought to bear during the 1960s has continued merrily along in astrodynamics, unheeding the fundamental discoveries of nonlinear dynamics. Today we take for granted the great power and scope of modern computers, treating them as  the supreme intelligence imagined by Laplace, and the construction of increasingly more `accurate' and grandiloquent dynamical models and simulation capabilities has become the central task of the field.

As long as our thought processes are limited along the inflexibilities of determinism, we will remain forever ignorant of the possible range and vagaries of chaos in Earth-satellite orbits. An understanding of these chaotic phenomena is of fundamental importance for all efforts to assess debris mitigation measures --- efforts which may shed much light on the design and definition of optimal disposal strategies throughout all space regions (LEO, MEO, GEO, HEO, LPO), taking into account orbital interaction and environmental evolution. In this context, there has been considerable recent interest in designing novel deorbiting or re-orbiting solutions for the MEO navigation satellites \citep[qq.v.][]{eA15,jR15,dStY15}, since the operational constellations and recommended graveyard orbits have been found to be unstable \citep{cC00,aJaG05}.

The intent of this paper is to provide a case study on the European Galileo system that can be used as a reference for the other constellations, and to serve as a springboard for investigating new dynamical situations that may arise. We begin by reviewing a recent parametric numerical study on two end-of-life disposal strategies, based on the Laplacian paradigm, which investigates the role of the initial parameters of the disposal orbits (the semi-major axis, eccentricity, inclination, orientation phase angles, and epoch) on their long-term stability over centennial and longer timescales \citep[given in detail in][]{eA15}. We briefly summarise our findings from this extensive numerical experiment, as they pertain to Galileo, and show, based on our recent studies of the dynamical structure of MEO \citep{jDaR15}, why such general recommendations and guidelines should be taken with a grain of salt. We then tailor our results on the resonant and chaotic structures of the phase space near lunisolar secular resonances \citep{jDaR15,aR15} towards the analysis of the disposal options for Galileo. In this respect, we address many of the questions left open by the former numerical study of \citet{eA15}. We omit on this occasion any mathematical discussion and simply present the main results at which we have arrived.

\vspace{-3pt}
% -------------------------------------------------------------------------------------------------------------------------------------- 
%     PARAMETRIC STUDY
% -------------------------------------------------------------------------------------------------------------------------------------- 
\section{Parametric Study on Two Disposal Strategies}
\label{sec:parametric}

%  ****  Introduction and Experimental Setup
\subsection{Introduction and experimental setup}

Considerable attention is now being devoted to the problem of determining the long-term stability of medium-Earth orbits. The problem has been especially timely ever since the advent and launch of the European Galileo and the Chinese Beidou constellations. The main physical mechanisms that can lead to substantial variations in eccentricity, thereby affecting the perigee radius, are resonance phenomena associated with the orbital motion of artificial satellites. While the dynamics of MEOs, governed mainly by the inhomogeneous, non-spherical gravitational field of the Earth, is usually only weakly disturbed by lunar and solar gravitational perturbations, for certain initial conditions, appreciable effects can build up through accumulation over long periods of time. Such lunisolar resonances, which can drastically alter the satellite's orbital lifetime, generally occur when the second harmonic of the Earth's gravitational potential ($J_2$) causes nodal and apsidal motions which preserve a favorable relative orientation between the orbit and the direction of the disturbing force \citep[q.v.][and references therein]{aR15}. There is also another class of resonances that occurs when the satellite's mean motion is commensurable with the Earth's rotation rate, thereby enhancing the perturbing effects of specific tesseral harmonics in the geopotential. These tesseral resonances pervade the MEOs of the navigation satellites and their net effects are to produce small, localised instabilities in the semi-major axis \citep{tEkH97}. 

A proper understanding of the stability characteristics of the two main types of resonances in MEO is vital for the analysis and design of disposal strategies for the four constellations. This concerns particularly the question as to whether suitable stable orbits exist such that satellites in these graveyards will not interfere with the constellations, or whether strong instabilities exist, whose destabilizing effects manifest themselves on decadal to centennial timescales, that can be exploited to permanently clear this region of space from any future collision hazard. The process of dynamical clearing of resonant orbits is a new paradigm in post-mission disposal \citep{aJaG05}, but has not been hitherto rigorously studied. 

Accordingly, an ESA/GSP study was conducted to numerically examine this idea \citep{eA15}, using an accurate dynamical model accounting for the Earth's gravity field, lunisolar perturbations, and solar radiation pressure (Table \ref{tab:models}). \citet{eA15} particularly investigated to what extent the changes in initial parameters of storage orbits can affect the long-term stability of these orbits over long intervals of time. The study was based on integrations of averaged equations of motion, using a semi-analytic model suitable for all dynamical configurations, which has been approved as the reference model for the French Space Operations Act (through the software, STELA, and its Fortran prototype\footnote{STELA (Semi-analytic Tool for End of Life Analysis) can be downloaded from the CNES website:\\ https://logiciels.cnes.fr/content/stela}).

\begin{table}
	\caption{Gravitational perturbations added to the central part of the geopotential for the 
	numerical stability analysis. Model 4 (which also includes SRP perturbations with Earth shadow effects) 
	is used for the MEM maps of the ESA study and Model 1 for the FLI and Lyapunov time stability maps 
	of Section \ref{sec:FLI}}
	\label{tab:models} 
	\vspace{6pt}
	\centering
	\tabulinesep=0.25em
	\setlength{\tabcolsep}{10pt}
	\begin{tabu}{ccccc} \Xhline{3\arrayrulewidth}
	\multicolumn{1}{c}{} &
	\multicolumn{1}{c}{\textsc{Zonal}} &
	\multicolumn{1}{c}{\textsc{Tesseral}} &
	\multicolumn{1}{c}{\textsc{Lunar}} &	
	\multicolumn{1}{c}{\textsc{Solar}} \\\Xhline{3\arrayrulewidth}
		\rowcolor{LemonChiffon1}  
		model $1$ 	&  $J_{2}$ &  not considered &  up to degree $2$	
			&  up to degree $2$ \\
		model $2$ 	&  $J_{2}, J_{2}^2, J_{3}, \cdots ,J_{5}$ &  not considered &  up to degree $4$	
			&  up to degree $3$ \\
		model $3$ 	& $J_{2}, J_{2}^2, J_{3}, \cdots ,J_{5}$ & up to degree \& order $5$ &  up to degree $4$
			& up to degree  $3$ \\
		\rowcolor{LemonChiffon1}  
		model $4$ 	& $J_{2}, J_{2}^2, J_{3}, \cdots ,J_{7}$ & up to degree \& order $5$ &  up to degree $3$ 
			& up to degree $3$ \\
		model $5$ 	& $J_{2}, J_{2}^2, J_{3}, \cdots ,J_{7}$ & up to degree \& order $5$ &  up to degree $4$ 
			& up to degree $3$ \\\Xhline{3\arrayrulewidth}			
	\end{tabu}
\end{table} 

An analysis of the historical practices of the GNSS constellations was performed in order to properly define the reference simulation scenario. The nominal initial conditions and values of area-to-mass ratio considered for each disposal strategy are displayed in Table \ref{tab:disp_GAL}. For the {\itshape graveyard orbit scenario}, it is important to ensure that the storage orbits have only small-amplitude orbital deformations over long periods of time, so that the inactive satellites cannot cross the orbital region of active GNSS components (and possibly collide). This in turn implies that we must minimise the long-term eccentricity growth in order to delay or prevent the penetration of the GNSS altitude shells. Alternatively, for the {\itshape eccentricity growth scenario}, the possibility of deorbiting satellites was explored by pushing them into unstable phase-space regions that would slowly decrease their perigee distances, leading to a long-term reduction in the combined constellation and intra-graveyard collision risks \citep{aJaG05}.

The numerical investigation consisted in propagating the initial conditions of Table \ref{tab:disp_GAL} for 200 years, under dynamical Model 4 in Table \ref{tab:models}, for a large variety of initial orientation phase parameters and analyzing the maximum eccentricity attained in each case. This maximum-eccentricity method (MEM) provides a straightforward indication of orbital `stability' and has been used in a number of astronomical contexts \citep{rD03,iN06,xR15}. Instinctively and historically, we expect that the orbits become more unstable as their eccentricities grow; yet, we note that this method is not necessarily an estimator of chaos and stability (since large amplitude variations of eccentricity could be due to regular motion, e.g., secular perturbations; and small oscillations could be the result of slow manifestations of chaotic behaviours, e.g., orbits with large Lyapunov times). Each initial point of the parameter plane was characterised by their maximum eccentricity value (or a closely related quantity) under the following initial conditions \citep{eA15}:

\begin{itemize}
\item 36 equally spaced values of $\omega\in [0^\circ:360^\circ]$; 
\item 36 equally spaced values of $\Omega\in [0^\circ:360^\circ]$;
\item 38 equally spaced initial epochs, starting from $t_0= $ 26 February 1998 (a solar eclipse epoch) to $t_f=2 $ Saros, where Saros indicates a period of 6\,585.321\,347~days.
\end{itemize}

The same analysis was performed by increasing and decreasing, respectively, the initial inclination by 1$^\circ$ with respect to the nominal value. The aim was not only to see if the known resonant harmonic $2 \omega + \Omega$ is actually the most significant, as suggested, justly or unjustly,\footnote{See \citet{aR15} for a detailed literature review on the MEO stability problem.} by many \citep[][to name but a few]{cC00,aJaG05,dStY15}, but also to gain insight on the role of the initial inclination and of the Earth-Moon-Sun dynamical configuration on the long-term evolution of the orbits.

\begin{table}
	\caption{Initial mean orbital elements considered for the disposal orbits of the Galileo constellations, 
	and the corresponding values of area and mass. The difference in semi-major axis $\Delta a$ with 
	respect to the nominal constellation is also shown.}
	\label{tab:disp_GAL} 
	\vspace{9pt}
	\centering
	\renewcommand{\tabcolsep}{6pt}
	\renewcommand{\arraystretch}{1.25}
	\begin{tabular}{ccccccc}
	\Xhline{3\arrayrulewidth}
	{\scshape Disposal Strategy} & $a$ (km) & $\Delta a$ (km) & $e$ & $i$ (deg) 
		& $A$ (m$^2$) & $m$ (kg) \\ 
	\Xhline{3\arrayrulewidth}
	Graveyard Orbit  & 30\,150 & 550 &0.001 & 56 & 9.3 & 665 \\ \hline
	Eccentricity Growth & 28\,086 & -1514 & 0.0539 & 56  & 9.3 & 665 \\ \hline
	\end{tabular}
\end{table}

%  ****  Simulation Results and Discussion
\subsection{Simulation results and discussion}

We present here only a subset of the results as the full scope of the study was given in \citet{eA15} and its relation to the other navigation constellations was formulated there more completely. No space will be devoted therefore to any comparison between the similar, albeit less systematic, efforts to tackle this problem by other groups of researchers \citep[e.g.,][]{jR15,dStY15}. 

\begin{figure}
	\centering
	\includegraphics[scale=0.2]{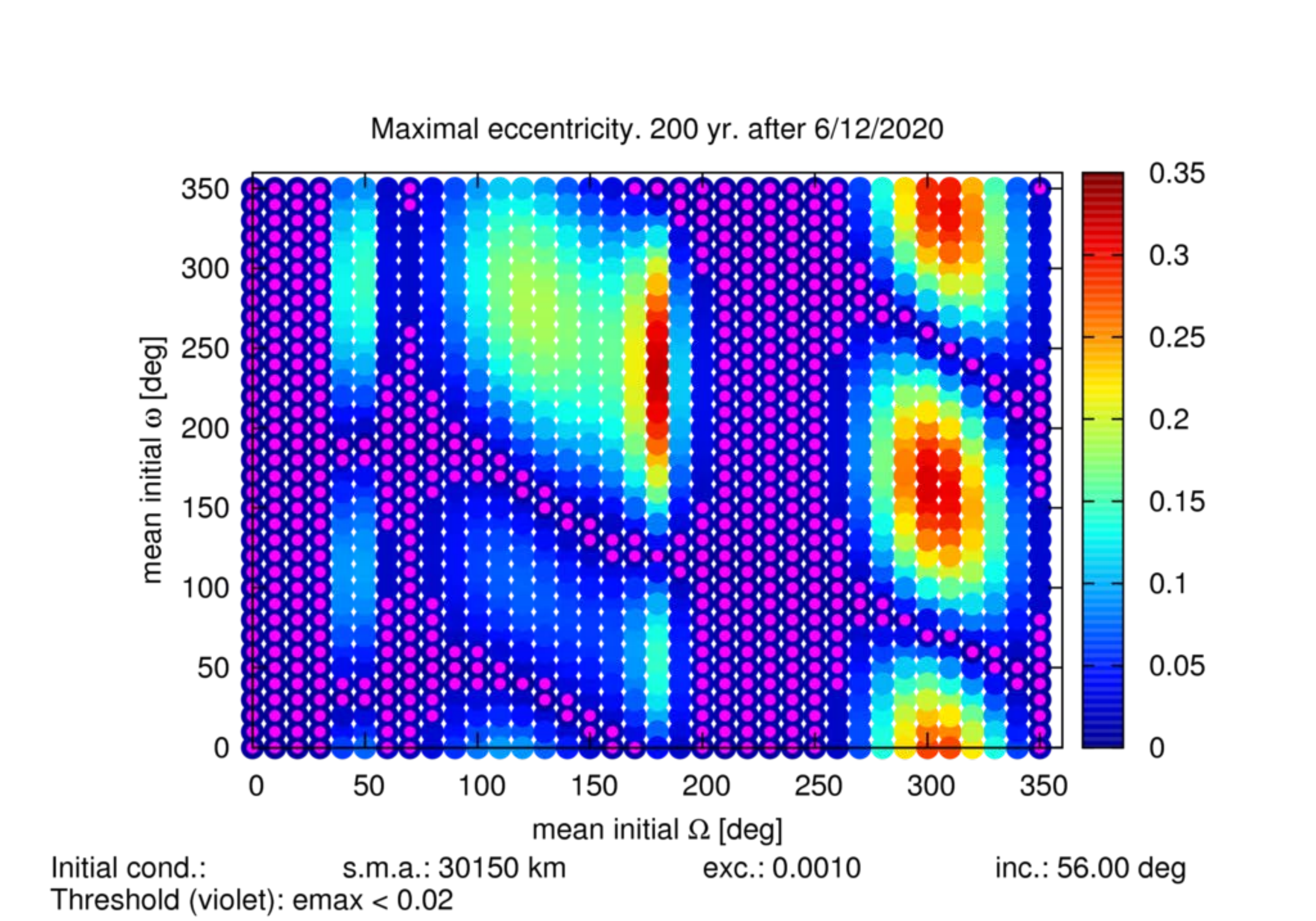} \hspace{-12pt}
	\includegraphics[scale=0.2]{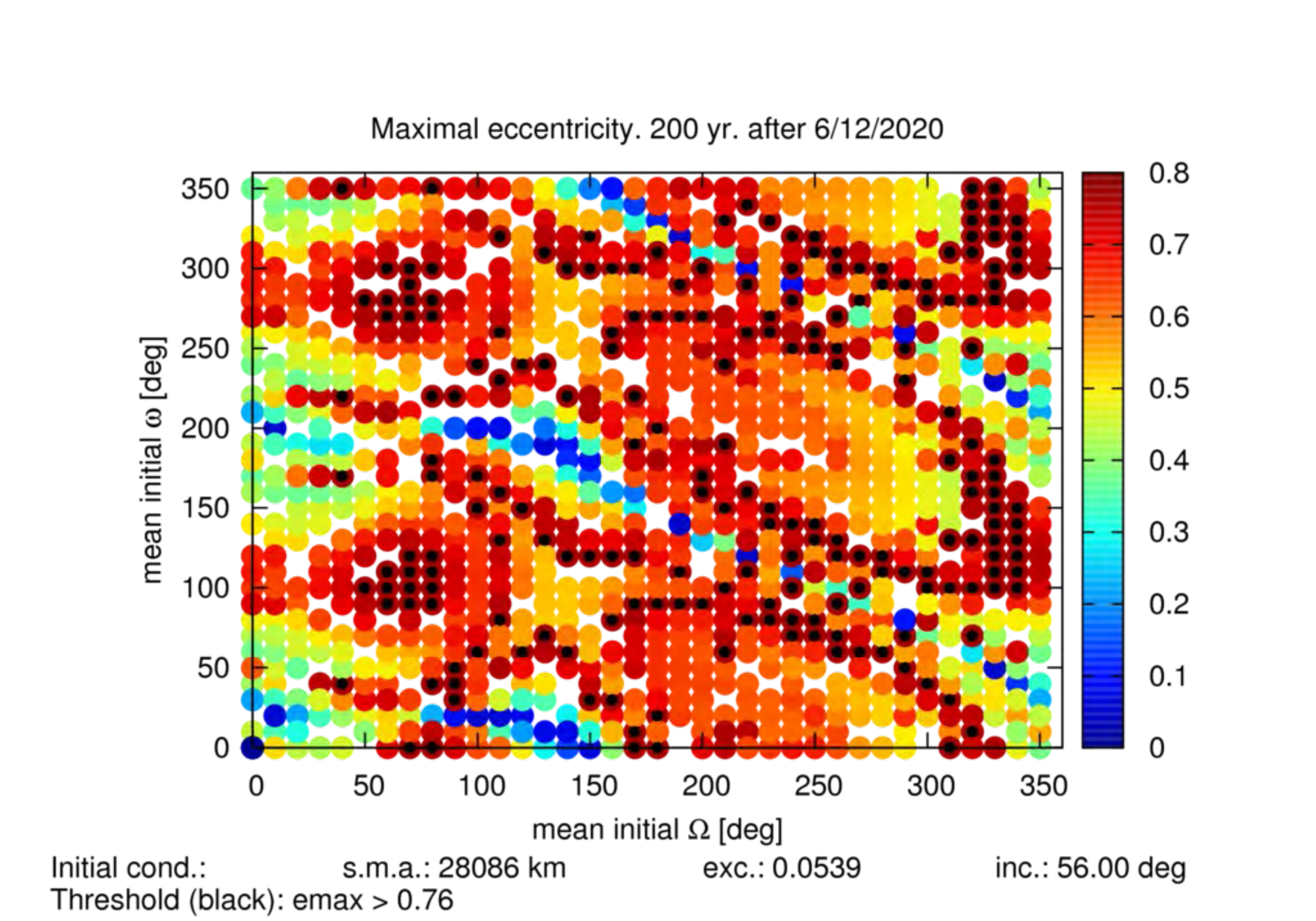} \\ \vspace{-12pt}
	\includegraphics[scale=0.2]{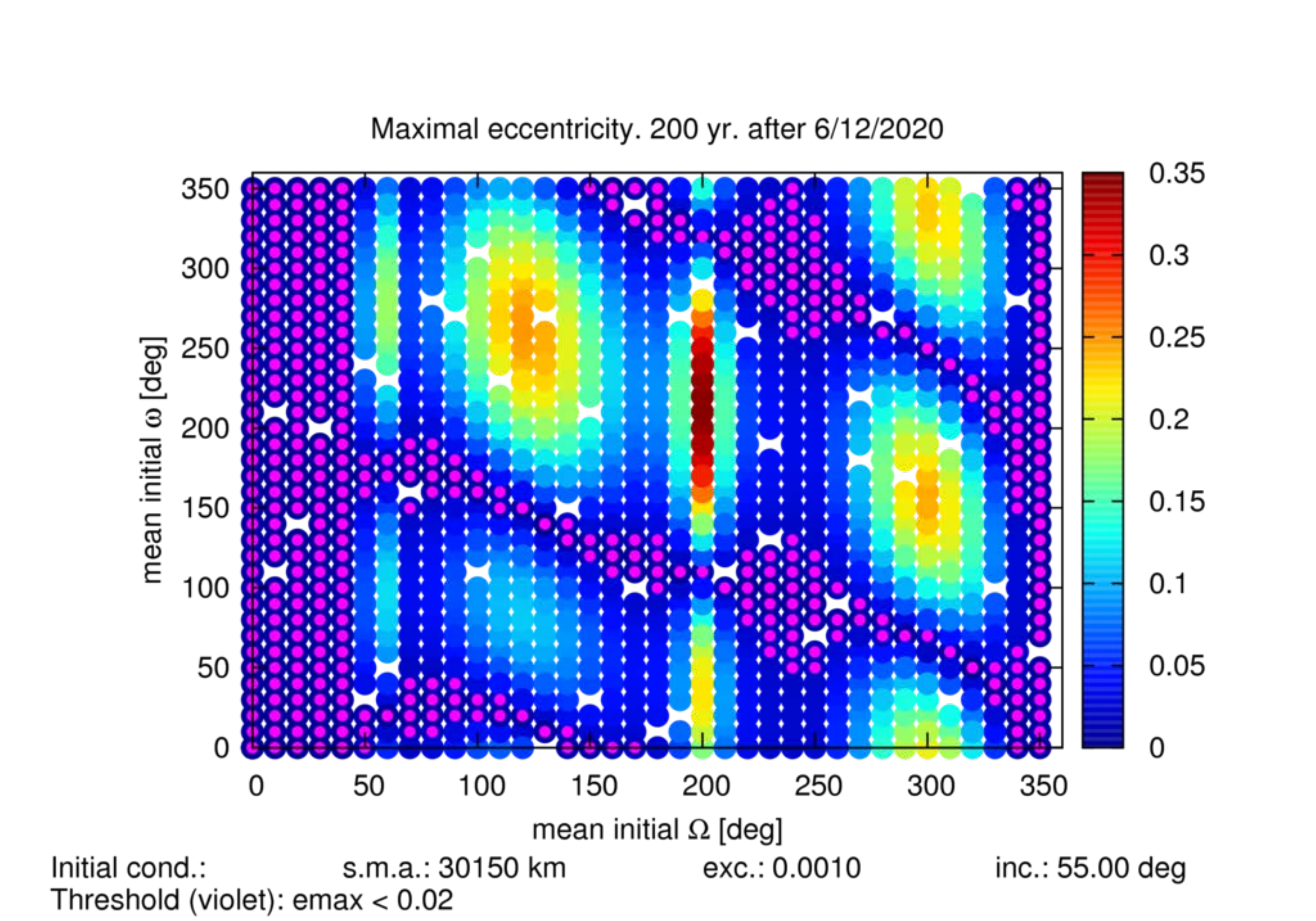} \hspace{-12pt}
	\includegraphics[scale=0.2]{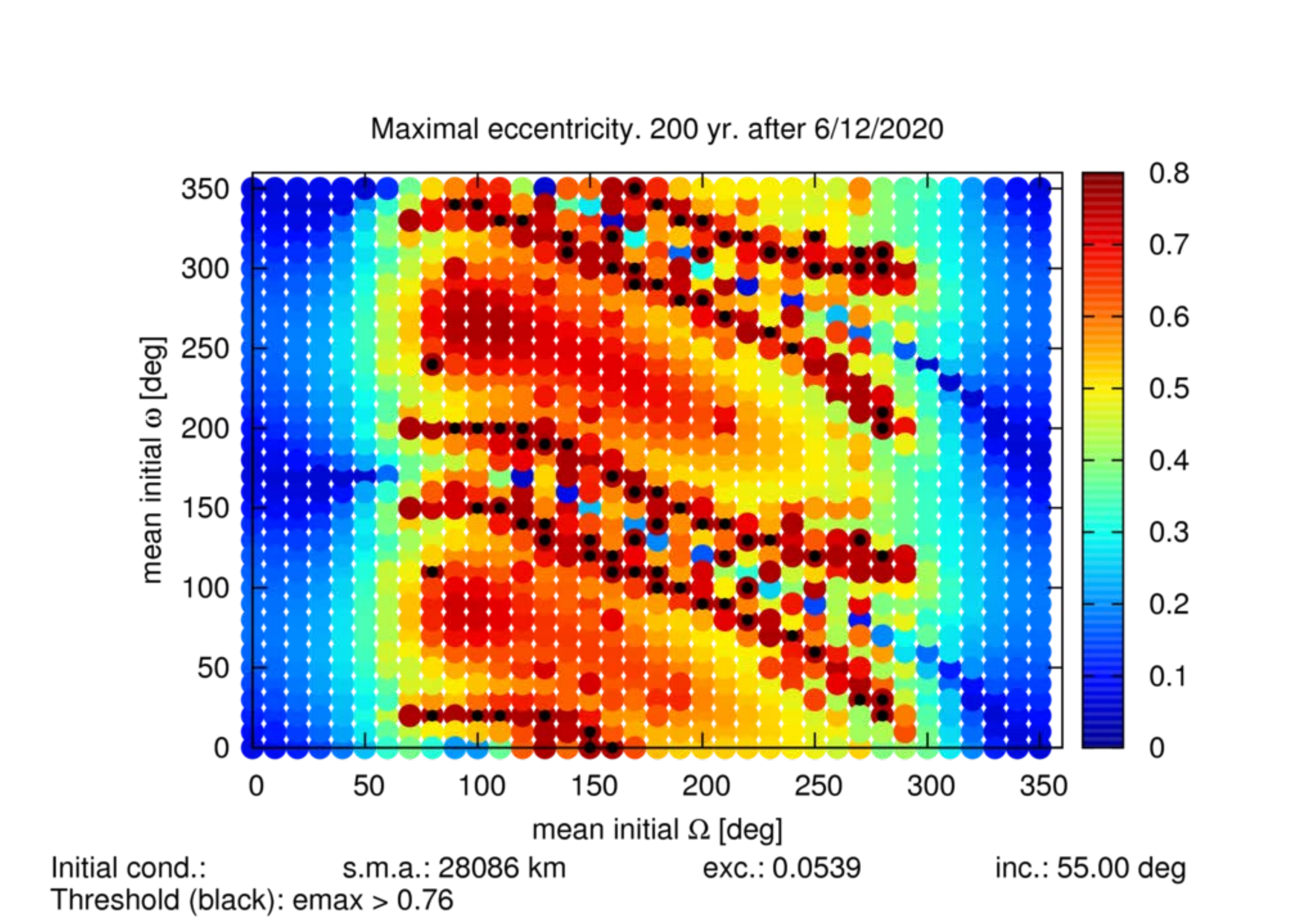} \\ \vspace{-12pt}
	\includegraphics[scale=0.2]{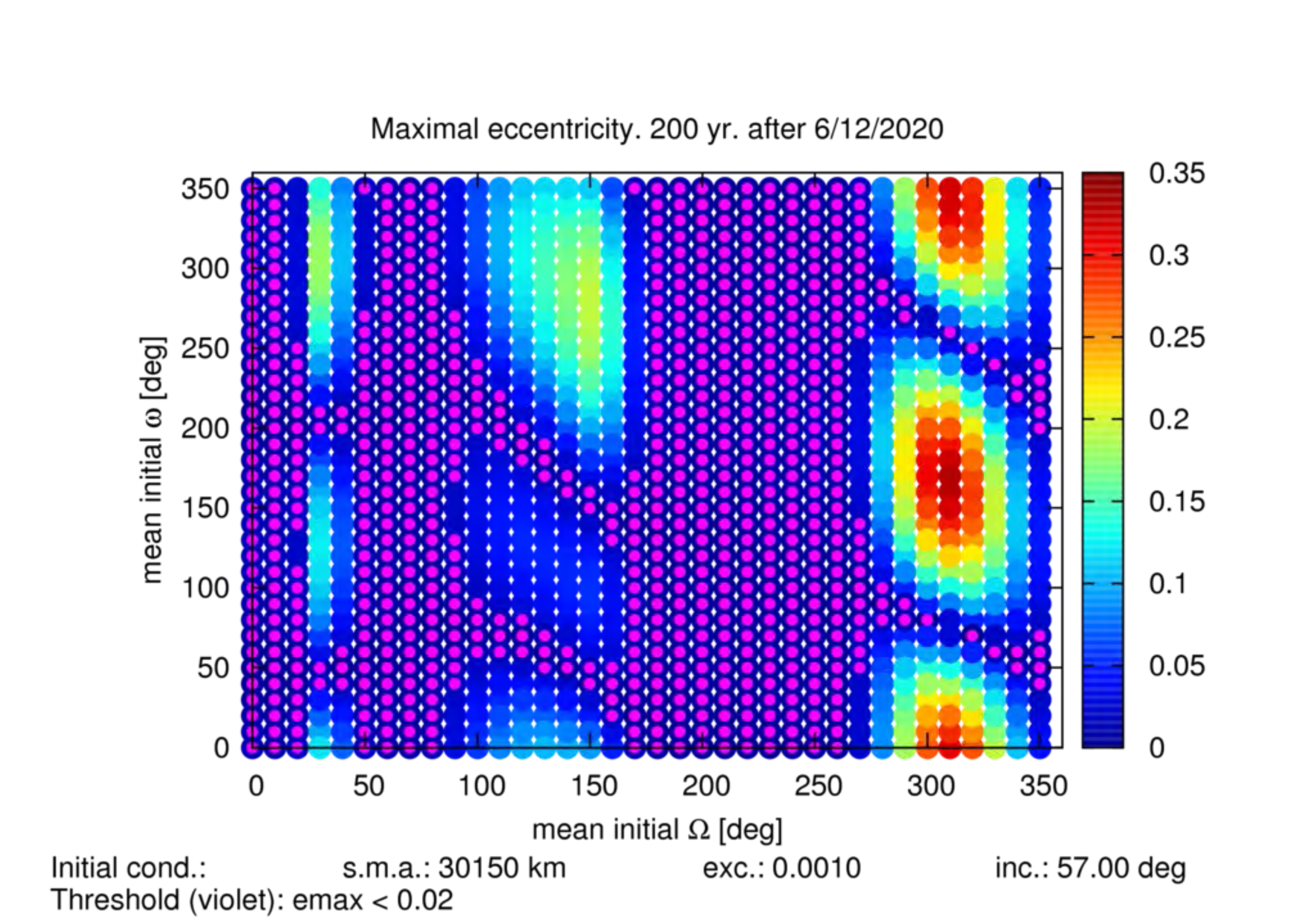} \hspace{-12pt}	
	\includegraphics[scale=0.2]{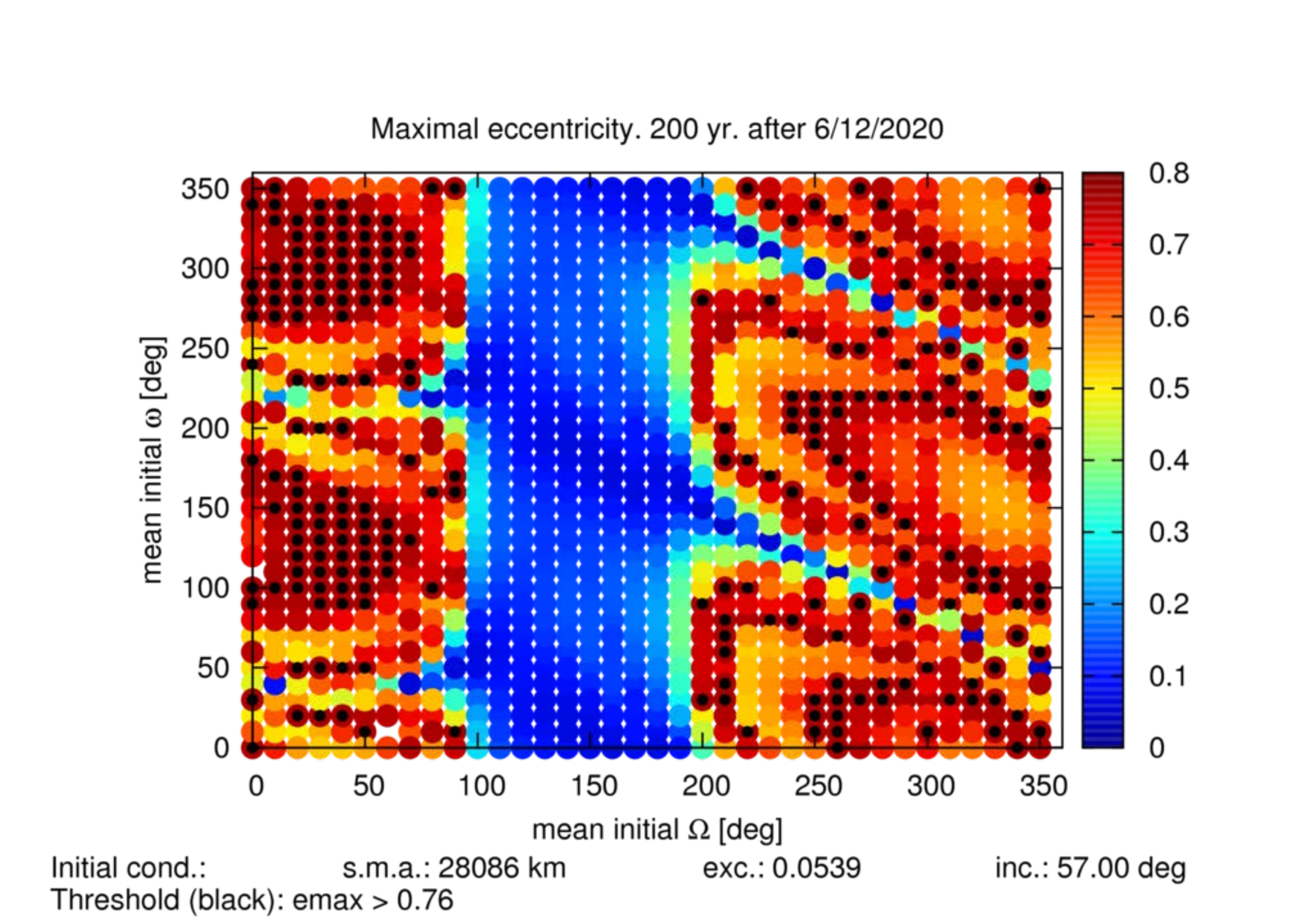}	
	\caption{The maximum eccentricity attained in 200 years (colourbar), as a function of the initial 
	longitude of ascending node and argument of perigee, at a given epoch, for the 
	graveyard orbit (left) and eccentricity growth (right) scenarios. Points that meet the various
	thresholds are indicated by violet ($e_{\max} < 0.02$) and black ($e_{\max} > 0.76$), respectively, 
	and the empty white spaces are locations where data is missing due to numerical issues.}
	\label{fig:ESA_sample}
\end{figure}

\begin{figure}
	\centering
	\includegraphics[scale=0.425]{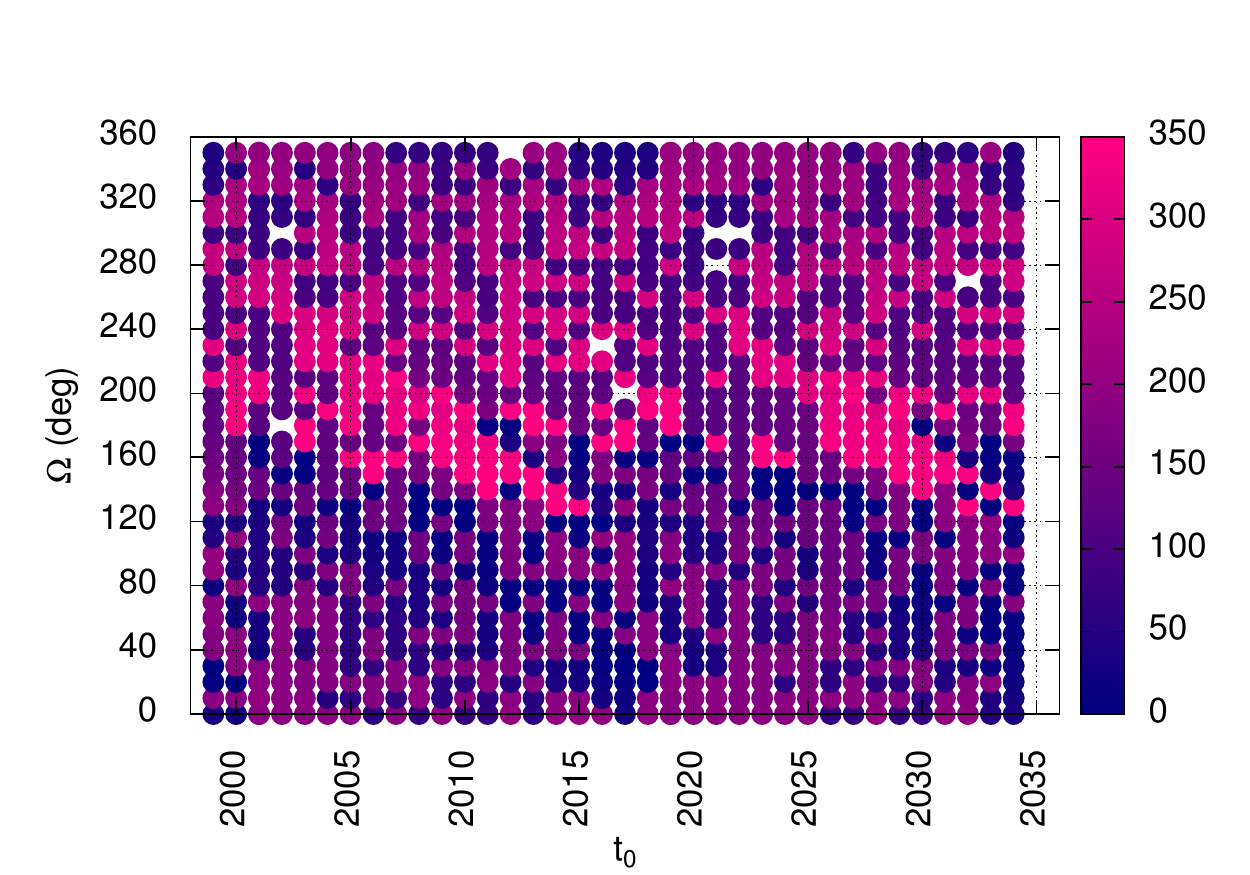}
	\includegraphics[scale=0.425]{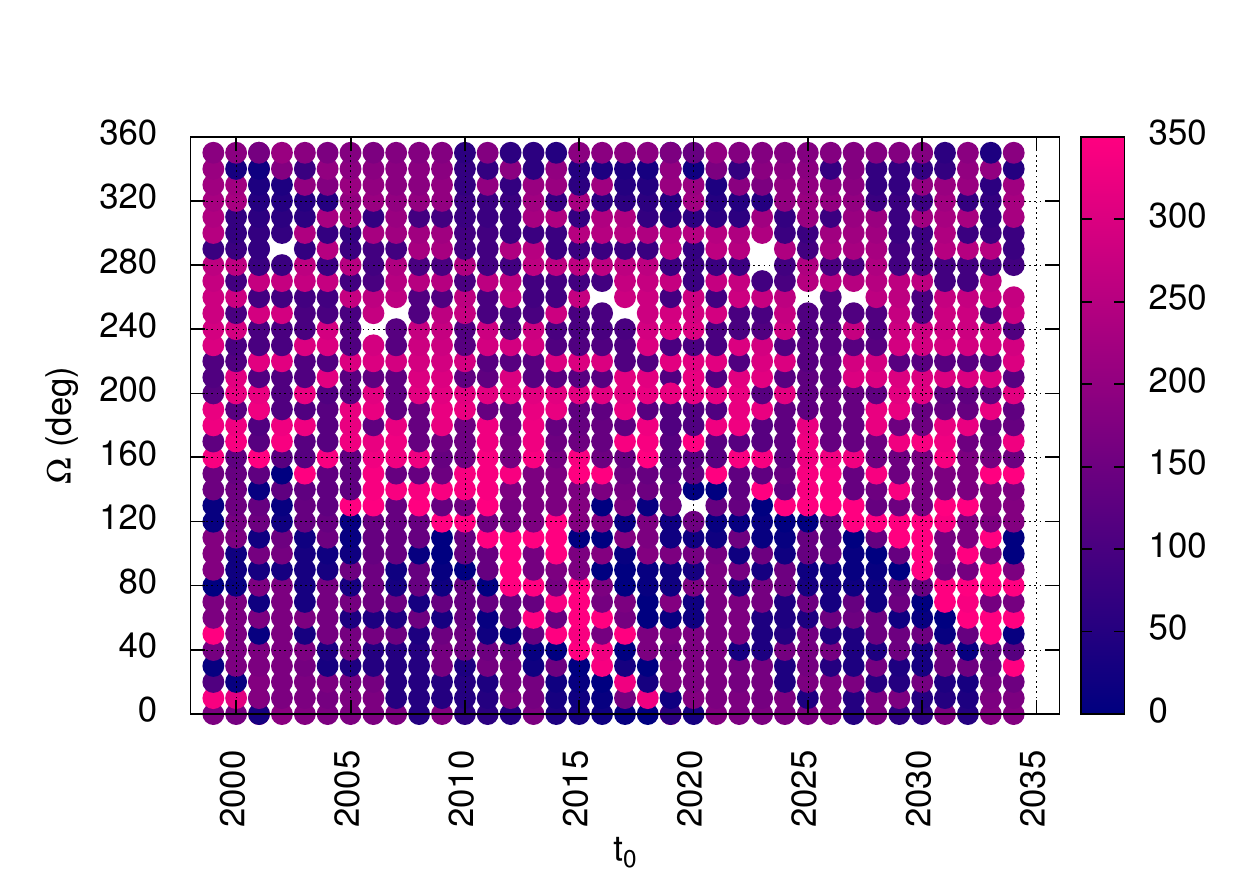}
	\includegraphics[scale=0.425]{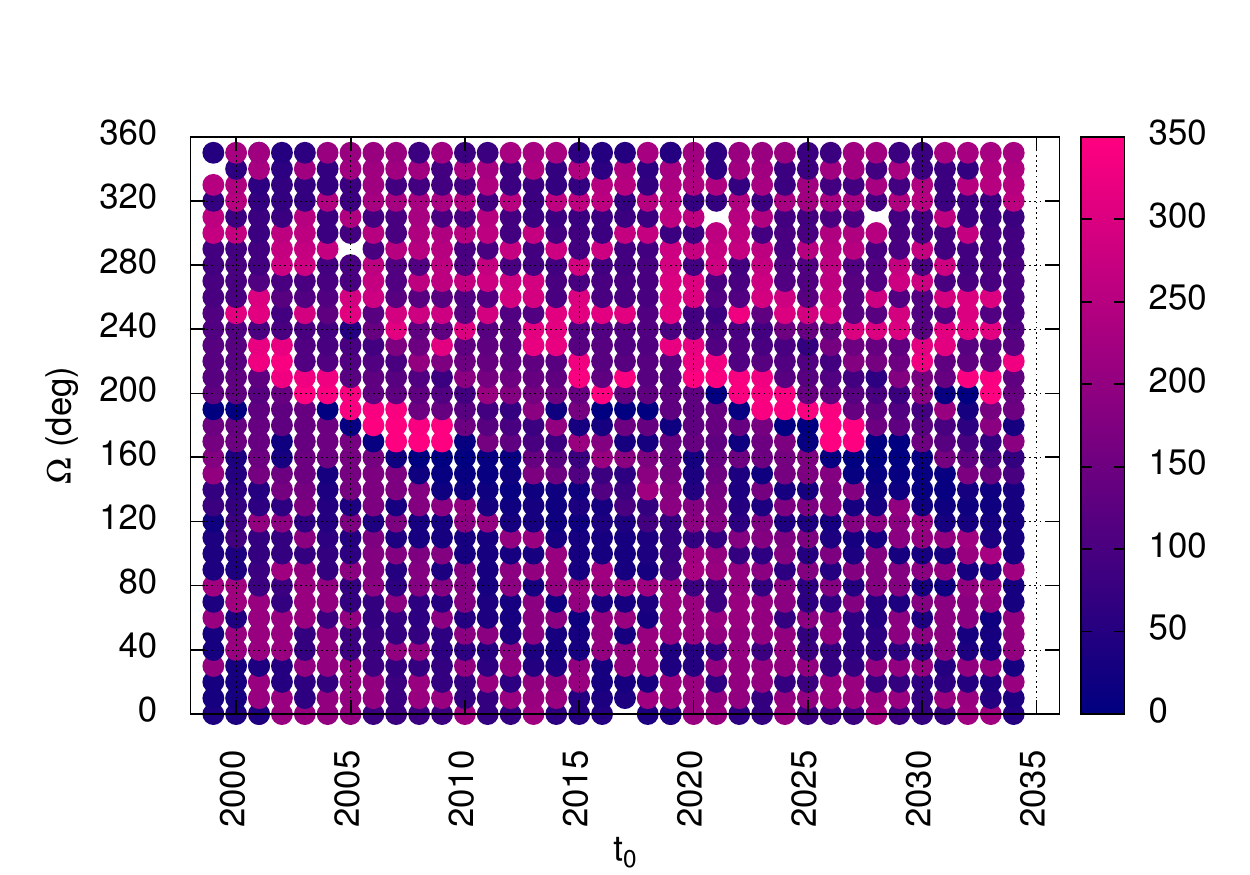} \\
	\includegraphics[scale=0.425]{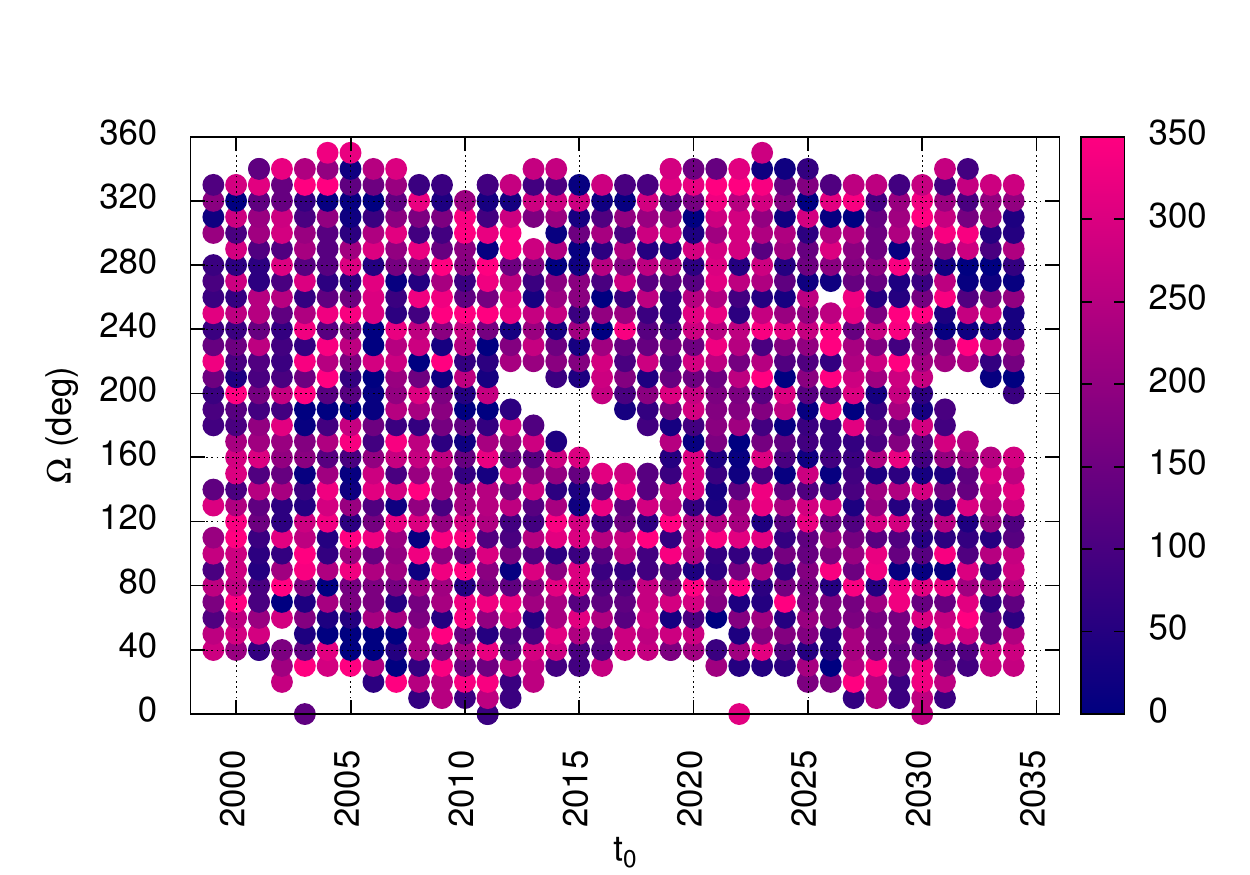}
	\includegraphics[scale=0.425]{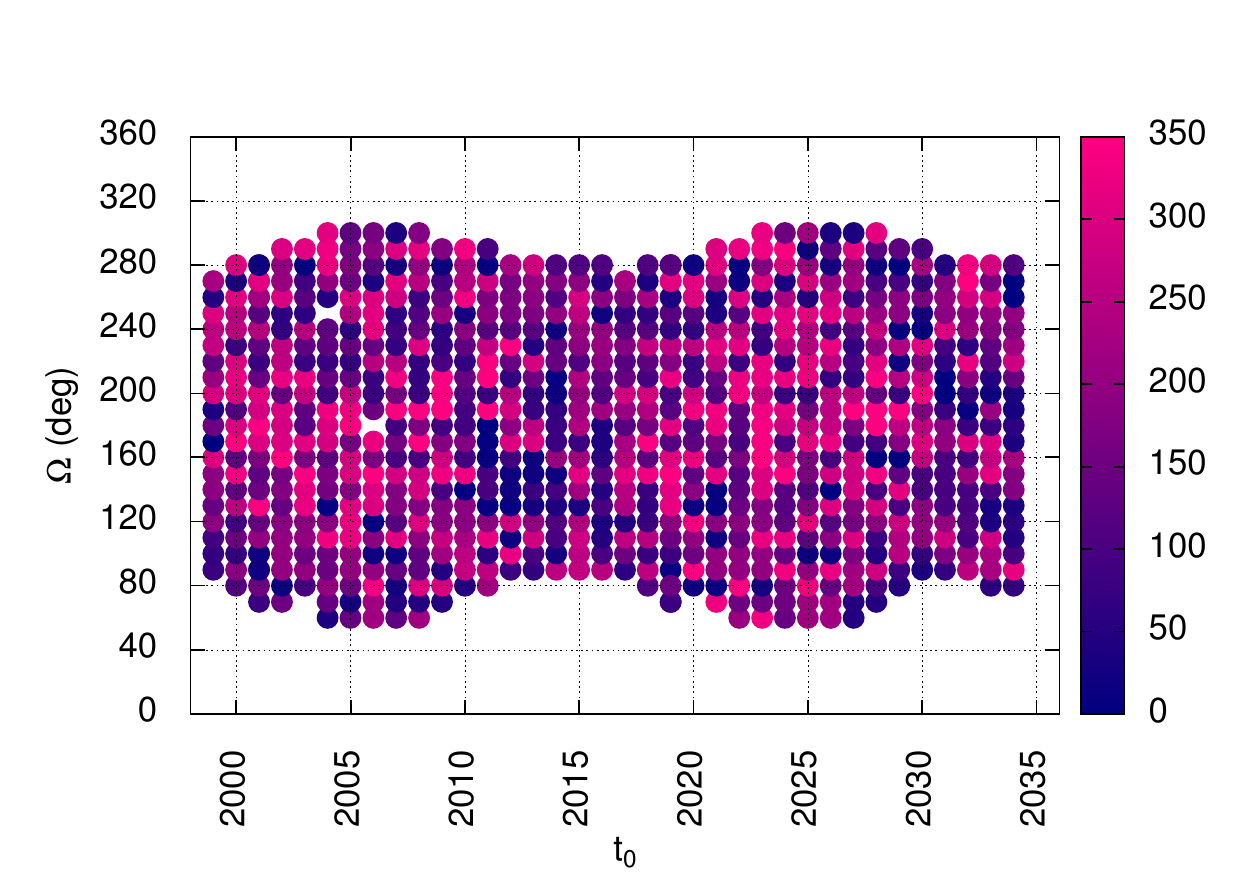}
	\includegraphics[scale=0.425]{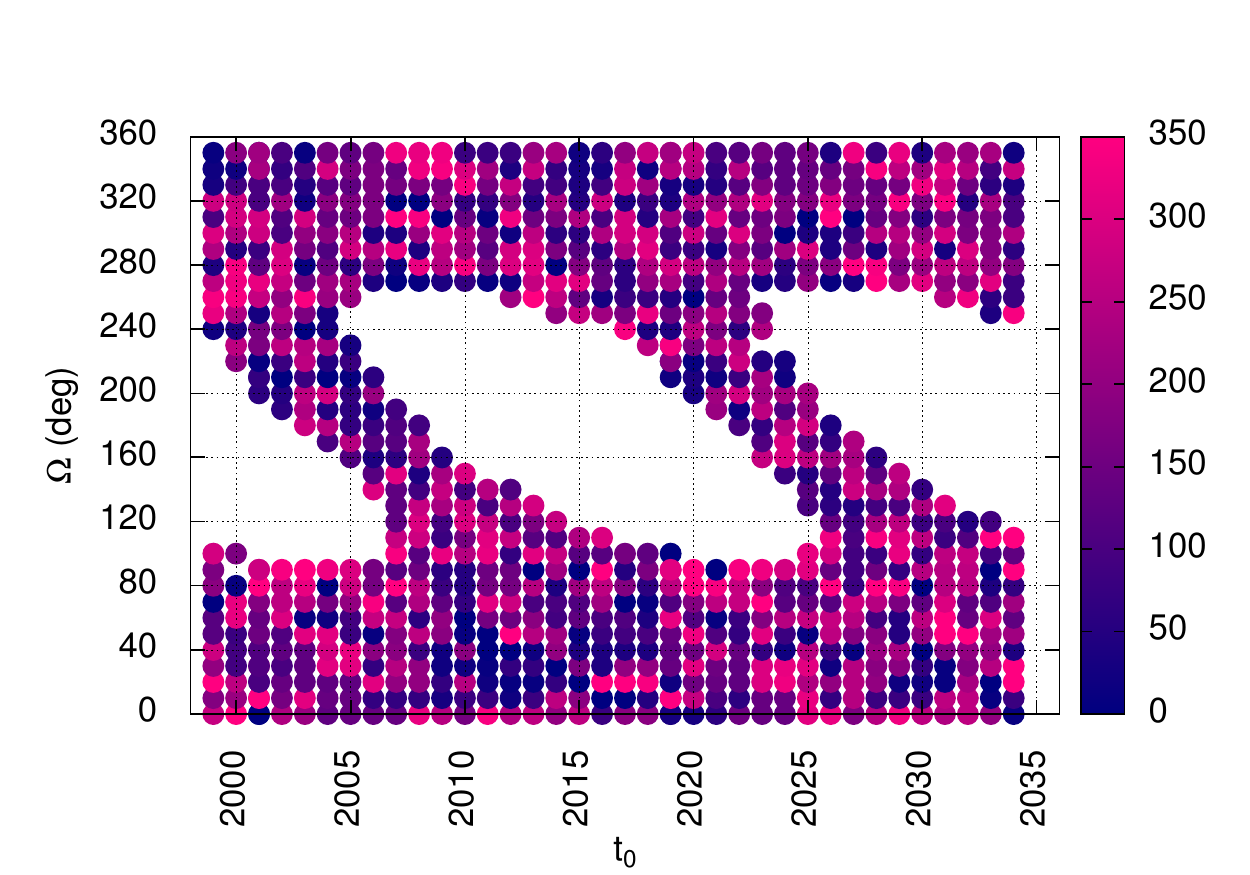} 	
	\caption{The $\omega$--targeting strategy: the value of argument of perigee (colourbar) which ensures 
	that the eccentricity will not exceed 0.02 in 200 years (top) or which ensures a re-entry (bottom), 
	as a function of the initial epoch and longitude of ascending node. 
	Left: nominal initial inclination; 
	middle: initial inclination decreased by $1^\circ$; 
	right: initial inclination increased by $1^\circ$.}
	\label{fig:omega_targeting}
\end{figure}

Figure~\ref{fig:ESA_sample} shows a sample of results from the numerical experiment, and Fig.~\ref{fig:omega_targeting} outlines an $\omega$--targeting strategy to achieve the desired outcome. Similar MEM maps were made for each eclipse year, and the variations in inclination and semi-major axis were tracked in addition to the eccentricity,\footnote{The inclination behaviour is presented in \citet{eA15} and will not be discussed here.} from which we can restate from \citet{eA15} the following general observations. The semi-major axis does not change significantly in 200 years (at most 70 km in absolute value) in any of the cases explored. Consequently, to avoid interferences with the operational constellation, the eccentricity should not exceed 0.02. The minimum eccentricity required to re-enter the atmosphere, assumed to occur whenever the altitude reaches at least 120 km, is about 0.76. 

For the {\itshape graveyard orbit} scenario, the eccentricity can reach about 0.4 for any of the considered initial inclinations, which depend in a complicated fashion on the various phase angles. Moreover, we note the vertical bands of stability (negligible eccentricity growth) in $(\Omega, \omega)$, Fig.~\ref{fig:ESA_sample}, which shift as a function of $t_0$ (not shown here). In general, we found that it was nearly always possible to target an argument of perigee ensuring `stability' (Fig.~\ref{fig:omega_targeting}); that is, for any given $(t_0, \Omega)$ there exists at least one initial $\omega$ corresponding to a safe disposal. The situation seems more favorable if the initial inclination is increased by $1^\circ$, in the sense that the stable vertical bands are wider. 

Concerning the {\itshape eccentricity growth} scenario, the eccentricity can increase up to 0.8, for the three initial values of inclination considered. In the nominal Galileo case, the eccentricity growth is remarkable in the entire $(t_0, \Omega, \omega)$ phase space; specifically, for any given epoch and ascending node, there exits always one (but generally more) initial $\omega$ leading to a re-entry (Fig.~\ref{fig:omega_targeting}). In the $-1^\circ$ case, re-entry values for $e$ can be achieved if $\Omega \in [50^\circ, 300^\circ]$, while in the $+1^\circ$ case, the $\Omega$ range depends on $t_0$. If the satellite's node does not match such values, then the eccentricity tends to stay below 0.1. Thus, while atmospheric re-entries were found to occur for the three cases, it requires at least 100 years.   

%  ****  Practical Implications of Chaos
\subsubsection{Practical implications of chaos}

\begin{figure}
  \centering
    \subfigure[Integrations of a nominally stable disposal orbit under model uncertainties. 
    The solid red line is the evolution from Model 4 in Table~\ref{tab:models}.]
    {\includegraphics[scale=0.55]{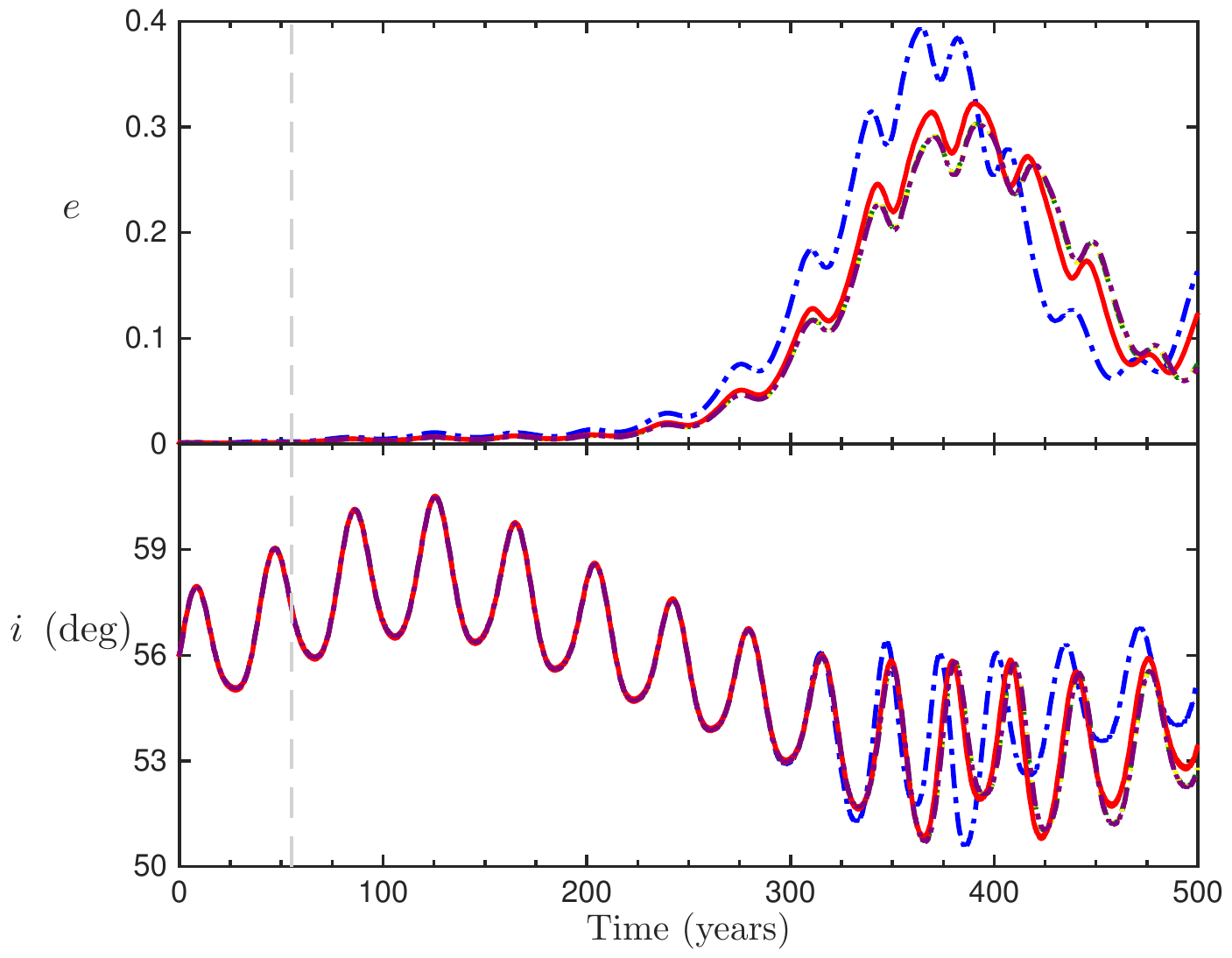} 
    \label{sfig:stable_models}} 
    \hspace{12pt}
    \subfigure[Integrations of the nominal re-entry disposal orbit under model uncertainties.
    The solid red line is the evolution from Model 4 in Table~\ref{tab:models}.]
    {\includegraphics[scale=0.55]{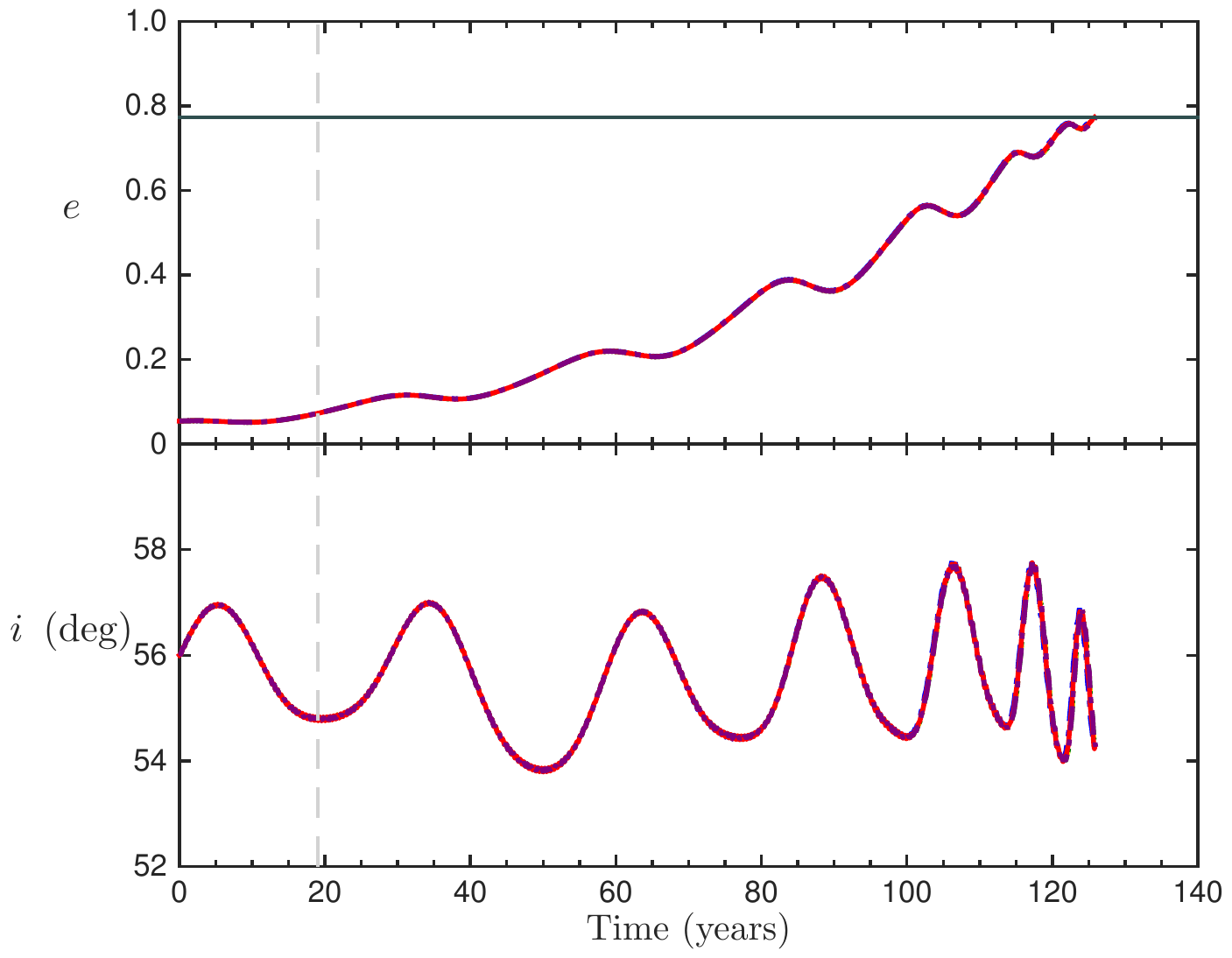}
    \label{sfig:unstable_models}}       		
	\subfigure[Integrations of 15 nearby orbits of the nominally stable case with random uncertainties
	of up to 0.1\% in $a_0$, $e_0$, $i_0$, $\Omega_0$, $\omega_0$, and $\Omega_M (t_0)$.
	The thick red line is the nominal orbit of \ref{sfig:stable_models}.] 
    {\includegraphics[scale=0.55]{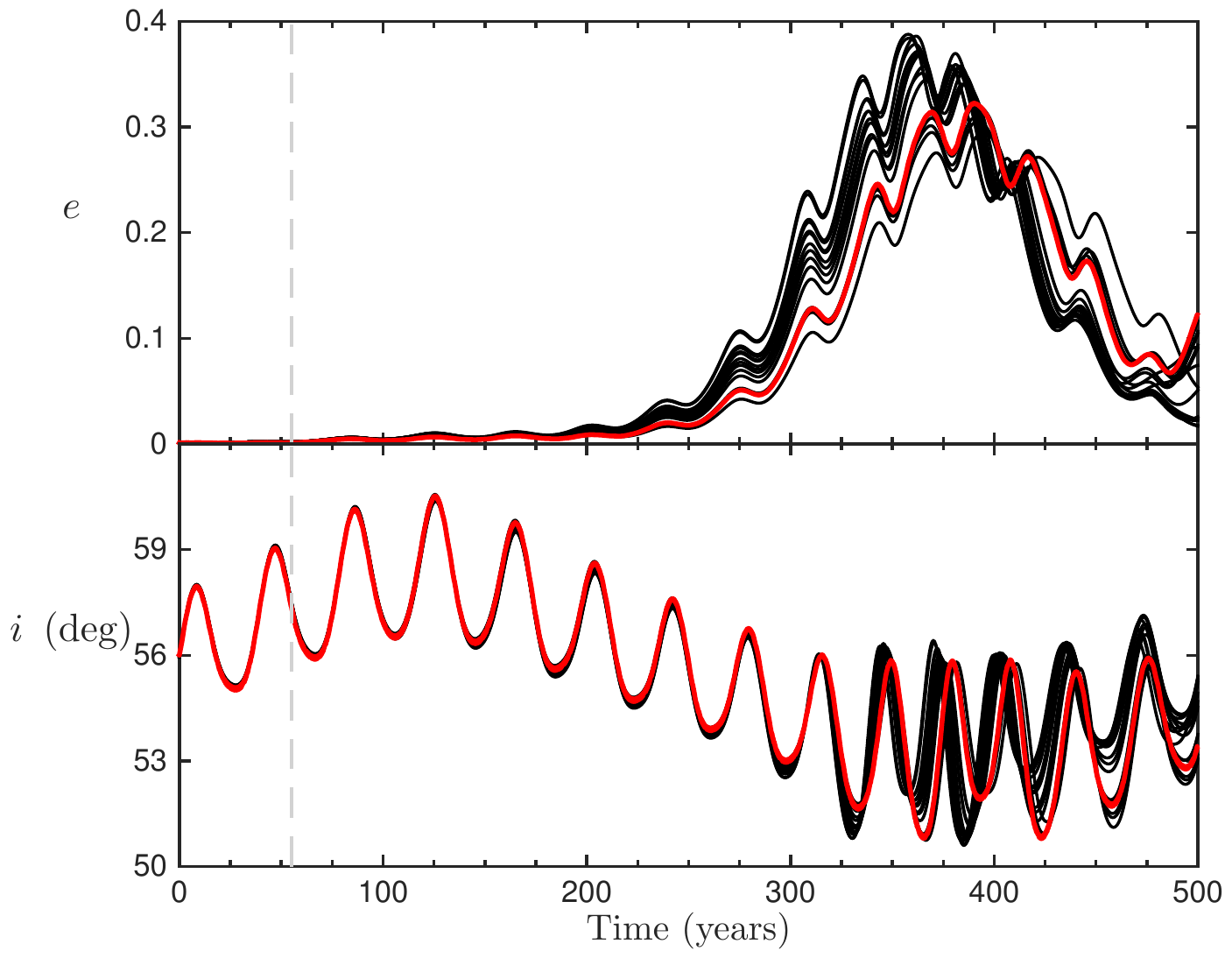}
    \label{sfig:stable_ICs}} 
    \hspace{12pt} 
	\subfigure[Integrations of 15 nearby orbits of the nominal re-entry case with random uncertainties 
	 of up to 0.1\% in $a_0$, $e_0$, $i_0$, $\Omega_0$, $\omega_0$, and $\Omega_M (t_0)$.
	The thick red line is the nominal orbit of \ref{sfig:unstable_models}.] 
    {\includegraphics[scale=0.55]{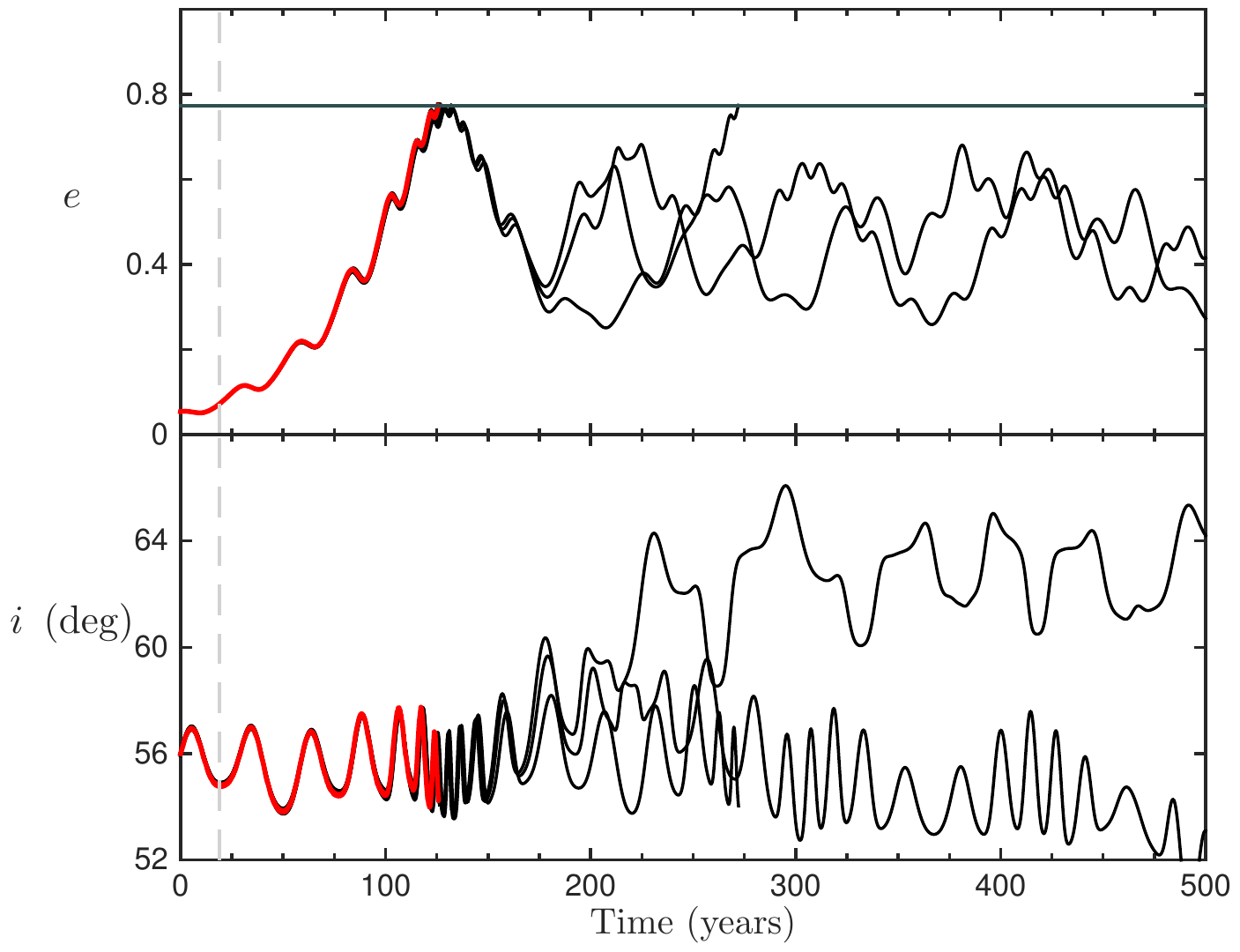}   
    \label{sfig:unstable_ICs}}              
  \caption{Numerical ensemble integrations according to the various dynamical models (top) in 
  	Table~\ref{tab:models} and of nearby orbits (bottom) for an apparently safe disposal ($a_0 = 30,150$ km, 
  	$e_0 = 0.001$, $i_0 = 56^\circ$, $\Omega_0 = 70^\circ$, $\omega_0 = 70^\circ$, epoch: 6 DEC 2020) 
	and re-entry orbit ($a_0 = 28\,086$ km, $e_0 = 0.0539$, $i_0 = 56^\circ$, $\Omega_0 = 60^\circ$, 
	$\omega_0 = 100^\circ$, epoch: 6 DEC 2020), where the nominal orbits were selected according to 
	Fig.~\ref{fig:ESA_sample}. The vertical lines indicate the Lyapunov times, corresponding to an average 
	limit of predictability of each orbit, and the horizontal line indicates the eccentricity value leading 
	to re-entry.} 
  \label{fig:chaos_implications}
\end{figure}

Any initial uncertainty in our knowledge of a chaotic system will have small consequences early but profound consequences late, often being rapidly amplified in time. While it is true that the verification of some criteria of stability to define the initial parameters of storage orbits requires long-term orbit propagation up to more than 100 years, most international recommendations and numerical studies seem fixated on 200-year forecasts. The 200-year timespan for future projections is not only arbitrary, but completely nonsensical from a dynamical perspective. Every distinct problem in orbital dynamics conditions its own particular scheme of computation, and the question of an appropriate timescale upon which to investigate cannot therefore be answered in a general manner; the answer depends largely on the problem in question and on the degree of knowledge aimed at. An improper assessment can lead to erroneous conclusions regarding stability and chaos. Consider, for example, one of the declared safe graveyard orbits of Fig.~\ref{fig:ESA_sample}, as shown in Fig.~\ref{fig:chaos_implications}. This orbit does not manifest any significant eccentricity growth for 200 years, and yet is revealed by our stability analysis (Section~\ref{sec:FLI}) to be chaotic with a Lyapunov time of 55 years. Alternatively, chaotic orbits which initially appear to re-enter (Fig.~\ref{sfig:unstable_models}) may follow evolutionary paths that lead to long-lasting eccentric orbits (Fig.~\ref{sfig:unstable_ICs}). Note that in Fig.~\ref{sfig:unstable_models}, the differences between the various models (i.e., mis-modelling effects) were too small to affect any appreciable change in the time-evolutions, over 126 years integration; yet, a 0.1\% change in the initial state can have a significant impact, as shown in Fig.~\ref{sfig:unstable_ICs}.

% -------------------------------------------------------------------------------------------------------------------------------------- 
%     RESONANCE OVERLAP
% -------------------------------------------------------------------------------------------------------------------------------------- 
\section{Resonance Overlap and the Origin of Chaos}

%  ****  Background
\subsection{Background}

Resonances are regions in the phase space of a dynamical system in which the frequencies of some angular variables become nearly commensurate. Such regions have a profound effect on the long-term dynamics of the system, giving rise to a rich spectrum of highly complicated behaviours \citep{aLmL92}. It is of great practical importance to understand the mechanisms behind these features, both qualitatively and quantitatively. Recently, it has been realised that lunisolar secular resonances (i.e., those caused by the Moon and the Sun on long timescales) are of particular importance in the medium-Earth orbit regime \citep{aR15,jDaR15}. We review in this section our investigations on the detection of regular structures and chaotic behaviours in the phase space near the navigation satellites. Studying the long-term effects of lunisolar secular resonances is crucial, not only because we need to understand their stability properties, but also because we would like to know whether they could be used (and how) for eventually deorbiting satellites, by forcing them to slowly drift towards high eccentricities and different inclinations.    

Despite the variety and complexity of the nature of the dynamics near resonances, we can build an initial intuitive understanding using the mechanics of a pendulum. Pendulum-like behaviour is fundamental to the mathematics of resonance: phase-space structure, separatrices of a periodic motion, and stability \citep{nM01}. The principal effect of the interaction of two resonances is to produce qualitative changes in the separatrix of the perturbed resonance, producing a stochastic layer in its vicinity. The onset of deterministic chaos and the loss of stability is predicted to occur when the separation between the resonances is of the order of their resonance widths \citep{aLmL92}. Nearly all chaos in the Solar System and beyond has been attributed to the overlapping of resonances \citep{aM02}.\footnote{Note that while this is the main physical mechanism for the generation of chaos, two overlapping resonances may lead to regular motion sometimes; see, e.g., \citet{jW86}.}

%  ****  Lunisolar Resonant Skeleton
\subsection{Lunisolar resonant skeleton}

Focusing on the MEO region located between three and five Earth radii, namely in a region for which the variation of the argument of perigee $\omega$ and longitude of ascending node $\Omega$ may be estimated by considering only the effect of $J_2$ (the second zonal harmonic coefficient of the geopotential) and for which the lunar and solar potentials may be approximated with sufficient accuracy by quadrupole fields, the centre of each lunisolar secular resonances (for prograde orbits) may be defined in the inclination--eccentricity ($i$--$e$) phase space by the curves \citep{aR15,jDaR15}
\begin{equation}
	\label{eq:res_center}
	\mathcal{C}_{\bm n} = \left\{ (i, e) \in [0, \frac{\pi}{2}] \times [0, 1] \ : \ 
		\dot\psi_{\bm n} = n_1 \dot\omega + n_2 \dot\Omega + n_3 \dot\Omega_\M = 0 \right\}, 
\end{equation}
for integer coefficients $\ n_1 = \big\{ {-2}, 0 , 2 \big\}$, $n_2 = \big\{ 0, 1, 2 \big\}$, $n_3 \in \llbracket {-2}, 2 \rrbracket$ (not all zero), where
\begin{equation}
	\dot\omega (i, e) = \frac{3}{4} \frac{J_2 R^2 \sqrt{\mu}}{a^{7/2}} \frac{5 \cos^2 i - 1}{(1 - e^2)^2}, \quad
	\dot\Omega (i, e) = -\frac{3}{2} \frac{J_2 R^2 \sqrt{\mu}}{a^{7/2}}  \frac{\cos i}{(1 - e^2)^2}, \quad
	\dot\Omega_\M = -0.053^\circ/\rmn{day}. 
\end{equation}
Here the semi-major axis $a$ is a parameter,\footnote{The lunar and solar perturbation parameters are proportional to $a$ as $\varepsilon_\M = \varepsilon_\M (a/a_\M)$ and $\varepsilon_\S = \varepsilon_\S (a/a_\S)$; see, for example, \citet{aCaR15}.} $R$ is the mean equatorial radius of the Earth, and $\mu$ its gravitational parameter. Using the full machinery for pendulums, it can be shown that the curves delimiting the maximum separatrix width of each resonance (i.e., the maximum amplitude inside the libration zone, when each resonance is treated in isolation) are defined by \citep{jDaR15}  
\begin{equation}
	\label{eq:res_width}
	\mathcal{W}_{\bm n}^\pm \equiv \left\{ (i, e) \in [0, \frac{\pi}{2}] \times [0, 1] \ : \ 
		\dot\psi_{\bm n} = \pm \Delta_{\bm n} \right\}, 
\end{equation}
in which
\begin{equation}
	\label{eq:half_width}
	\Delta_{\bm n} = 2 \sqrt{ \frac{3}{2} \frac{J_2 R^2}{a^4}
		\left| \frac{n_1^2 \left( 2 - 15 \cos^2 i_\star \right) 
		+ 10 n_1 n_2 \cos i_\star - n_2^2}{(1 - e_\star^2)^{5/2}} h_{\bm n} (i_\star, e_\star) \right|}, 
\end{equation}
where $h_{\bm n}$ is the harmonic coefficient in the lunar and solar disturbing function expansions, associated with the harmonic angle which is in resonance,\footnote{Explicit expressions for $h_{\bm n}$ for each of the 31 distinct curves of secular resonances are given in \citet{jDaR15}.} and $(i_\star, e_\star)$ are the `actions' at exact resonance; namely, the inclinations and eccentricities that satisfy equation~(\ref{eq:res_center}).

\begin{figure}
	\centering
	\includegraphics[scale=0.75]{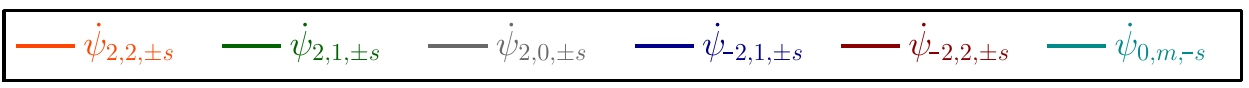} \\[0.15em]
	\includegraphics[scale=1]{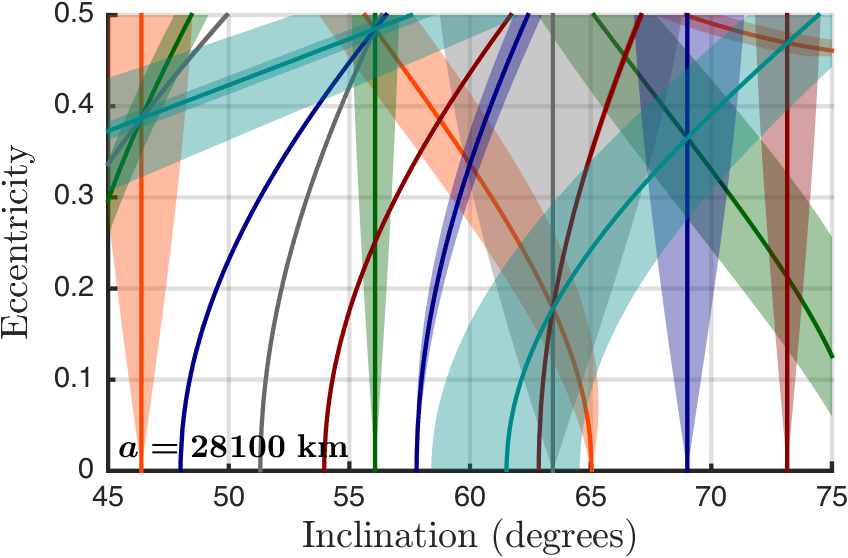}	
	\includegraphics[scale=1]{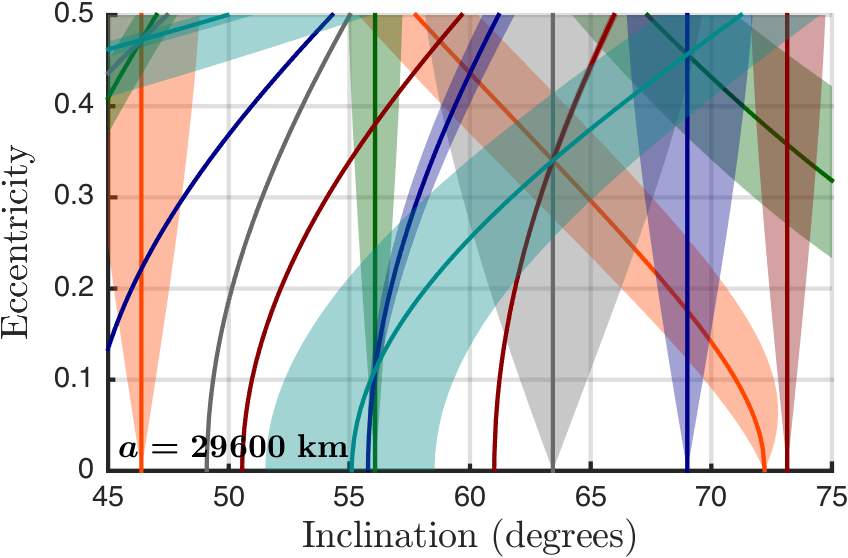}	
	\includegraphics[scale=1]{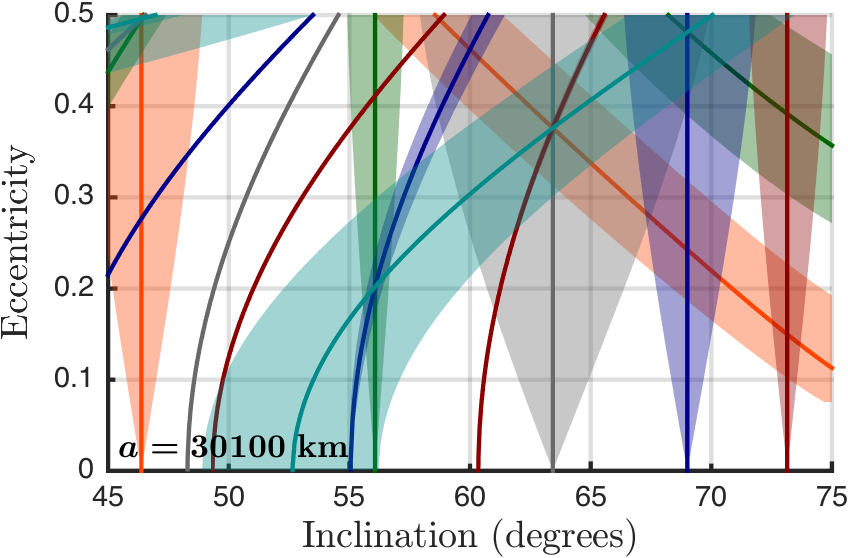}	
	\caption{Lunisolar resonance centres $\mathcal{C}_{\bm n}$ (solid lines) and 
	widths $\mathcal{W}_{\bm n}^\pm$ (transparent shapes) for various values of the 
	satellite's semi-major axis near Galileo. This plot shows the regions of overlap between 
	distinct resonant harmonics. (Adapted from \citet{jDaR15}, to which we refer for the omitted details.)} 
	\label{fig:apertures}
\end{figure}

Figure~\ref{fig:apertures} shows that resonances fill the phase space near the Galileo constellation. These resonances form in some sense the skeleton or dynamical backbone, organising and governing the long-term orbital motion. The resulting dynamics can be quite complex, and it has been shown that chaos ensues where resonances overlap \citep{aR15,jDaR15,aCaR15}. When such overlapping occurs, only the central part of the resonances, near their elliptic fixed points, might be expected to host regular motion. Chaotic motion can also exist in the vicinity of the perturbed separatrices of isolated resonances, but the absence of overlapping generally guarantees the local confinement of the motion \citep{jDaR15,aM02}. It is particularly noteworthy that the nominal inclination of Galileo lies right at the cusp of three distinct and dynamically significant resonant harmonics. Such naivety in the placement of these important assists reflects the need of a real dynamical assessment in constellation design. What is more to the point is that the conclusions drawn from the computationally expensive parametric study of Section~\ref{sec:parametric} are easily corroborated here. In the graveyard orbit scenario, increasing the inclination by $1^\circ$ moves the storage orbits outside of the overlapping regime, and thus we would naturally expect this to be the more dynamically stable case. For the eccentricity growth scenario, the nominal Galileo case is the more unstable situation because the orbits lie at the primary $\dot\psi_{2,1,0}$ resonance, while the instabilities in the other inclination cases are likely due to the generation of secondary resonances (commensurabilities of the libration and circulation frequencies of primary resonances) that expand the size of the chaotic zones about the $\dot\psi_{2,1,0}$ resonance. Rather ironically, the targeting of such a low semi-major axis for this disposal strategy appears inappropriate, as keeping the constellation at or near the Galileo semi-major axis would have resulted in greater instabilities with the interaction of the three distinct primary resonances. This basic understanding reached, using pen-and-paper calculations in the manner of Lagrange and Laplace, is a strong testimony to the enduring power of analytical theories in celestial mechanics. 

%  ****  Disposal Criteria Based on Resonant Fixed Points
\subsection{Disposal criteria based on resonant fixed points}

Orbital resonances can be a source of both chaos and stability, the nature of the dynamics depending sensitively upon the initial orientation angles of the satellite and the initial lunar node. The asteroid and trans-Neptunian belts of small bodies offer an abundance of instructive examples that illustrate how small zones of stability can persist in the vicinity of resonances, of which the most famous is Pluto \citep{aM92,rM95}. Pluto's orbit is chaotic with a Lyapunov time of about 20 Myr, yet it remains macroscopically stable over billion-year timescales, in the sense that the action variables do not show significant changes. For Pluto, the Kozai-Lidov effect occurs embedded inside a mean-motion resonance with Neptune, giving rise to an argument-of-perihelion libration and a libration of a 3:2 resonant angle, which provide a dynamical protection mechanism against close encounters \citep{rM95}. 

The salient feature of a resonance is the existence of an elliptic fixed point, with regular phase space trajectories encircling it, and of hyperbolic fixed points, connected by a separatrix trajectory. As chaos first develops around the hyperbolic equilibria and separatrices of resonances, their identification (in the pendulum model) will provide a natural definition for the eccentricity growth disposal scenario. Conversely, the elliptic fixed points of the resonances would represent stable phase-space regions for the definition of the graveyard disposal orbits. It must be emphasised that all of the foregoing results and statements hold good only when the resonances admit pendulum-like structures in the phase space; that is, when the associated Hamiltonians can be reduced to the First Fundamental Model of resonance \citep{sB03}. Setting this caveat aside, the small stability islands can also be completely destroyed, depending on the strength of the interaction between resonances.

There are three principal resonances affecting Galileo-like orbits and their disposal regions: $\dot\psi_{2,1,0} = 2 \dot\omega + \dot\Omega \approx 0$, $\dot\psi_{-2,1,-1} = -2\dot\omega + \dot\Omega - \dot\Omega_\M \approx 0$, and $\dot\psi_{0,2,-1} = 2 \dot\Omega - \dot\Omega_\M \approx 0$. Their associated harmonic coefficients and fixed points are given in Tables~\ref{tab:fixed_points}, and follow from the pendulum reduction of \citet{jDaR15}. These elliptic and hyperbolic equilibria lead to simple criteria for the definition of the initial parameters of the disposal orbits: the problem thus reduces to a trivial resonance phase matching scheme, as will be discussed in detail in Section~\ref{sec:DispCrit}. 

\begin{table}
	\caption{Lunisolar resonance conditions, harmonic coefficients, and equilibria. Here we use the abbreviations $\rmn{s} = \sin$, $\rmn{c} = \cos$, $C = \cos (\epsilon/2)$, and $S = \sin (\epsilon/2)$.}
	\label{tab:fixed_points} 
	\vspace{6pt}
	\centering
	\tabulinesep=0.6em 
	\setlength{\tabcolsep}{5pt}
	\begin{tabu}{cccc} \Xhline{3\arrayrulewidth}
    \multicolumn{1}{c}{$\dot\psi_{\bm n}$} &            
    \multicolumn{1}{c}{\textsc{Harmonic Coefficient}, $h_{\bm n}$} &
    \multicolumn{1}{c}{\textsc{Elliptic}} &
	\multicolumn{1}{c}{\textsc{Hyperbolic}} \\\Xhline{3\arrayrulewidth}
		\rowfont{\color{Green4}} 
    		$\dot\psi_{2, 1, 0}$ 
		& $\ds \frac{15 a^2}{16} \left[ \frac{\mu_\M (2 - 3 \rmn{s}^2 i_\M) C S^{-1} (-2 C^4 + 3 C^2 - 1)}
		{a_\M^3 (1 - e_\M^2)^{3/2}} -\frac{\mu_S \rmn{s} i_S \rmn{c} i_S}
		{a_S^3 (1 - e_S^2)^{3/2}} \right] e^2 \rmn{s} i (1 + \rmn{c} i)$
		& $2 \omega + \Omega = \pm \pi$ & 
		$2 \omega + \Omega = 0$ \\
		\rowfont{\color{Blue4}}      
   		$\dot\psi_{^{\_}2, 1, \!^{\_}1}$ 	
    		& $\ds \frac{15 \mu_\M a^2 \rmn{s} i_\M \rmn{c} i_\M C^2 (4 C^2 - 3)}
		{16 a_\M^3 (1 - e_\M^2)^{3/2}} e^2 \rmn{s} i (1 - \rmn{c} i)$
		& $-2 \omega + \Omega - \Omega_\M = \pm \pi$ 
		& $-2 \omega + \Omega - \Omega_\M = 0$ \\	
		\rowfont{\color{Cyan4}}
		$\dot\psi_{0, 2, \!^{\_}1}$ 
    		& $\ds \frac{3 \mu_\M a^2 \rmn{s} i_\M \rmn{c} i_\M C^3 S^{-1} (C^2 - 1)}
		{8 a_\M^3 (1 - e_\M^2)^{3/2}} (2 + 3 e^2) \rmn{s}^2 i$ 
		& $2 \Omega - \Omega_\M = 0$ 
		& $2 \Omega - \Omega_\M = \pm \pi$ \\\Xhline{3\arrayrulewidth}			
	\end{tabu}
\end{table} 

\section{Numerical Stability Analysis}
\label{sec:FLI}

Figure~\ref{fig:apertures} gives a crude, global picture of the basic regions in the 2D inclination--eccentricity phase space for which chaotic orbits can be found, but gives no information about which initial angles ($\omega$, $\Omega$, and $\Omega_\M$) will lead to chaos, nor does it account for secondary resonances that arise from the nonlinear beating between primary resonances. For this, we turn to the numerical detection of chaotic and regular motion through FLI stability and Lyapunov time maps, which furthermore provide both a global and local visualisation of the curious symbiosis of these two fundamental types of behaviours. 

It was shown in \citet{jDaR15} that model 1 in Table~\ref{tab:models} captures, qualitatively and quantitatively, all of the dynamical structures revealed by the more realistic and more complicated models. We cannot show here how abundant and fruitful the consequences of this realisation have proved. The application of this basic physical model leads to simple and convincing explanations of many facts previously incoherent and misunderstood. Here we tailor the recent results of \citet{jDaR15}, to which we refer for omitted details, to the evaluation of the proposed disposal strategies of \citet{eA15} and of our new disposal criteria based on use of the resonance equilibria.  

\begin{figure}
  \centering
    \subfigure[FLI map.]
    {\includegraphics[scale=0.24]{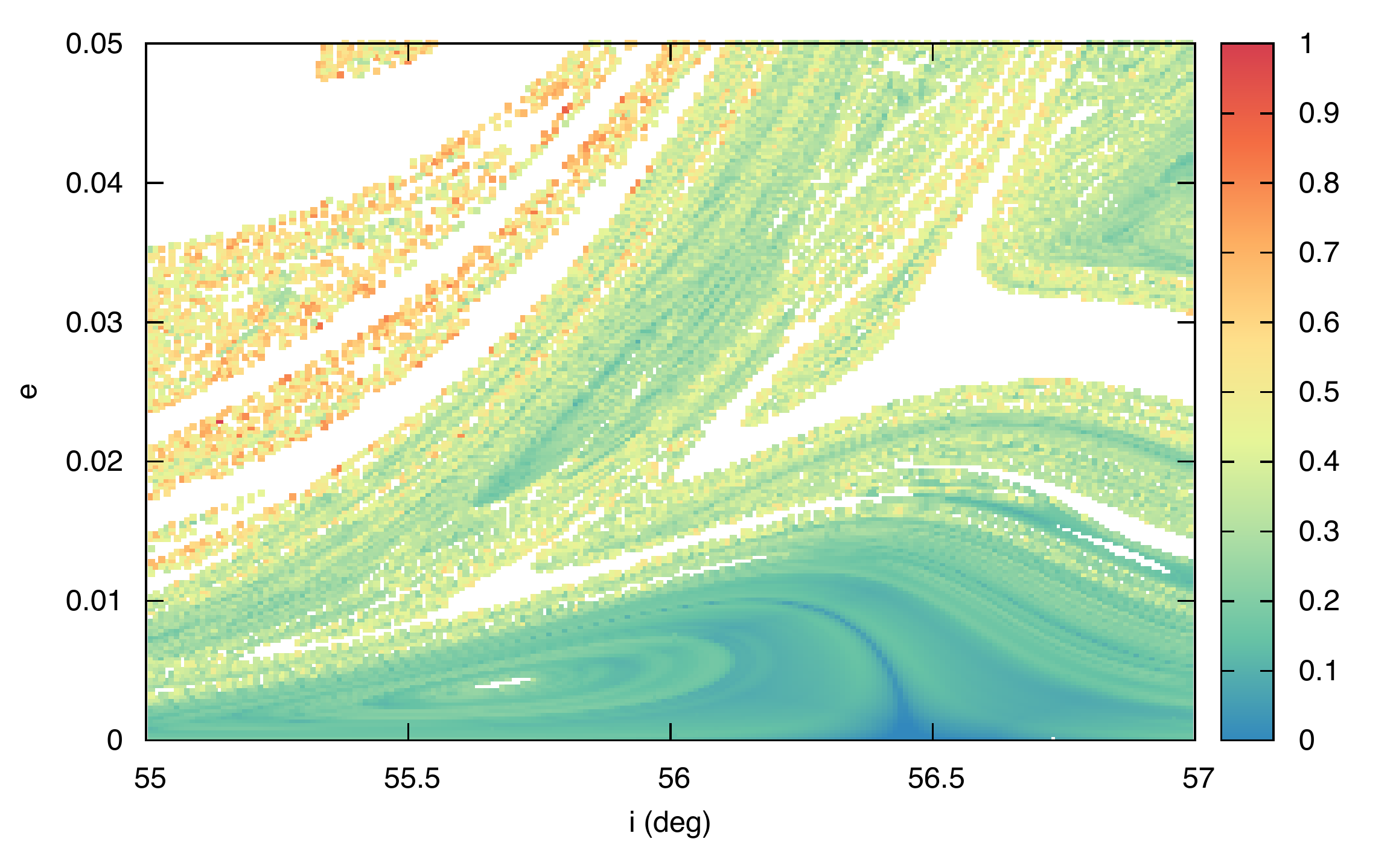}} 
    \subfigure[Lyapunov time map.]
    {\includegraphics[scale=0.24]{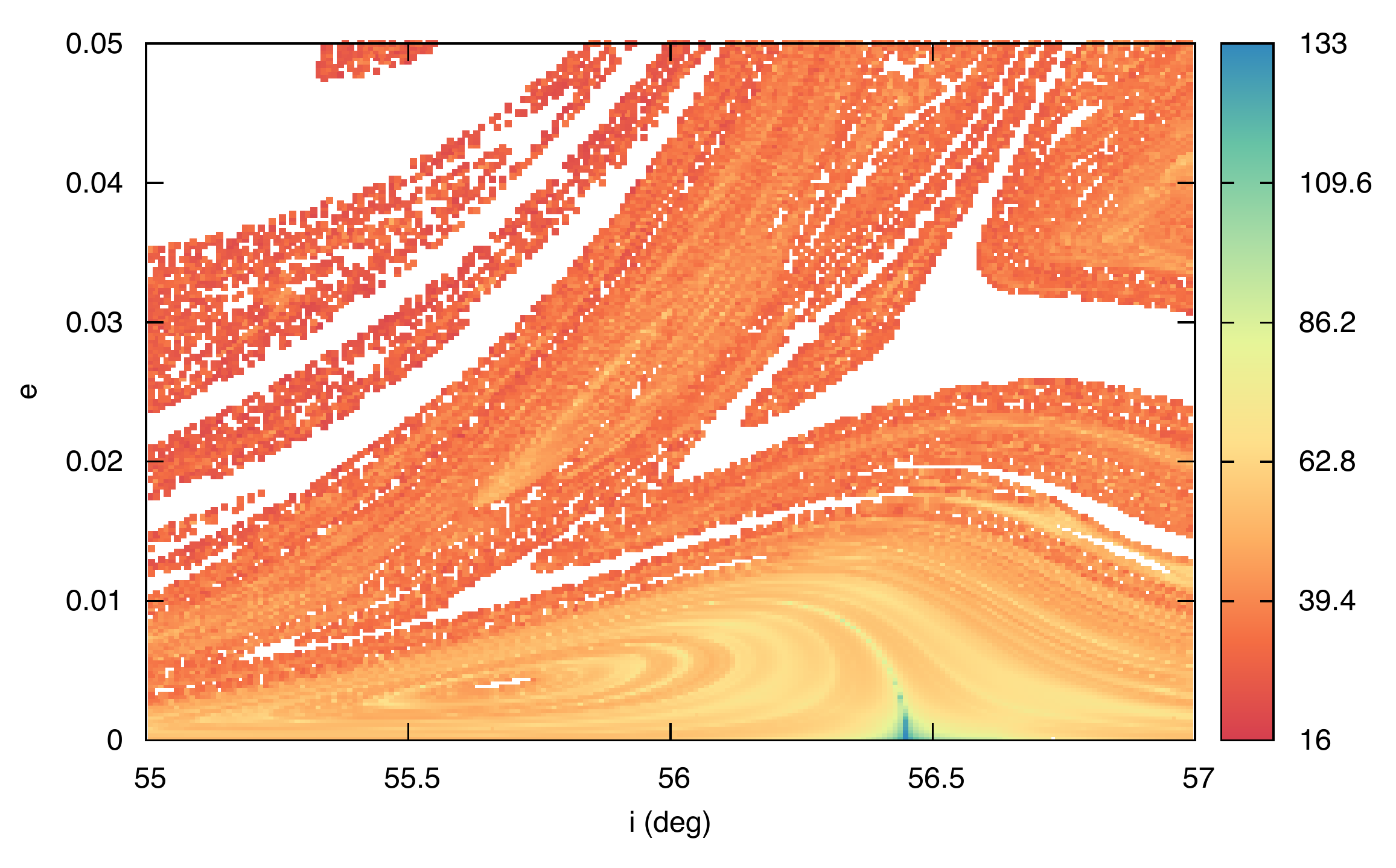}} 
    \subfigure[Collision time map.]
    {\includegraphics[scale=0.24]{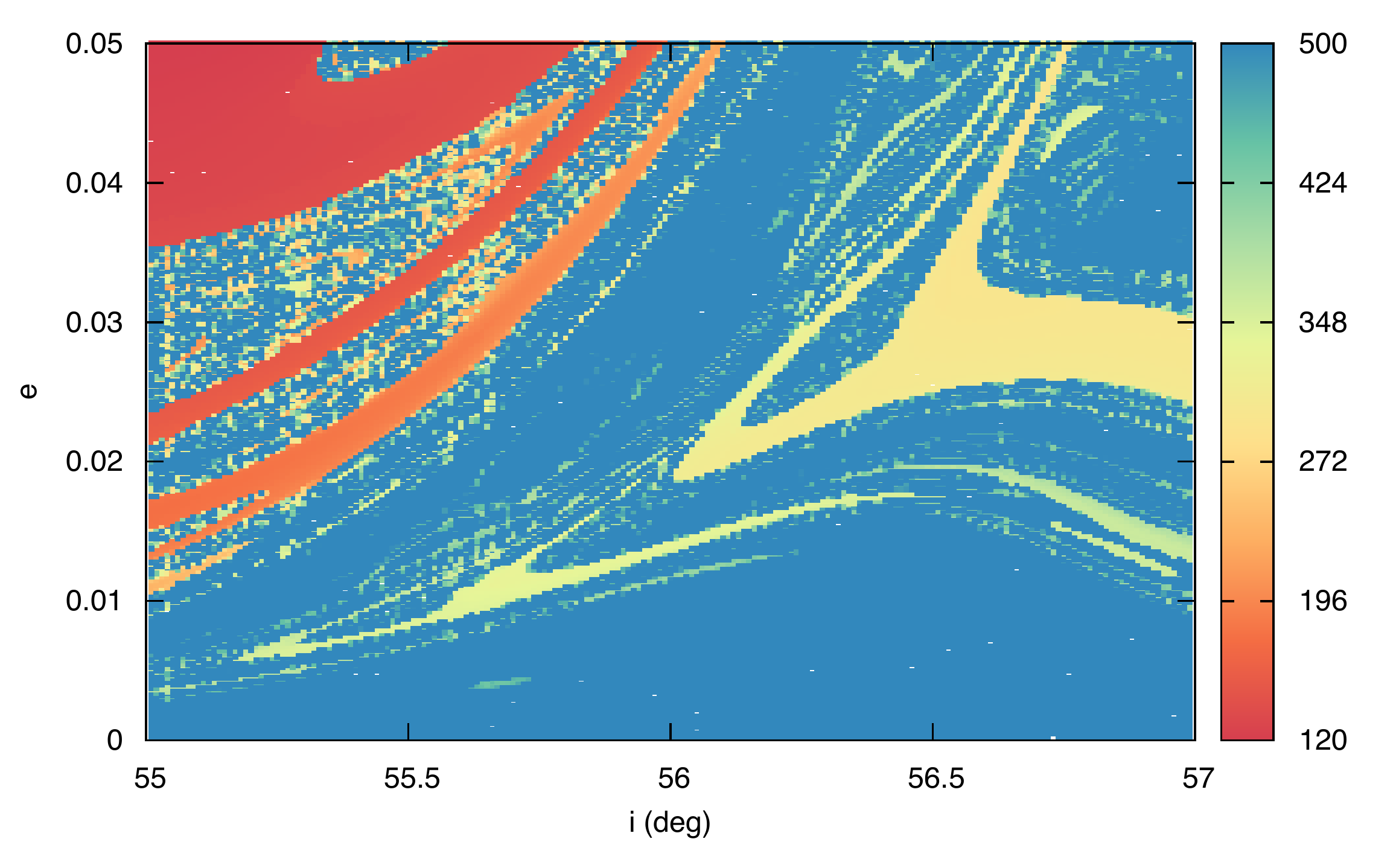}}        
  \caption{Stability maps characterizing the local hyperbolicity (normalized to 1) and the barrier of 
  predictability (in years) in the vicinity of a proposed graveyard orbit case 
  ($a_0 = 30\,100$ km, $\Omega_0 = \omega_0 = 70^\circ$, epoch: 6 DEC 2020). 
  The collision time map is provided to illustrate the period of time (in years) after which 
  atmospheric re-entry occurs, and completes the variational maps.} 
  \label{fig:stable_FLI_TL_TC}
\end{figure}

\begin{figure}
  \centering
    \subfigure[FLI map.]
    {\includegraphics[scale=0.238]{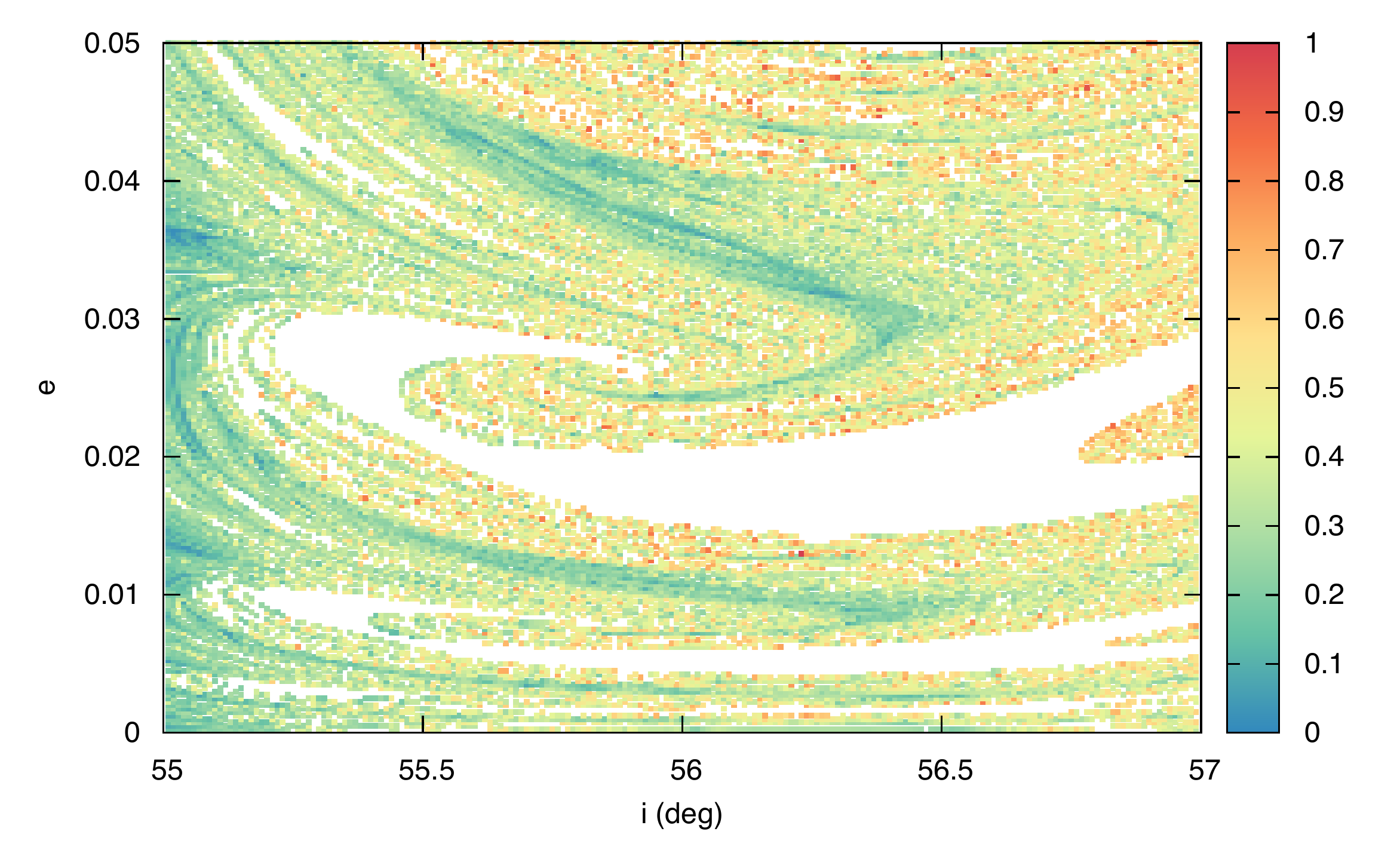}}        
    \subfigure[Lyapunov time map.]
    {\includegraphics[scale=0.238]{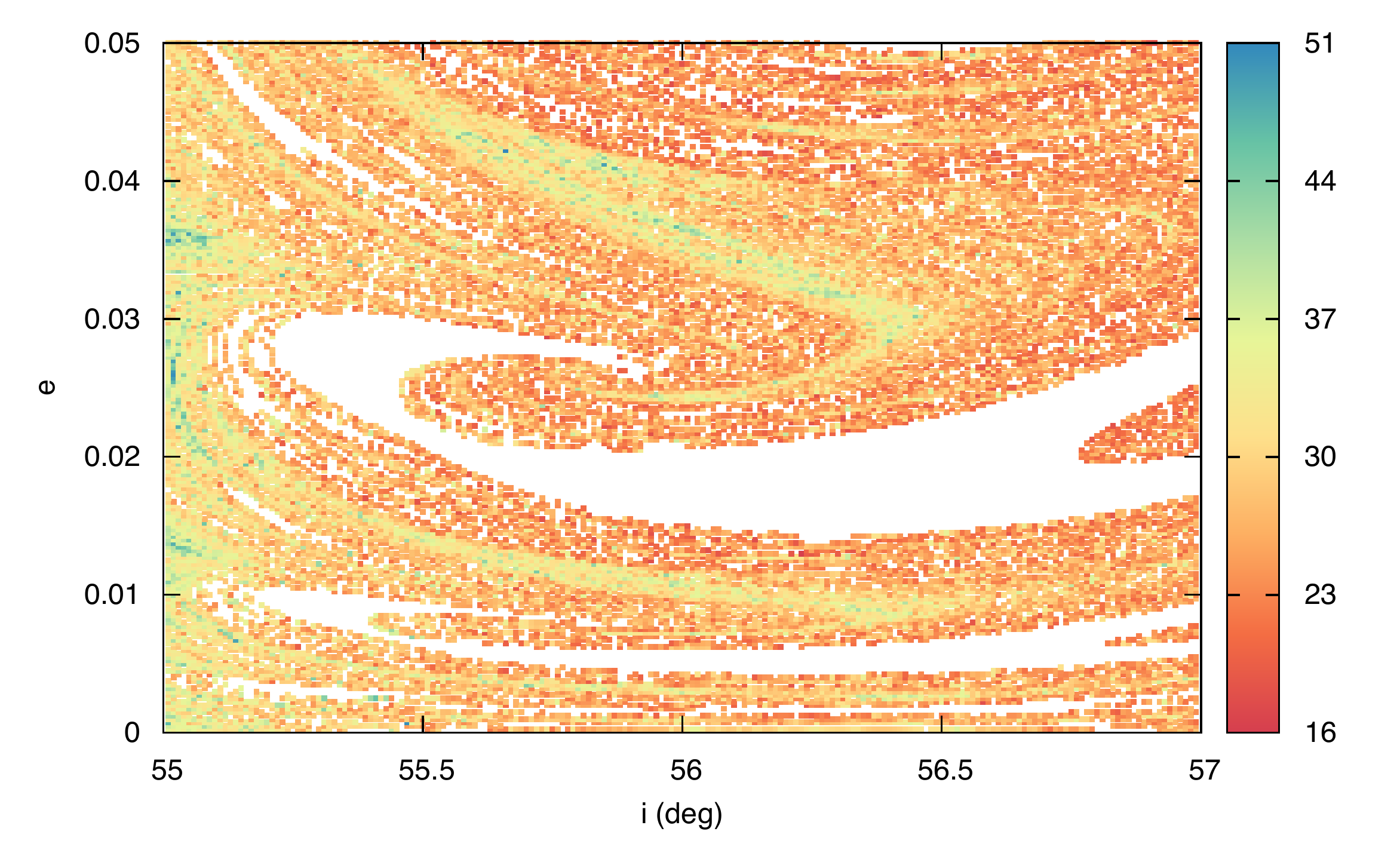}} 
    \subfigure[Collision time map.]
    {\includegraphics[scale=0.238]{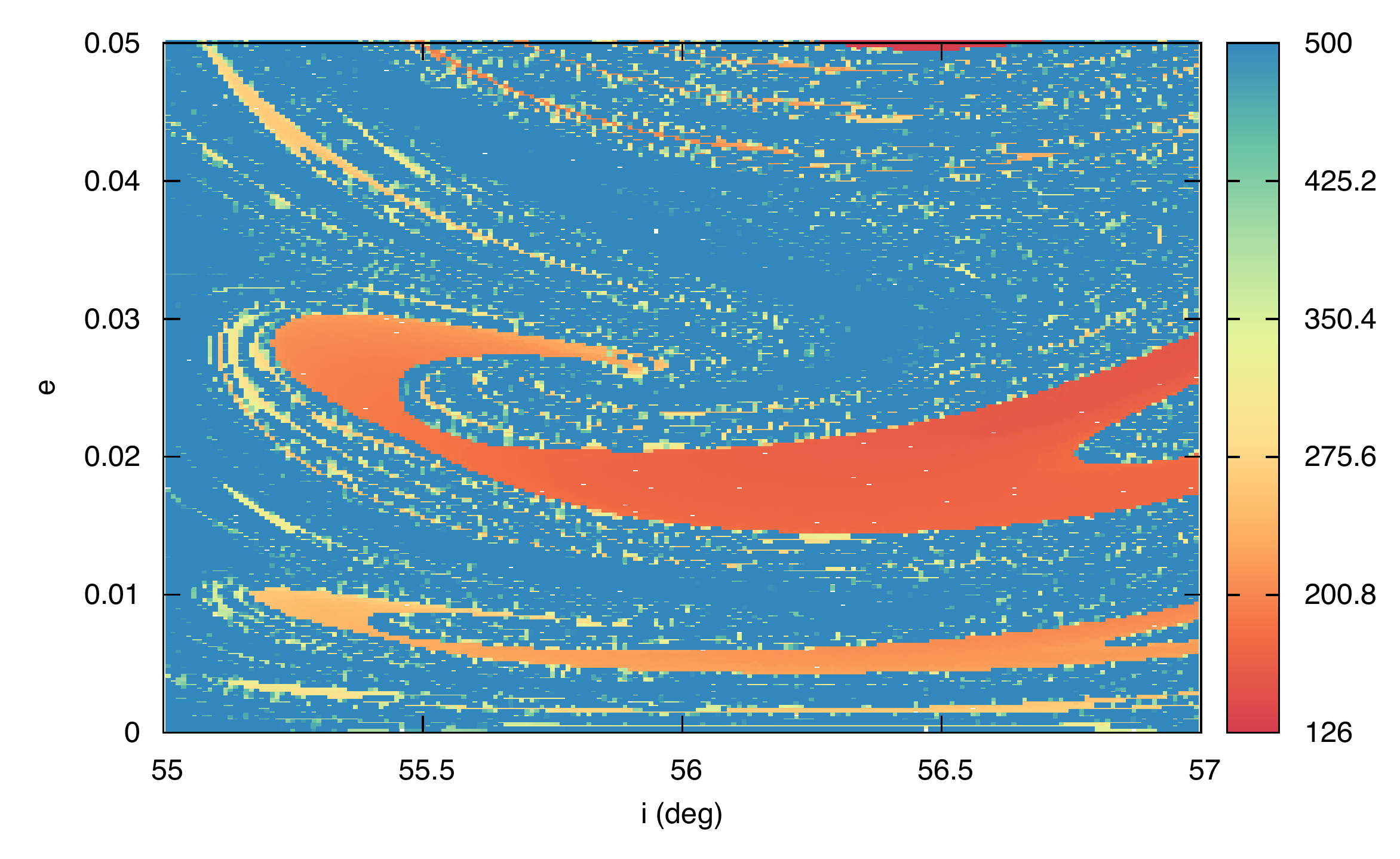}}  
  \caption{Same as Fig.~\ref{fig:stable_FLI_TL_TC}, but for a proposed eccentricity growth case
  ($a_0 = 28\,100$ km, $\Omega_0 = 60^\circ$, $\omega_0 = 100^\circ$, epoch: 6 DEC 2020).} 
  \label{fig:unstable_FLI_TL_TC}
\end{figure}

Figures~\ref{fig:stable_FLI_TL_TC} and \ref{fig:unstable_FLI_TL_TC} present several dynamical quantities of interests, in a series of maps,\footnote{To produce the various stability maps, the initial conditions were distributed in a regular grid of $200 \times 200$ resolution, and the model was propagated for 500 years.} for semi-major axes and parameters near two sample disposal orbits of Section~\ref{sec:parametric}: the FLIs \citep{cF00,nTbN15}, characterizing the degree of hyperbolicity; the Lyapunov time, an estimate of the prediction horizon \citep{jLsir86}; and collision time. The FLIs of all regular orbits appear with the same dark blue colour, while light blue corresponds to invariant tori, yellow and red to chaotic regions, and white to collision orbits. We find that the volume of collision orbits is roughly the same for the stable and unstable semi-major axes, but that the volume of chaotic obits is indeed larger for the eccentricity growth scenario (where we also find highly unstable and re-entry orbits even for quasi circular orbits). Inside the collision orbit structures, the re-entry time is nearly constant, and the shortest dynamical lifetime was almost identical in both cases ($\sim 120$ years). For each scenario, the estimated values of the Lyapunov times imply a very short timescale for reliable predictability, with many orbits having values on the order of only a few decades.  

\begin{figure}
  \centering
    \subfigure[FLI map.]
    {\includegraphics[scale=0.29]{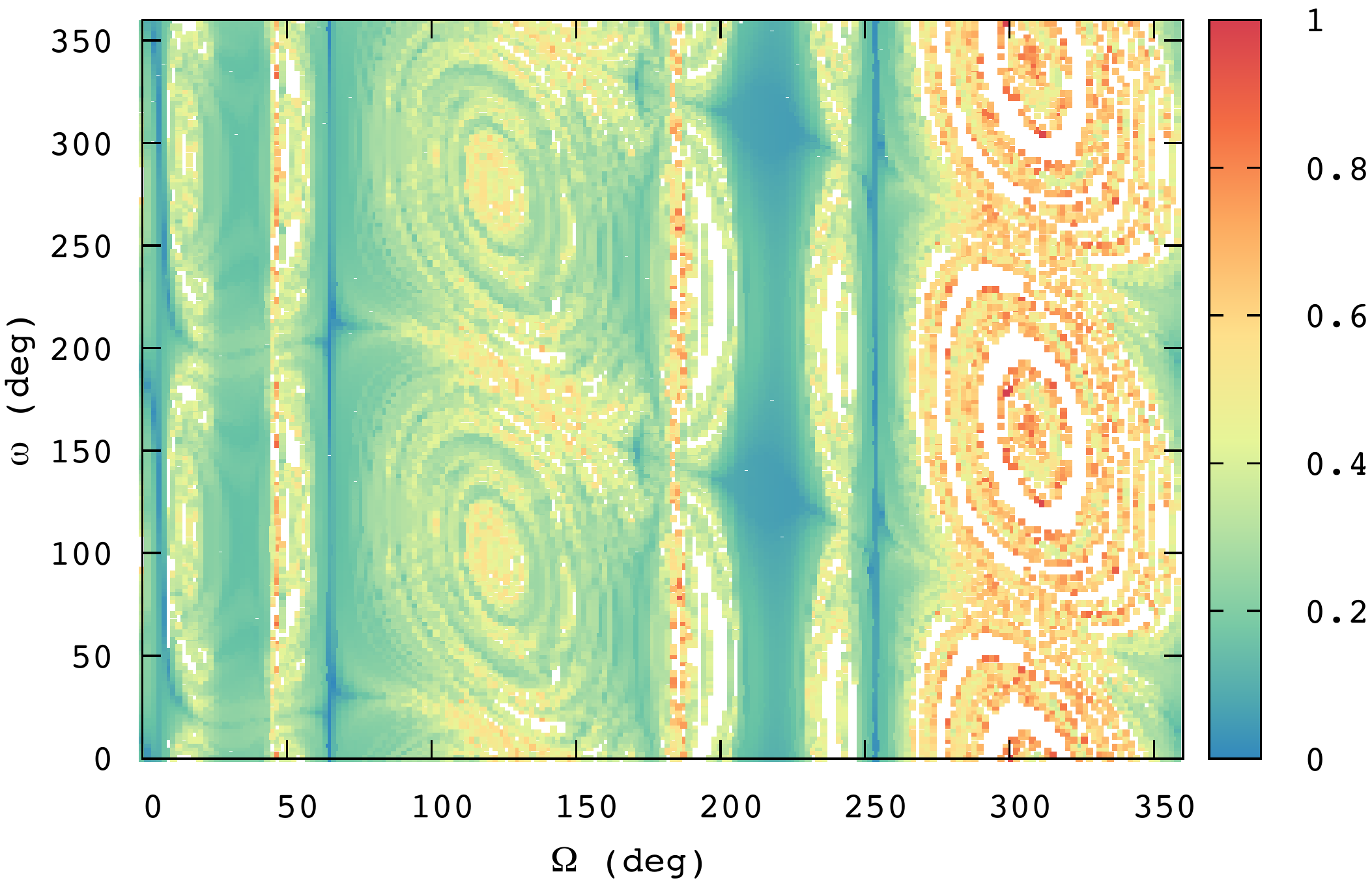}}
    \subfigure[FLI map.]
    {\includegraphics[scale=0.29]{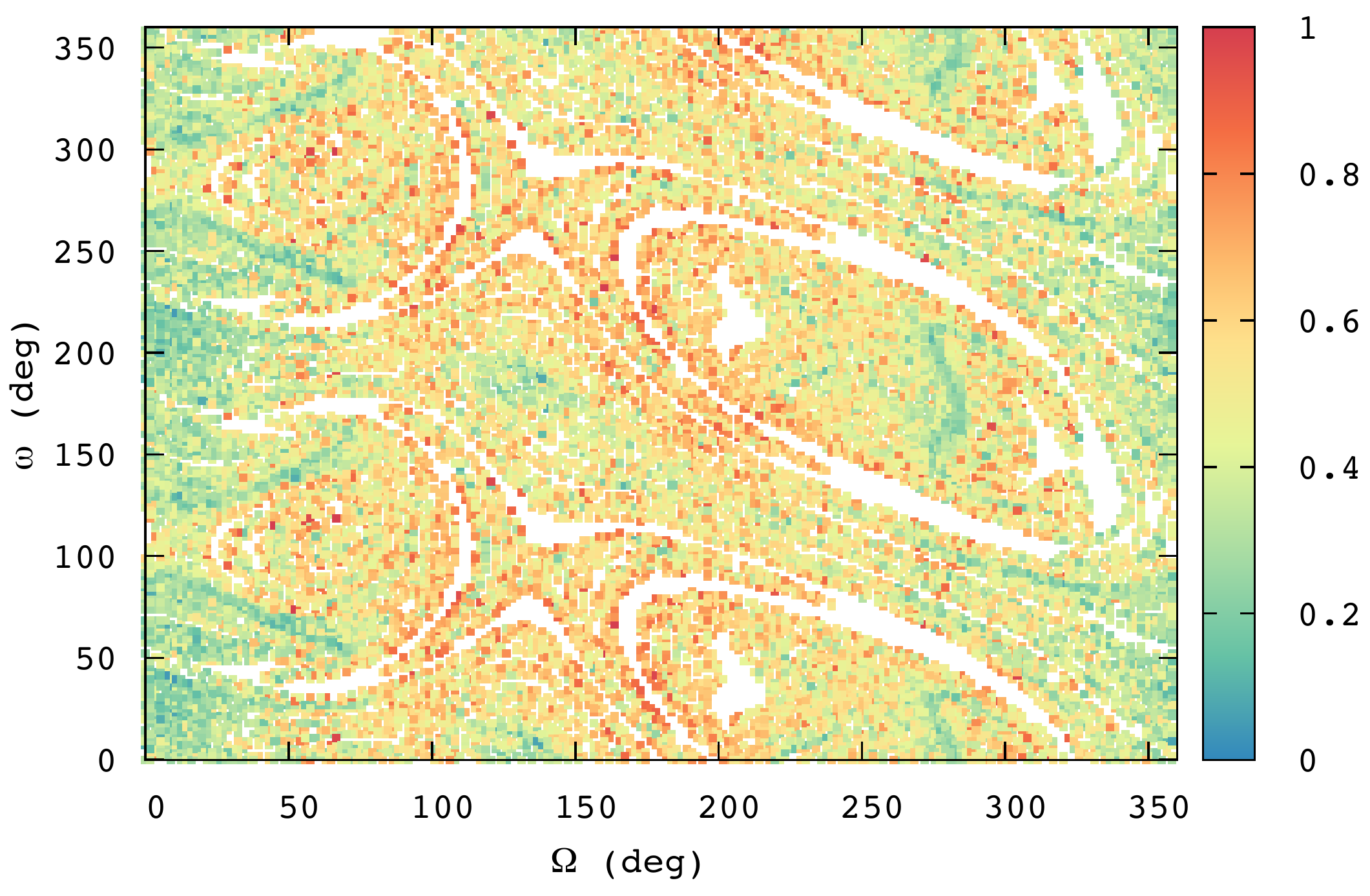}} 
    \subfigure[Lyapunov time map.]
    {\includegraphics[scale=0.29]{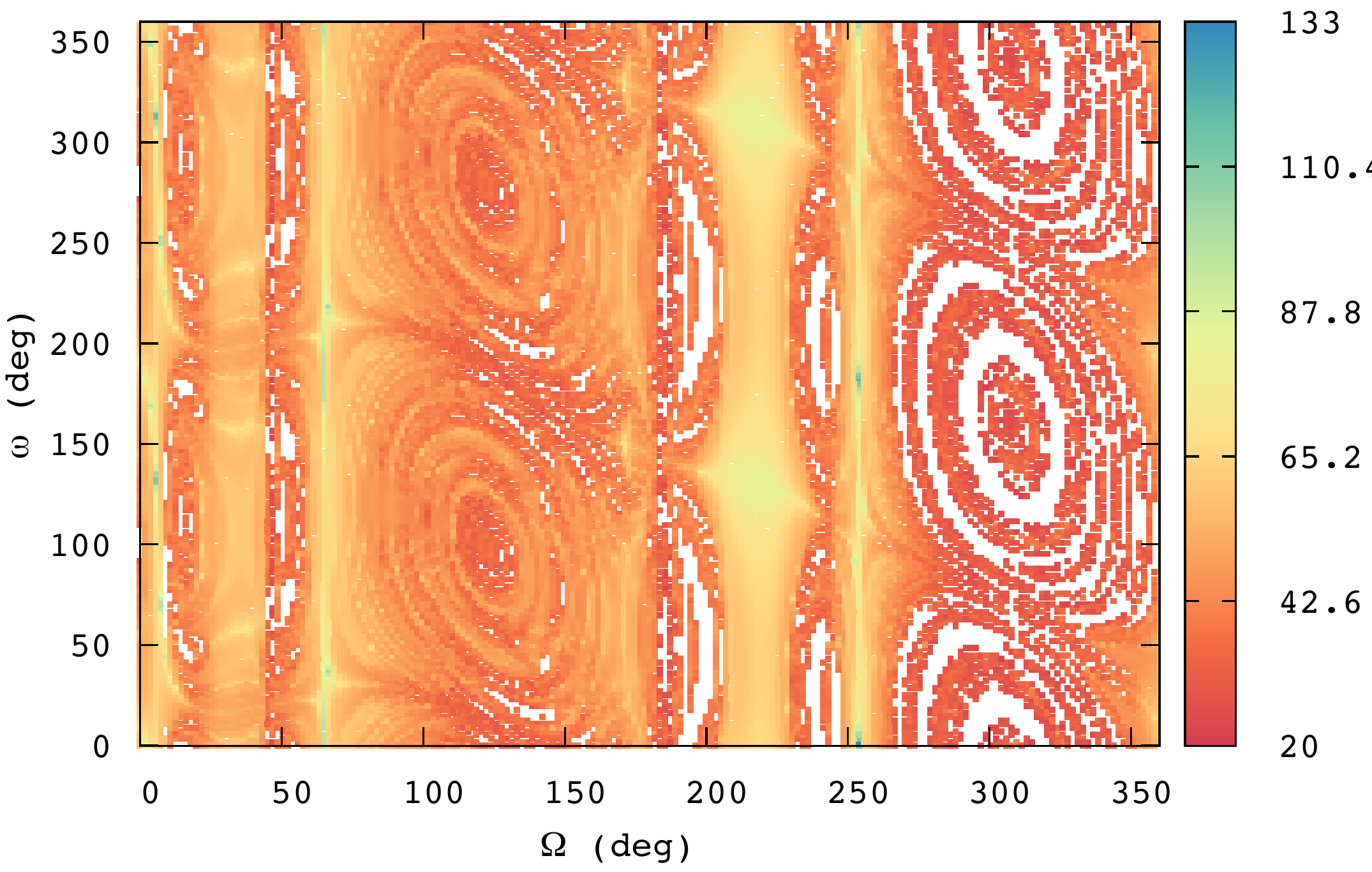}}        
    \subfigure[Lyapunov time map.]
    {\includegraphics[scale=0.29]{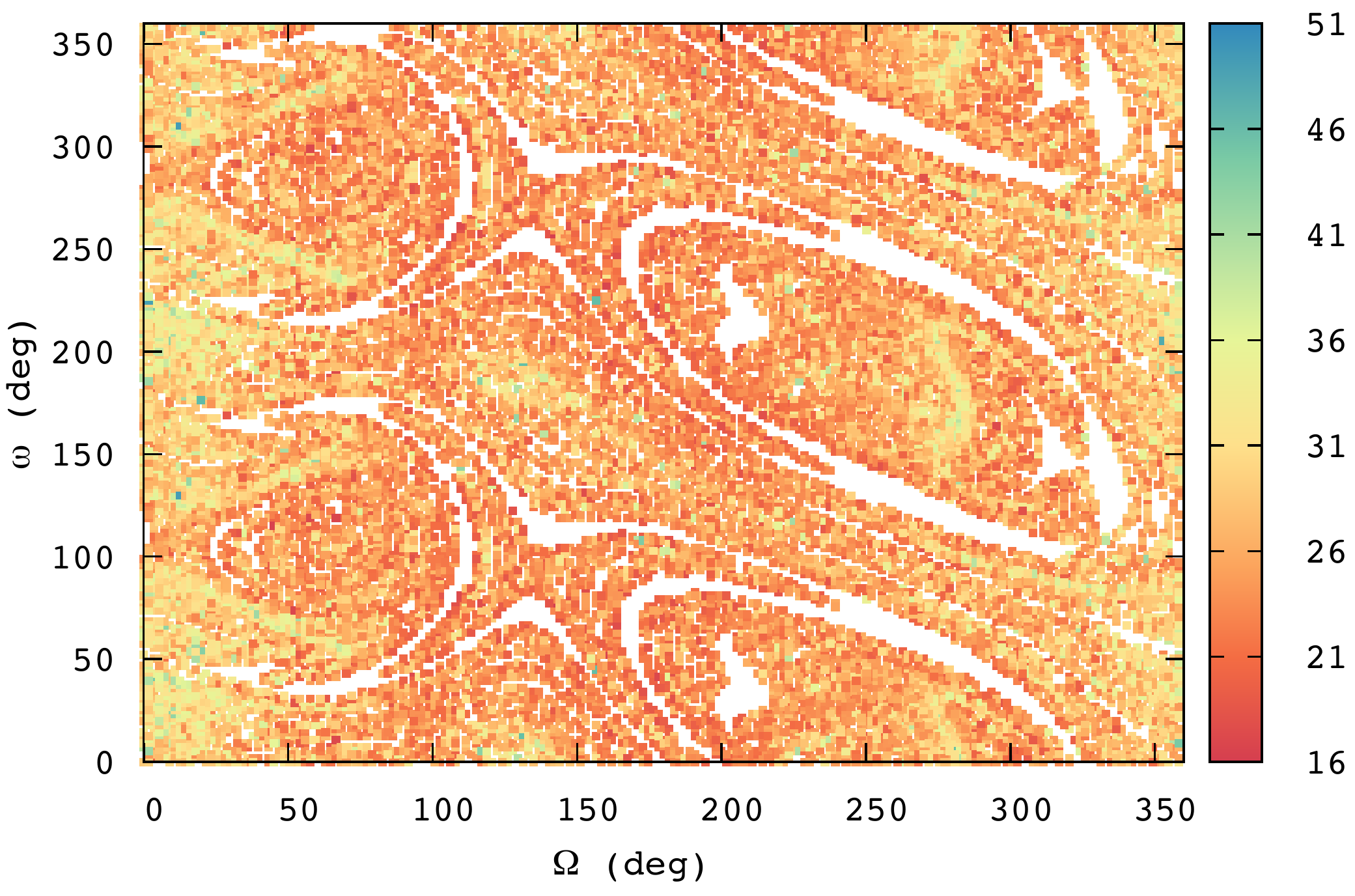}} 
  \caption{Dynamical structures of the stable (left: $a_0 = 30\,100$ km, $e_0 = 0.001$, $i_0 = 56^\circ$)
  and unstable (right: $a_0 = 28\,100$ km, $e_0 = 0.05$, $i_0 = 56^\circ$) cases in the 
  node--perigee phase space. The colourbar for the FLI maps is normalized to 1 and that for
  the Lyapunov time maps represents number of years.} 
  \label{fig:stable_unstable_Omega_omega}
\end{figure}

We must stress here that these charts, Figs.~\ref{fig:stable_FLI_TL_TC} and \ref{fig:unstable_FLI_TL_TC}, have been obtained by varying only the initial inclination and eccentricity, with the initial phases $(t_0, \Omega, \omega)$ being fixed for all computed FLI. Given the inherent difficulty to capture the dynamics of the whole six-dimensional phase space in a plane of dimension two, we must settle for only a partial insight into the dynamical structure \citep{nTbN15,mR14}. We now fix the action-like quantities to their approximate nominal values, along with the epoch date, and investigate the geometrical organisation and coexistence of chaotic and regular motion in the $\Omega$--$\omega$ phase space (Fig.~\ref{fig:stable_unstable_Omega_omega}). Note the similarity between the MEM maps of Fig.~\ref{fig:ESA_sample} (top), computed over a 200 year timespan; yet, the FLI and Lyapunov time maps, besides providing much finer detail for the proper detection of invariant structures and chaotic regions, give actual physical information on these unpredictable orbits, whereas the MEM maps provide only one trajectory realisation. In the stable case, we point out again how the structures seem to be aligned along vertical bands, and can observe a highly stable region, relatively speaking, near $\Omega = 220^\circ$ (notice how the misleadingly wide bands of stable orbits in Fig.~\ref{fig:ESA_sample} disappear in a proper resolution and computational time). The volume of escaping orbits is larger for the unstable case, and it becomes much more difficult to identify stable regimes. Finally, we note that in both cases, there exists a strong symmetry in the argument of perigee in both maps, which follows naturally from the fact that the secular equations governing quadrupolar gravitational interactions are invariant under the transformation $\omega \rightarrow \omega + 180^\circ$ \citep[qq.v.,][]{pM61,sT09}. 

\begin{figure}
  \centering
    \subfigure[$a_0 = 30\,100$ km, $e_0 = 0.001$, $i_0 = 56^\circ$.]
    {\includegraphics[scale=0.475]{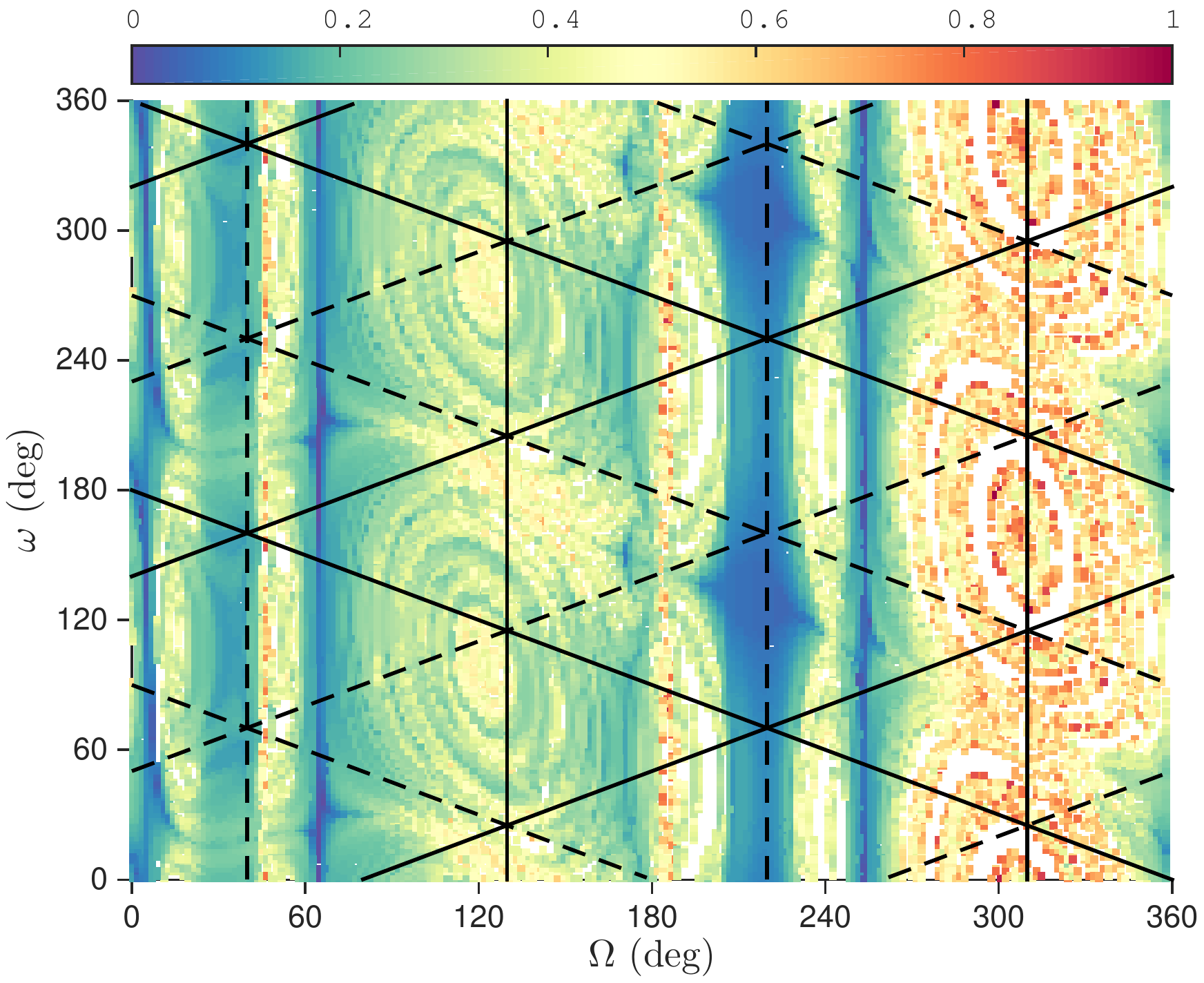} 
    \label{sfig:fxpts_GalStab}}
    \subfigure[$a_0 = 28\,100$ km, $e_0 = 0.05$, $i_0 = 56^\circ$.]
    {\includegraphics[scale=0.475]{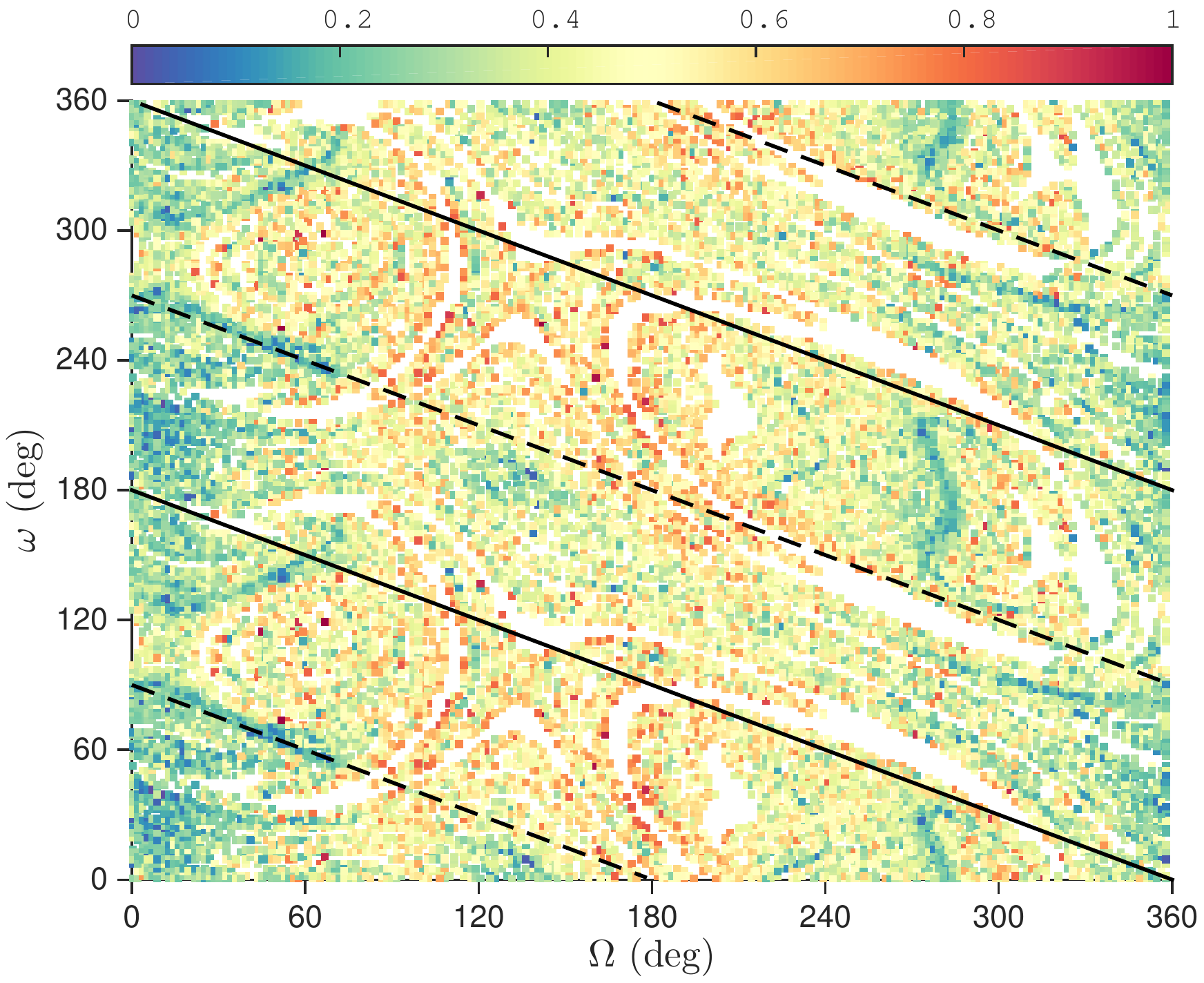}
    \label{fli_fxpts_GalUnstab}} 
  \caption{Elliptic (dashed) and hyperbolic (solid) equilibria of the relevant resonances of 
  Fig.~\ref{fig:apertures} superimposed on the FLI maps of Fig.~\ref{fig:stable_unstable_Omega_omega}.
  The fixed-point lines are defined in Table~\ref{tab:fixed_points}, where $\Omega_\M = 79.68^\circ$
  for these particular maps.} 
  \label{fig:equilib_criteria}
\end{figure}

From Fig.~\ref{fig:apertures}, it is easy to identify the main resonances that organise the global structures in the stability maps of Fig.~\ref{fig:stable_unstable_Omega_omega}. The proposed stable graveyard case at $a_0 = 30\,100$ km, $e_0 = 0.001$, and $i_0 = 56^\circ$ is primarily affected by the $\dot\psi_{2,1,0} = 2 \dot\omega + \dot\Omega \approx 0$ {\itshape apsidal} resonance and the $\dot\psi_{0,2,-1} = 2 \dot\Omega - \dot\Omega_\M \approx 0$ {\itshape nodal} resonance, though the weaker $\dot\psi_{-2,1,-1} = -2\dot\omega + \dot\Omega - \dot\Omega_\M \approx 0$ resonance is present nearby. We can obtain a partial analytical description of the dynamical structures in the maps though the computed stable and unstable fixed points of the resonances (Table.~\ref{tab:fixed_points}). The epoch date determines the initial geometry of the Earth-Moon-Sun system and thus the initial location of the lunar ascending node. Figure~\ref{fig:equilib_criteria} shows the equilibria conditions superimposed on the background FLI maps. The location of the strip of relative stability, apparent only in the graveyard case, is clearly related to the resonant geography of the $\dot\psi_{0,2,-1}$ nodal resonance, and the wide vertical band occurs precisely along the line $\Omega = (2 \pi + \Omega_\M)/2$, where $\Omega_\M \sim 79.68^\circ$ at the epoch 06 December 2020 UTC. In isolation, this resonance affects only the orbital inclination, and it seems that its elliptic fixed points provide a kind of protection mechanism against the large-scale eccentricity transport induced by the $\dot\psi_{2,1,0}$ resonance. The hyperbolic fixed points of this nodal resonance ($2 \Omega - \Omega_\M = \pm \pi$) identify the approximate location along $\Omega$ of the patches of chaotic and collision orbits. In both the graveyard and eccentricity growth case, the patterns and geometry are aligned along the equilibria curves of the $\dot\psi_{2,1,0} = 2 \dot\omega + \dot\Omega \approx 0$ resonance, something crudely pointed out by \citet{dStY15}. We should note that any attempt to describe the phase-space topology induced by the resonances in a rigorous way requires more sophisticated analytical methods that treat resonance interactions; yet, despite this formidable problem, we find that we can still achieve an intuitive understanding through the mathematical study of the resonant equilibria. 

\begin{figure}
  \centering
    \subfigure[$a_{0}=28,600$ km.]
    {\includegraphics[scale=0.24]{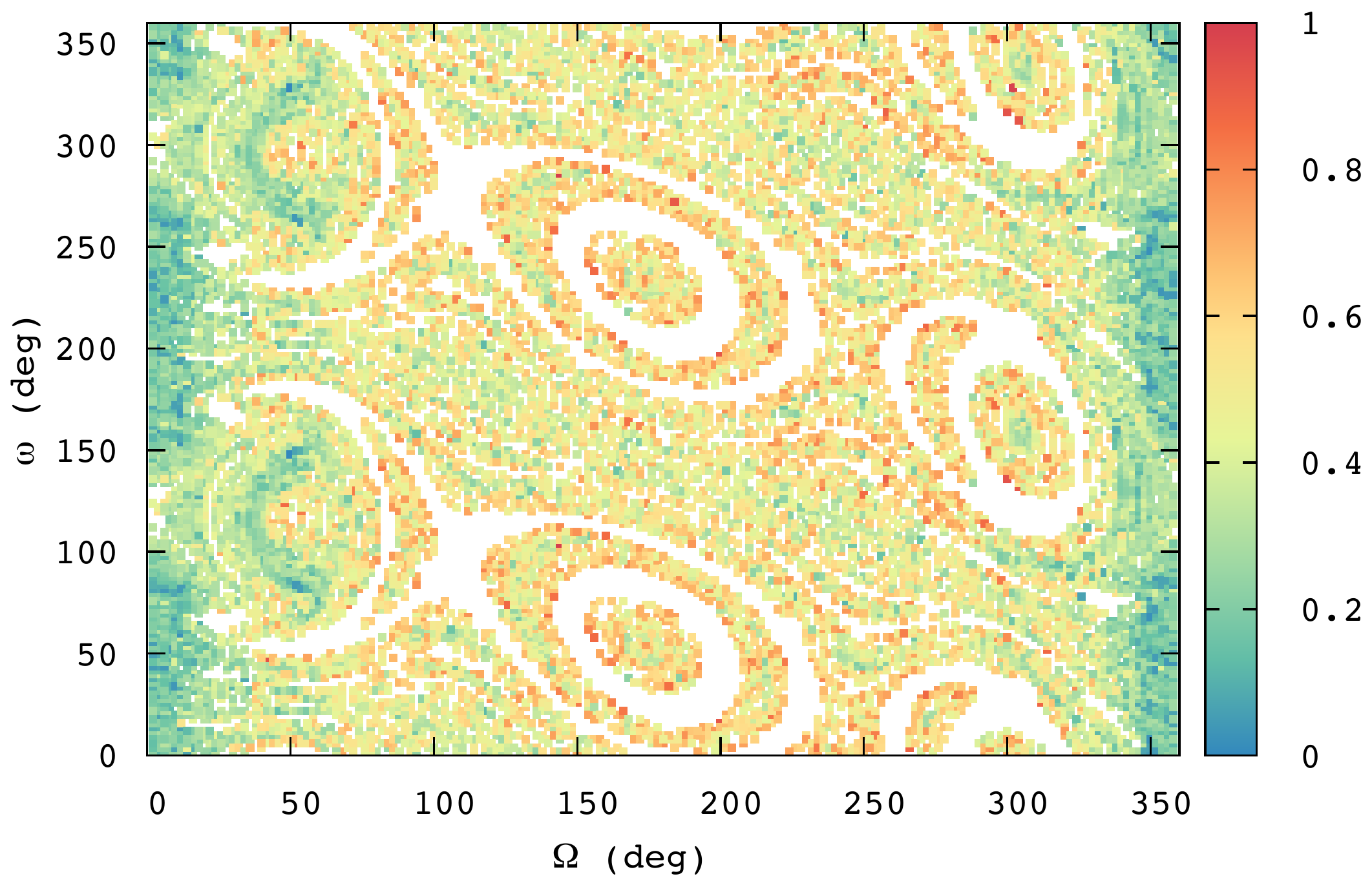}} 
      \subfigure[$a_{0}=29,100$ km.]
    {\includegraphics[scale=0.24]{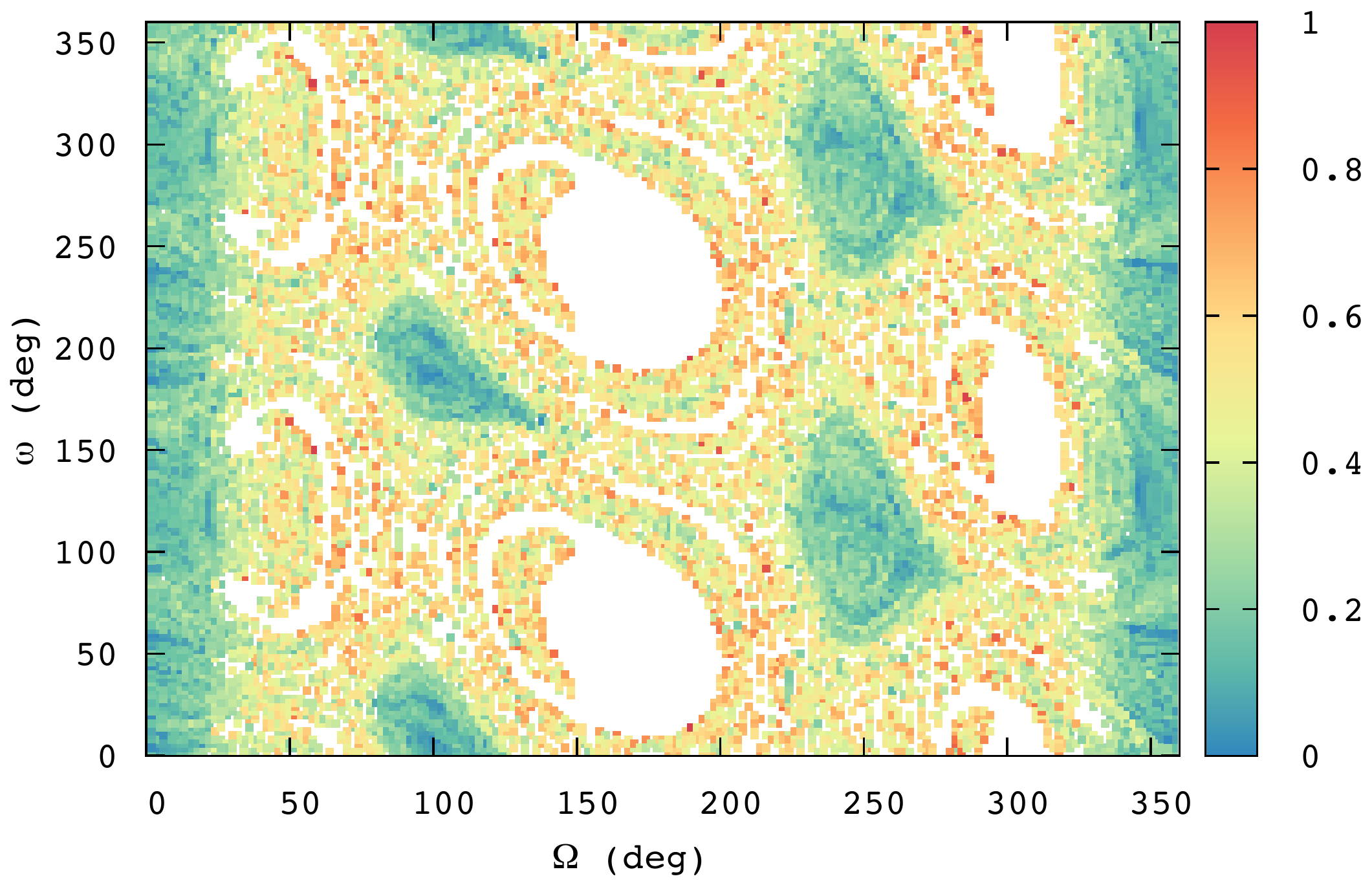}} 
     \subfigure[$a_{0}=29,600$ km.]
    {\includegraphics[scale=0.24]{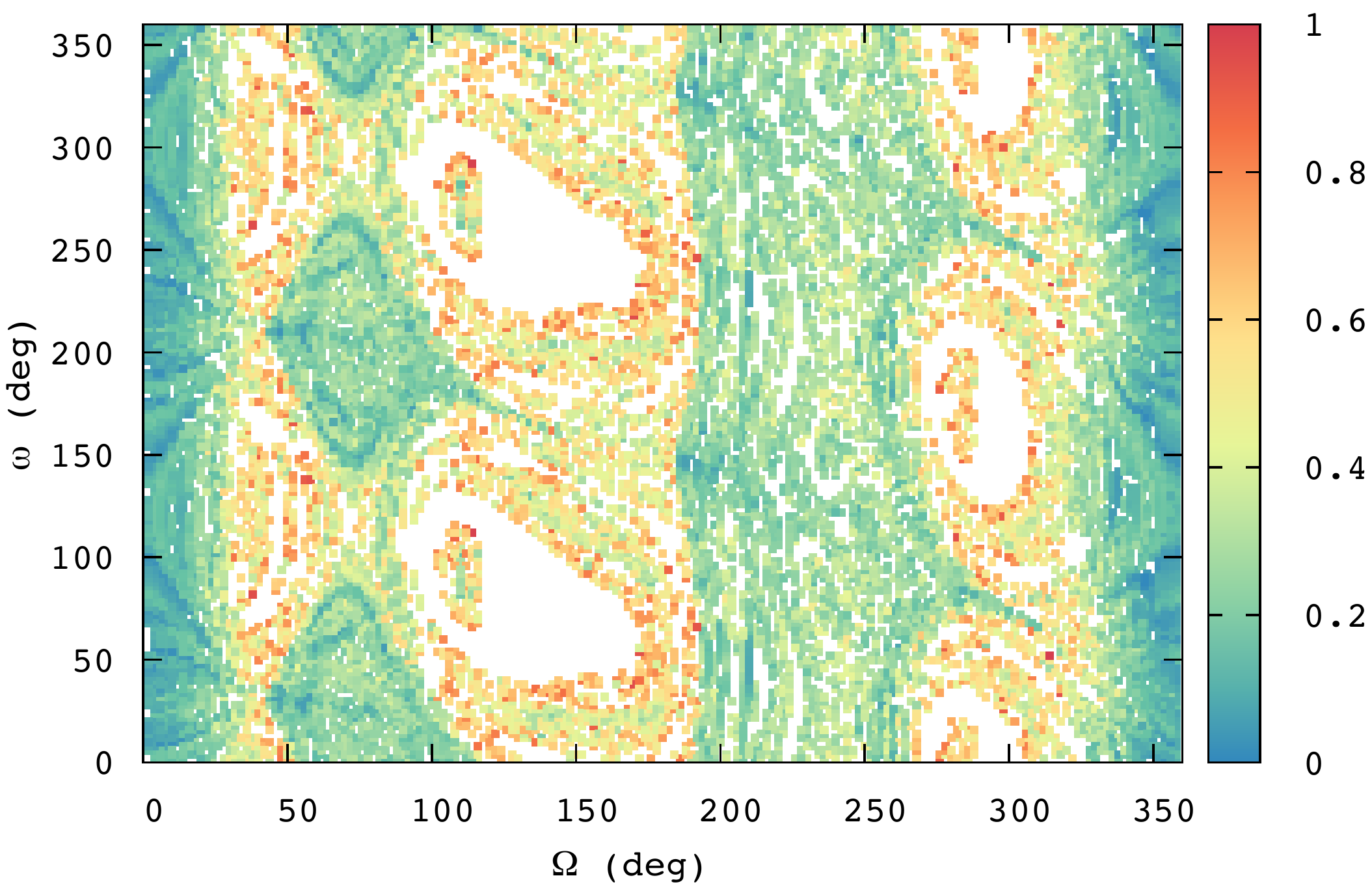}}
    \subfigure[$a_{0}=30,100$ km.]
    {\includegraphics[scale=0.24]{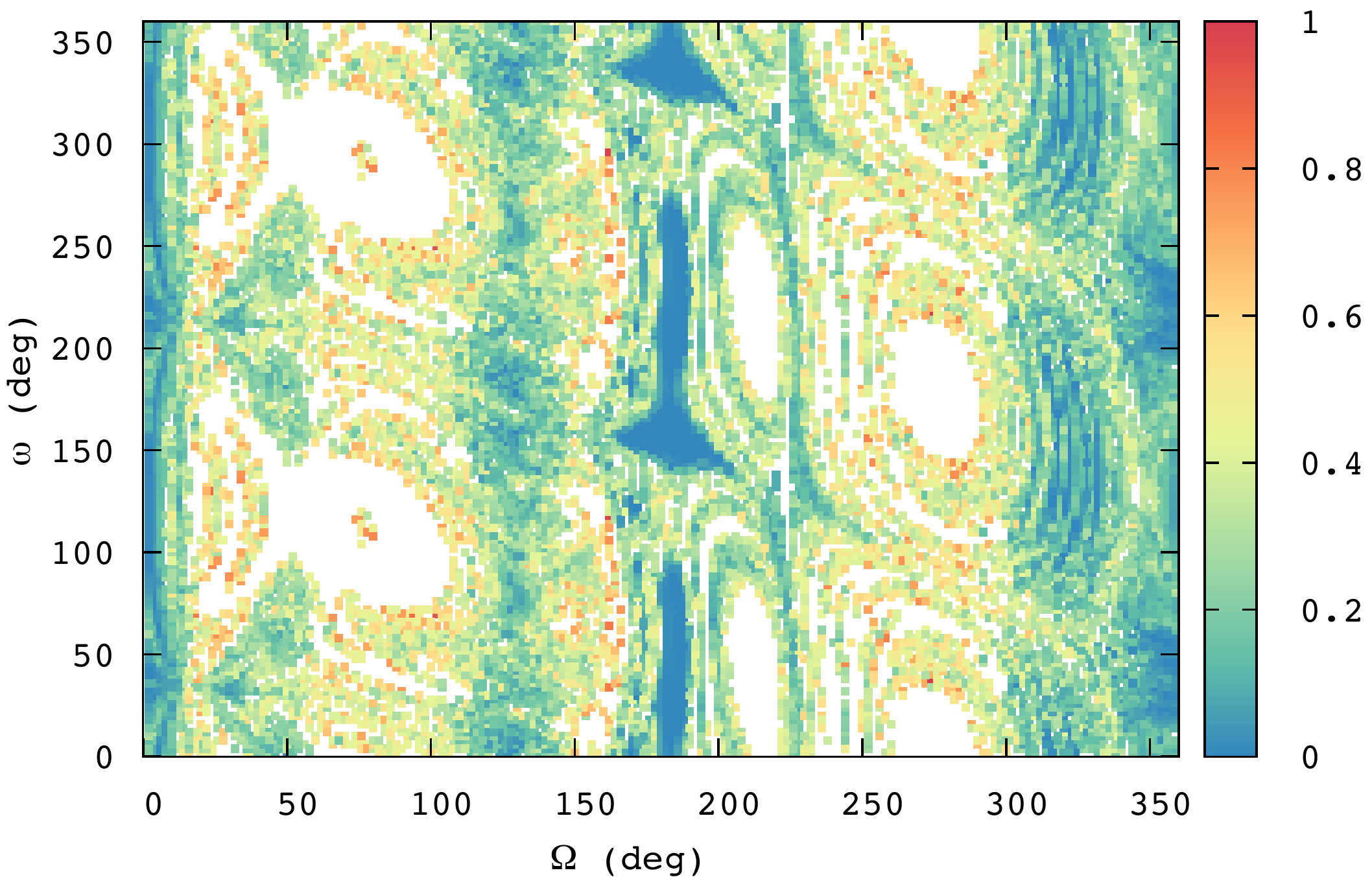}} 
    \subfigure[$a_{0}=30,600$ km.]
    {\includegraphics[scale=0.24]{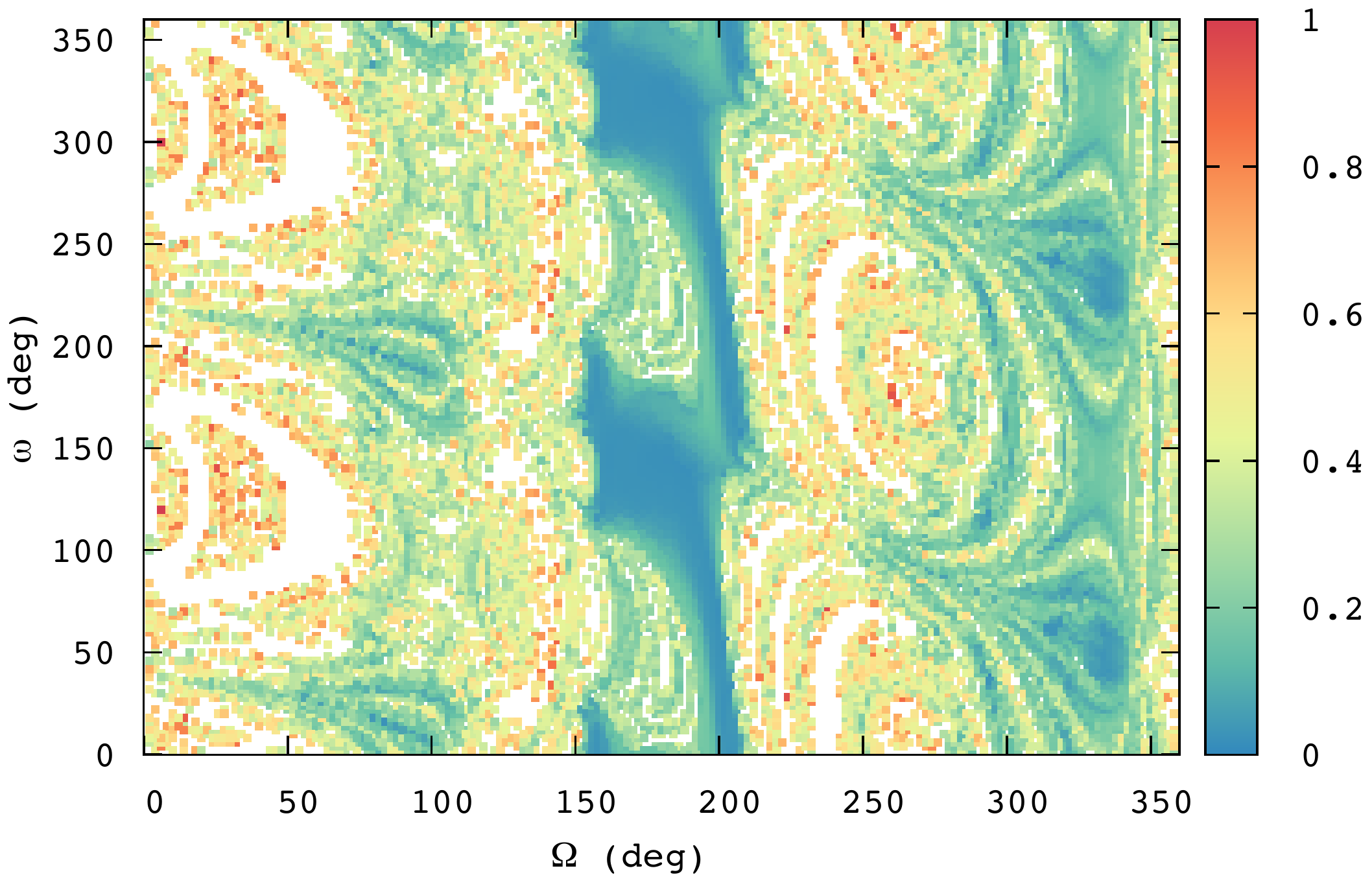}}        
  \caption{Influence of the strength of the perturbation on the dynamical structures 
  near Galileo's semi-major axis ($e_0 = 0.02$, $i_0 = 56.1^\circ$, epoch: 2 MAR 1969).} 
  \label{fig:Omega_omega_pert_param}
\end{figure}

Figure~\ref{fig:Omega_omega_pert_param} presents the evolution of the FLI maps in the node--perigee phase space, exploring the sensitivity to the initial semi-major axis near the nominal Galileo value. It is particularly noteworthy that the volume of stable orbits is found to increase with increasing semi-major axis, as with the width of the vertical band of stability, occurring near $\Omega = 180^\circ$ (the location of the elliptic fixed point of the $2 \dot\Omega - \dot\Omega_\M \approx 0$ resonance). On the contrary, decreasing the initial semi-major axis from the Galileo constellation, the $\Omega$--$\omega$ phase-space is nearly globally populated by unstable orbits that surround collisions orbits, the latter organised in pendulum-like structures along the slope defined by the fixed points of the  $2 \dot\omega + \dot\Omega \approx 0$ resonance. 

\begin{figure}
	\centering
	 \subfigure[$\Omega=240^{\circ}$, $\omega=30^{\circ}$, epoch: 2 MAR 1969.]
	{\includegraphics[scale=0.285]{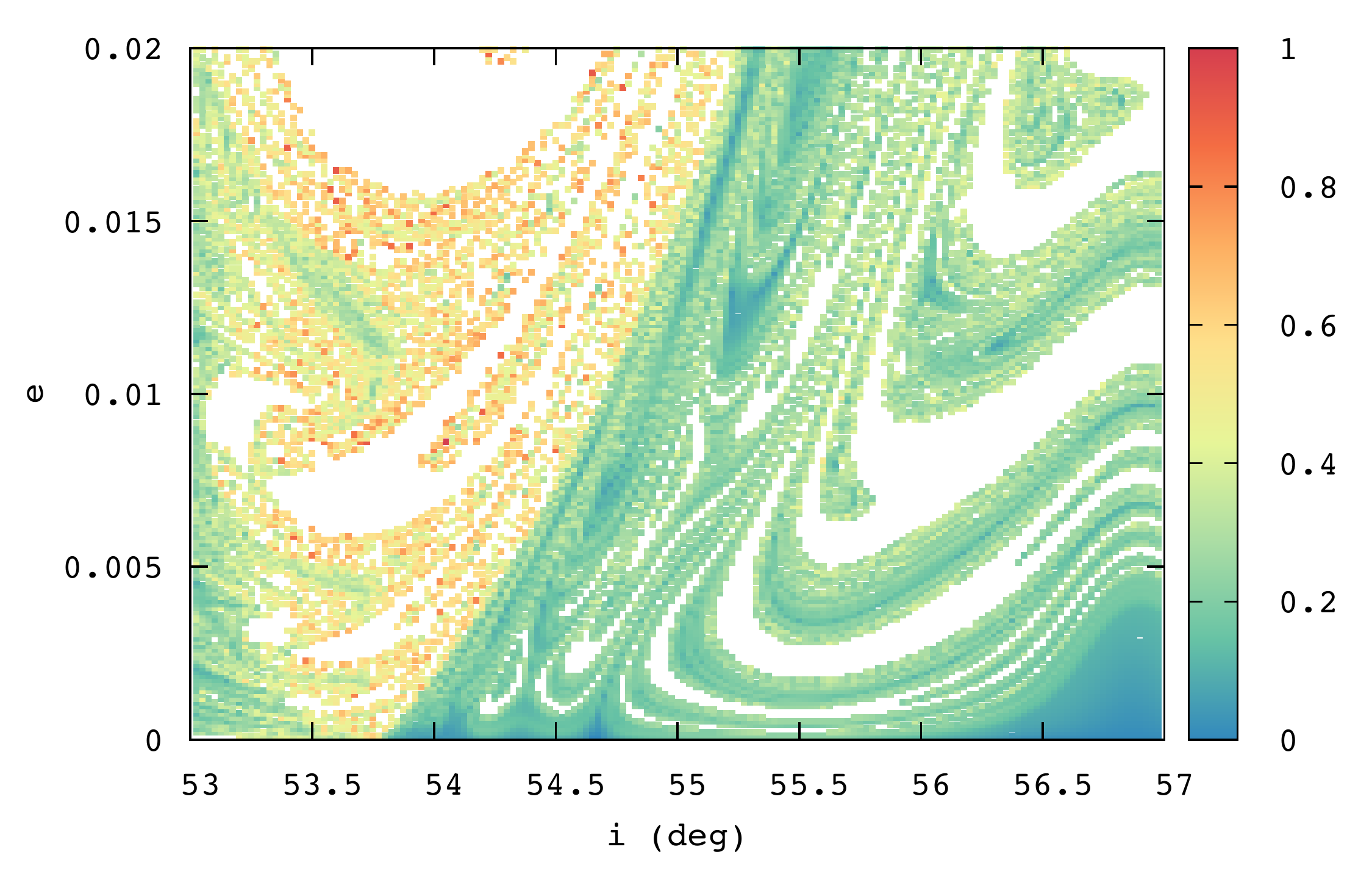}}
	\hspace{12pt}
	 \subfigure[$\Omega=120^{\circ}$, $\omega=120^{\circ}$, epoch: 2 MAR 1969.]
	{\includegraphics[scale=0.285]{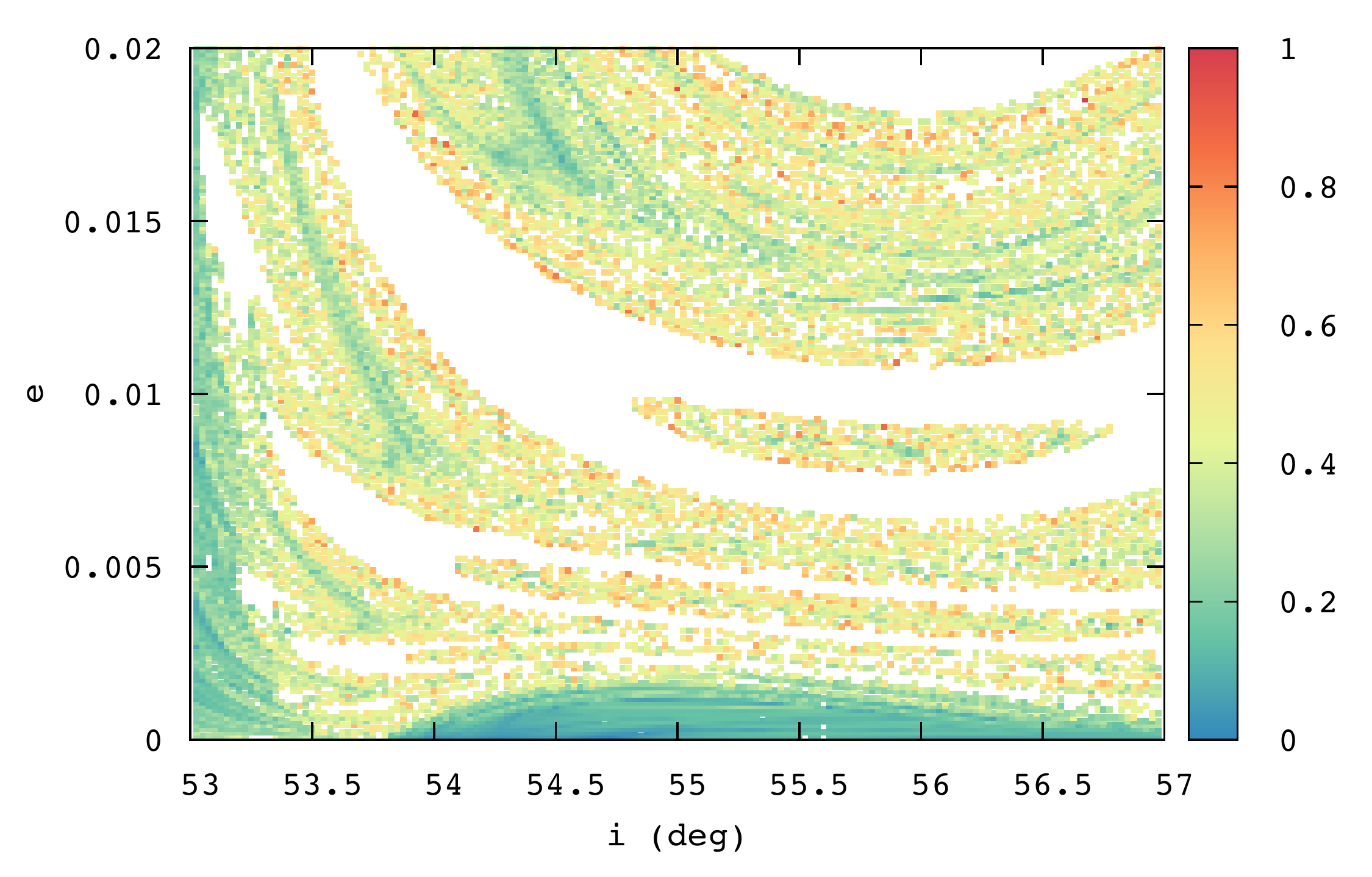}}
	 \subfigure[$\Omega=120^{\circ}$, $\omega=120^{\circ}$, epoch: 23 AUG 1974.]
	{\includegraphics[scale=0.285]{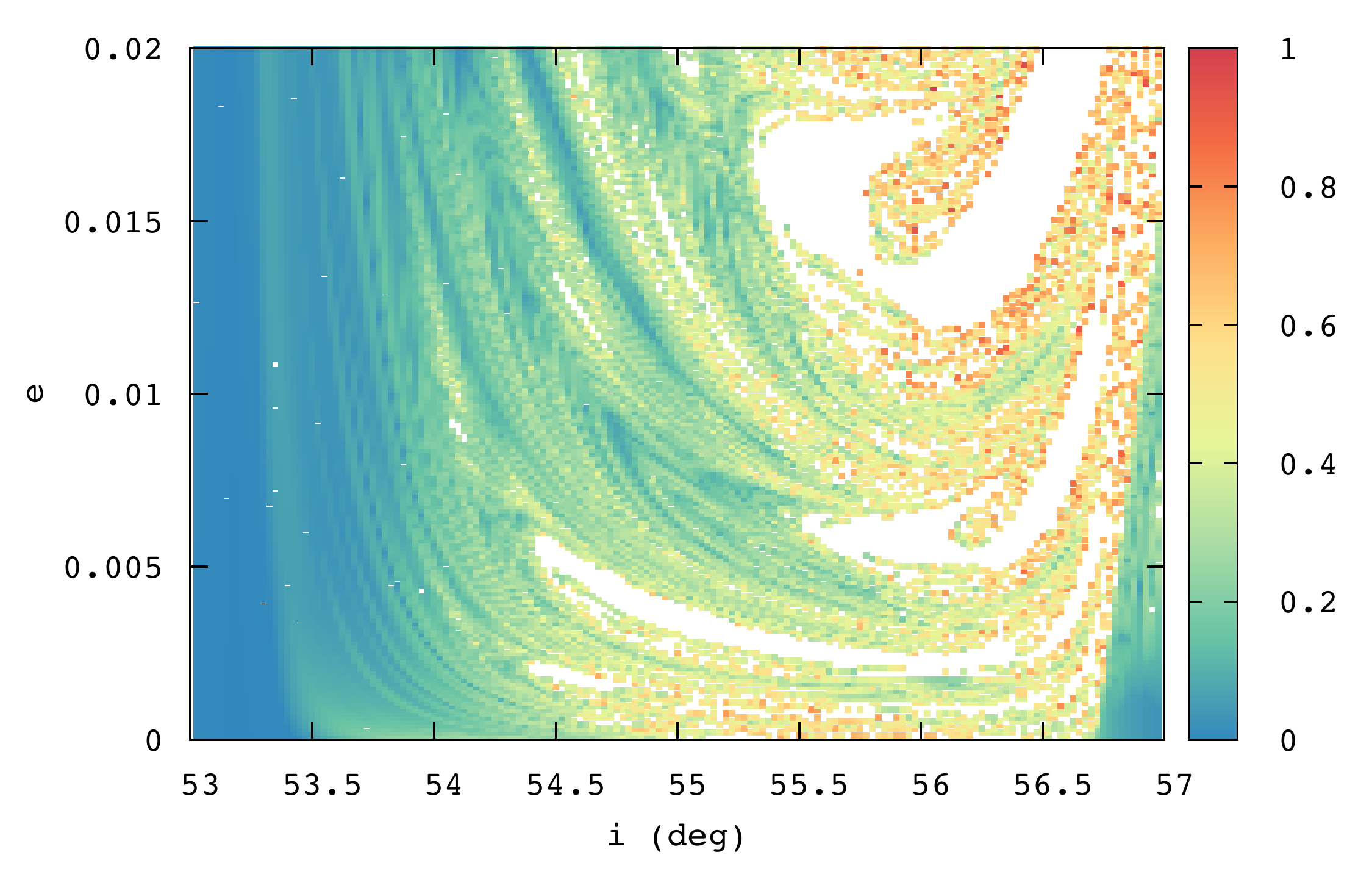}}
	\caption{Influence of the initial phases and the initial configuration of the Earth-Moon-Sun system 
	for Galileo's semi-major axis ($a_{0}=29,600$ km) as a representation of a dynamical system in a 
	lower dimensional phase space.} 
	\label{fig:FLI_phases}
\end{figure}

Figure~\ref{fig:FLI_phases} shows how the dynamical structures (stable, resonant, chaotic, or collision orbits) evolve by changing the initial phases $\Omega$ and $\omega$ or even the initial dynamical configuration of the Earth-Moon-Sun system (equivalent to changing the initial epoch). Of course, the FLI maps depend on the choice of initial angles because, as Todorovi\'c and Novakovi\'c write, ``$\ldots$ planes fixed at their different values cross the resonant islands at different positions, and in some special cases the crossing may not even occur. After all, the orbital space is 6D, while our plots are 2D, which certainly gives only a partial insight into the phase-space structure. However, we underline that this does not change the global dynamical pictures of the region, which is essential the same $\ldots$ \cite{nTbN15}''. To understand how such features evolve is clearly of remarkable practical application, and will require further study. 

%\clearpage
%  ****  Disposal Criteria Based on Resonant Fixed Points
\subsection{Robustness of the equilibria disposal criteria}
\label{sec:DispCrit}

In accordance with the results of Fig.~\ref{fig:stable_unstable_Omega_omega} that nearly all of the proposed disposal orbits of Section~\ref{sec:parametric} are inherently chaotic with Lyapunov times on the order of decades, the computation of individual trajectories becomes not only impractical but even irrelevant. The loss of long-time predictability implies that we must abandon the notion of individual orbits and instead focus on ensembles of trajectories. Rather than attempt an approximate, statistical description of the motion through transport theory \citep[q.v.,][]{jM92}, we content ourselves here with analyzing ensemble integrations only in order to test the robustness of the resonant fixed-point disposal scheme. 

\begin{figure}
  \centering
    \subfigure[Integrations of the elliptic and hyperbolic fixed points of the lunisolar secular resonances near 
    the proposed Galileo graveyard orbit ($a_0 = 30\,100$ km, $e_0 = 0.001$, $i_0 = 56^\circ$).]
    {\includegraphics[scale=0.55]{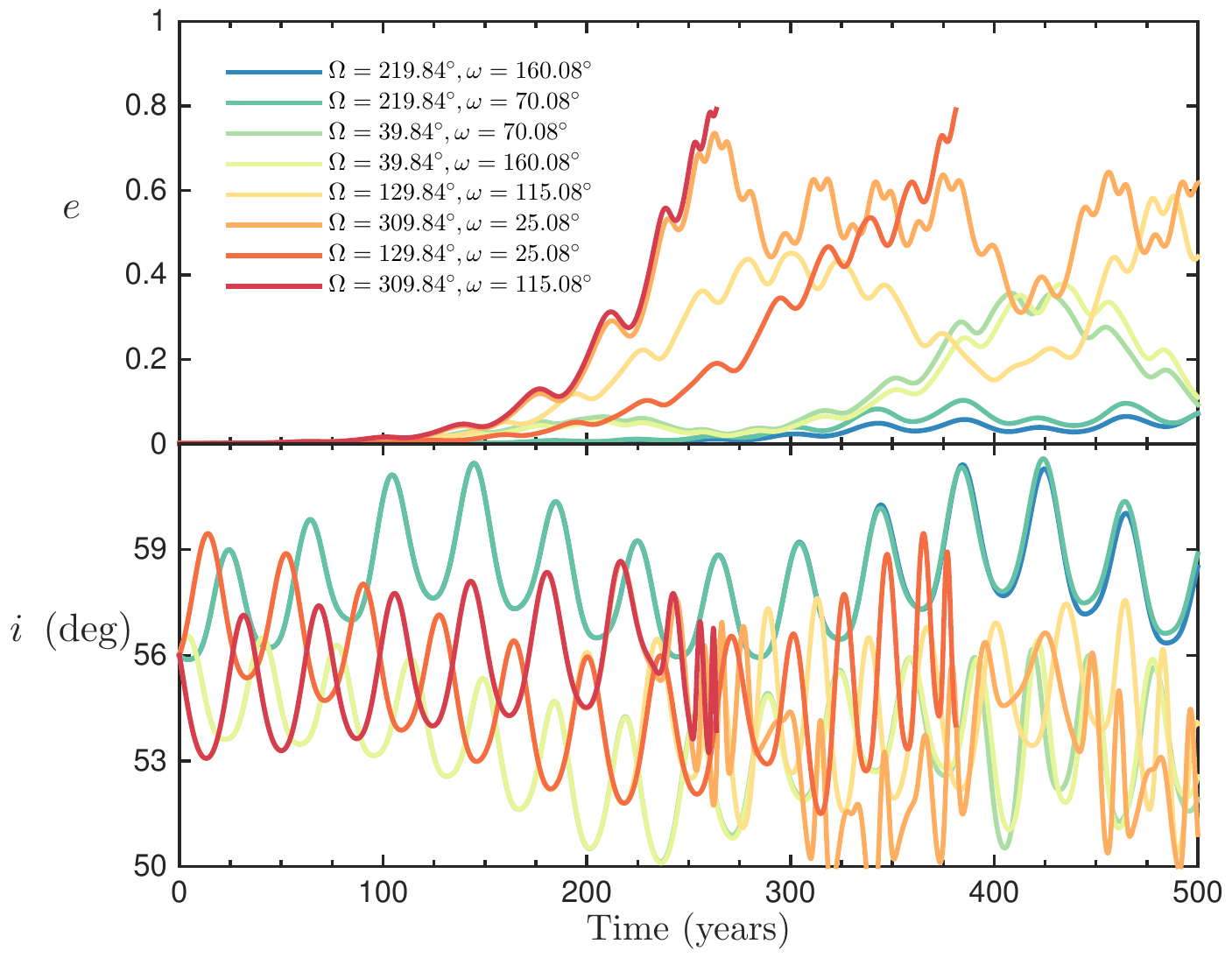} 
    \label{sfig:stab_fxtpts_traj}} 
    \hspace{12pt}
    \subfigure[One hundred ensemble integrations of the most {\itshape stable} orbit of 
    Fig.~\ref{sfig:stab_fxtpts_traj} ($\Omega = 219.84^\circ, \omega = 160.08^\circ$), 
    under the same errors as in Fig.~\ref{sfig:stable_ICs}.]
    {\includegraphics[scale=0.55]{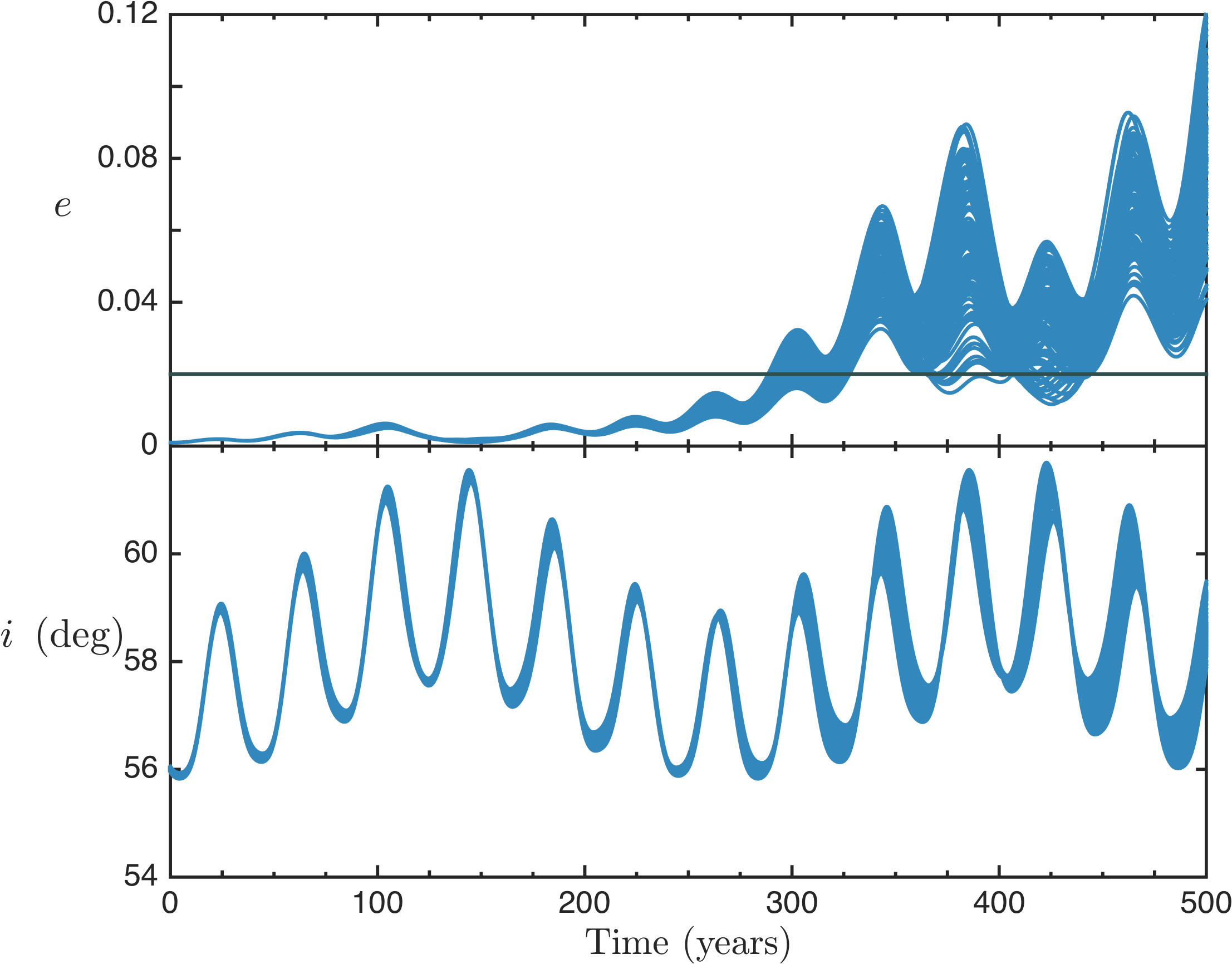}
    \label{sfig:stab_fxtpt_ensem}}
    \subfigure[One hundred ensemble integrations of the orbit located at the elliptic fixed point of
    nodal resonance and hyperbolic fixed points of the apsidal resonances
    at $\Omega = 219.84^\circ$ and $\omega = 70.08^\circ$, 
    under the same errors as in Fig.~\ref{sfig:stable_ICs}.]
    {\includegraphics[scale=0.55]{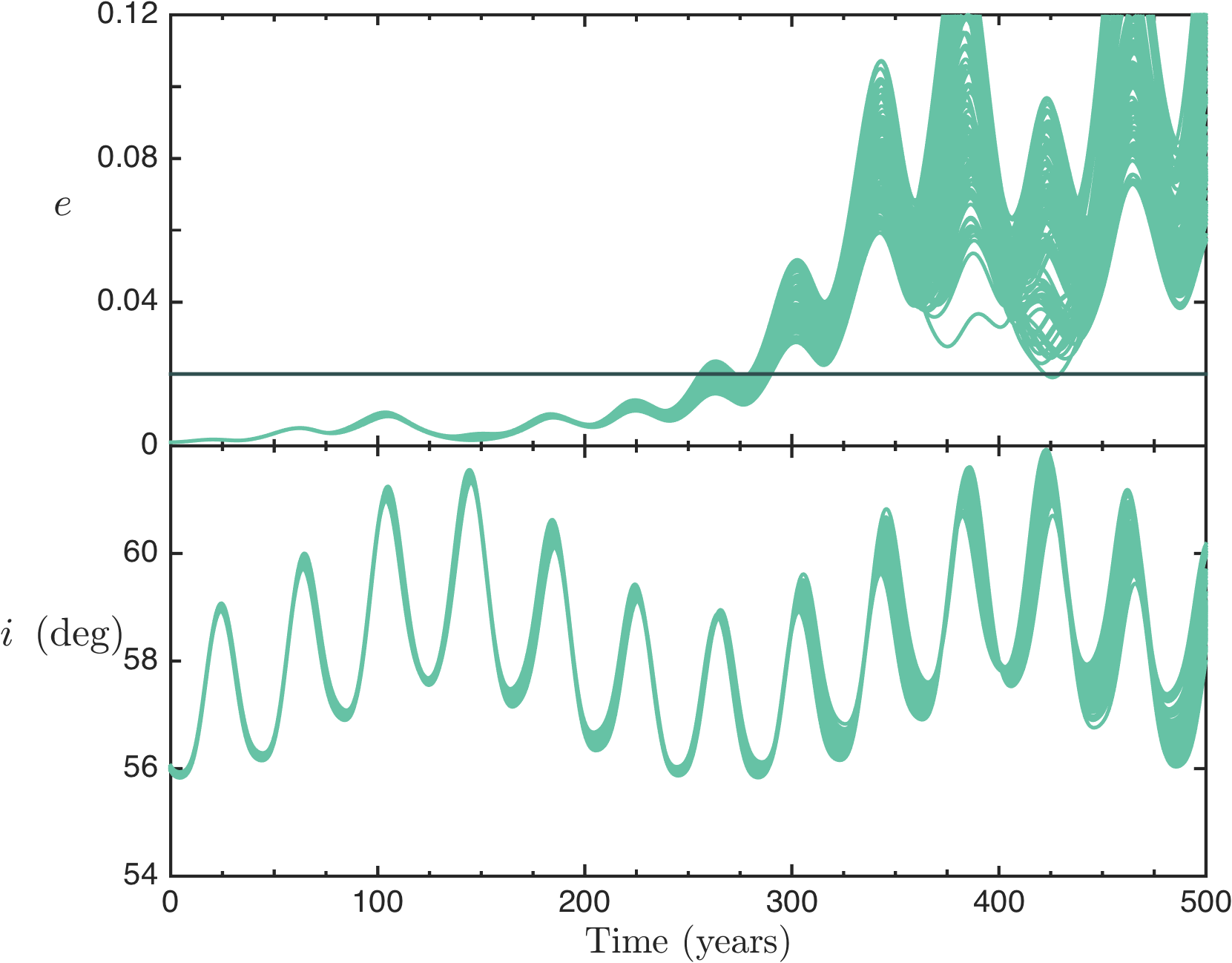}
    \label{sfig:stab_Om217w70_ensem}} 
    \hspace{12pt}       
    \subfigure[One hundred ensemble integrations of an orbit roughly located in the middle of the 
    stability pocket of the FLI map at $\Omega = 217^\circ$ and $\omega = 130^\circ$,
    under the same errors as in Fig.~\ref{sfig:stable_ICs}.]
    {\includegraphics[scale=0.55]{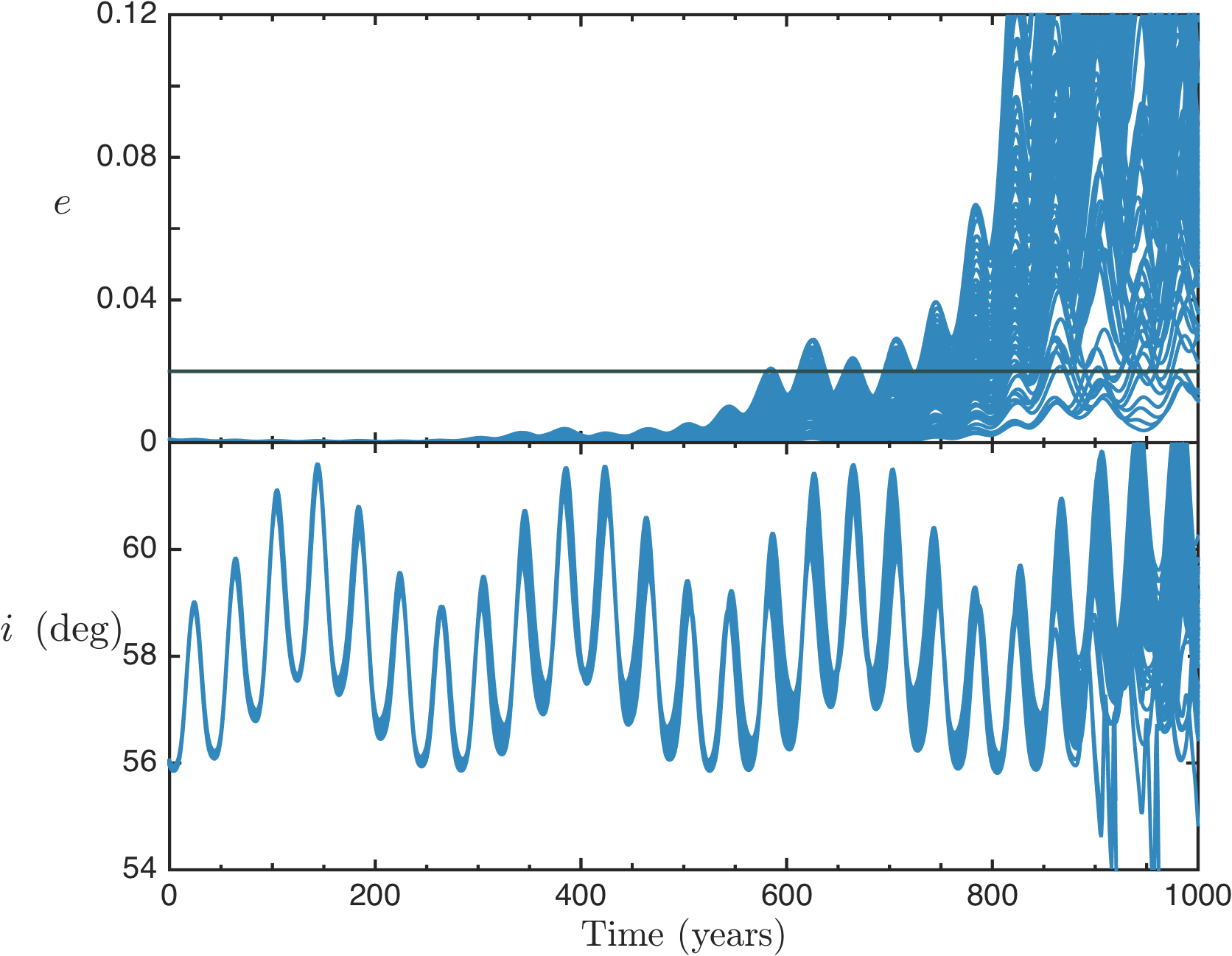}
    \label{sfig:stab_Om217w130_ensem}}           		            
  \caption{Numerical integrations of orbits selected according to the equilibria curves and FLI map of 
  Fig.~\ref{sfig:fxpts_GalStab} ($a_0 = 30\,150$ km, $e_0 = 0.001$, $i_0 = 56^\circ$, epoch: 6 DEC 2020).} 
  \label{fig:GalStab}
\end{figure}

Given the symmetry in the argument of perigee, there are essentially eight general points of interest in Fig.~\ref{sfig:fxpts_GalStab}; the intersection of the equilibrium curves of the primary resonances. Figure~\ref{sfig:stab_fxtpts_traj} presents their trajectory realizations over a 500-year timespan. As expected, the orbits along the elliptic equilibria of the $2 \dot\Omega - \dot\Omega_\M \approx 0$ nodal resonance yield more stable evolutions, while those along the hyperbolic equilibria are highly chaotic, some of which are dynamically short-lived (relatively speaking). The orbits located in the vertical band of Fig.~\ref{sfig:fxpts_GalStab} at $\Omega = 219.84^\circ$ appear to be the most stable (even if located at the hyperbolic fixed points of the apsidal resonances), and thus a simple graveyard disposal criterion naturally presents itself, which does not require the strict perigee targeting scheme of Section~\ref{sec:parametric}. The release epoch can be correlated with an initial lunar node, and as the satellite's node naturally precesses due to Earth oblateness perturbations, one must only wait for the appropriate lunar-satellite nodal phasing in order to ensure a stable graveyard (i.e., $e < 0.02$ for at least 200 years), as validated in Figs.~\ref{sfig:stab_fxtpt_ensem} and \ref{sfig:stab_Om217w70_ensem} on an ensemble level. Of course, a perigee and node around this orbit can be selected from the FLI maps themselves to yield an even more stable system (Fig.~\ref{sfig:stab_Om217w130_ensem}); yet, even here the orbit eventually succumbs to its chaotic nature. We note that the correlation between the Lyapunov time and an effective stability time is delicate to establish \citep{aM92}, and should be pursued in future work. 

\begin{figure}
  \centering
    \subfigure[$\Omega_\M = 160.56^\circ$ (epoch: 26 FEB 1998), \newline
    $\Omega = 260.28^\circ$, $\omega = 139.86^\circ$.]
    {\includegraphics[scale=0.38]{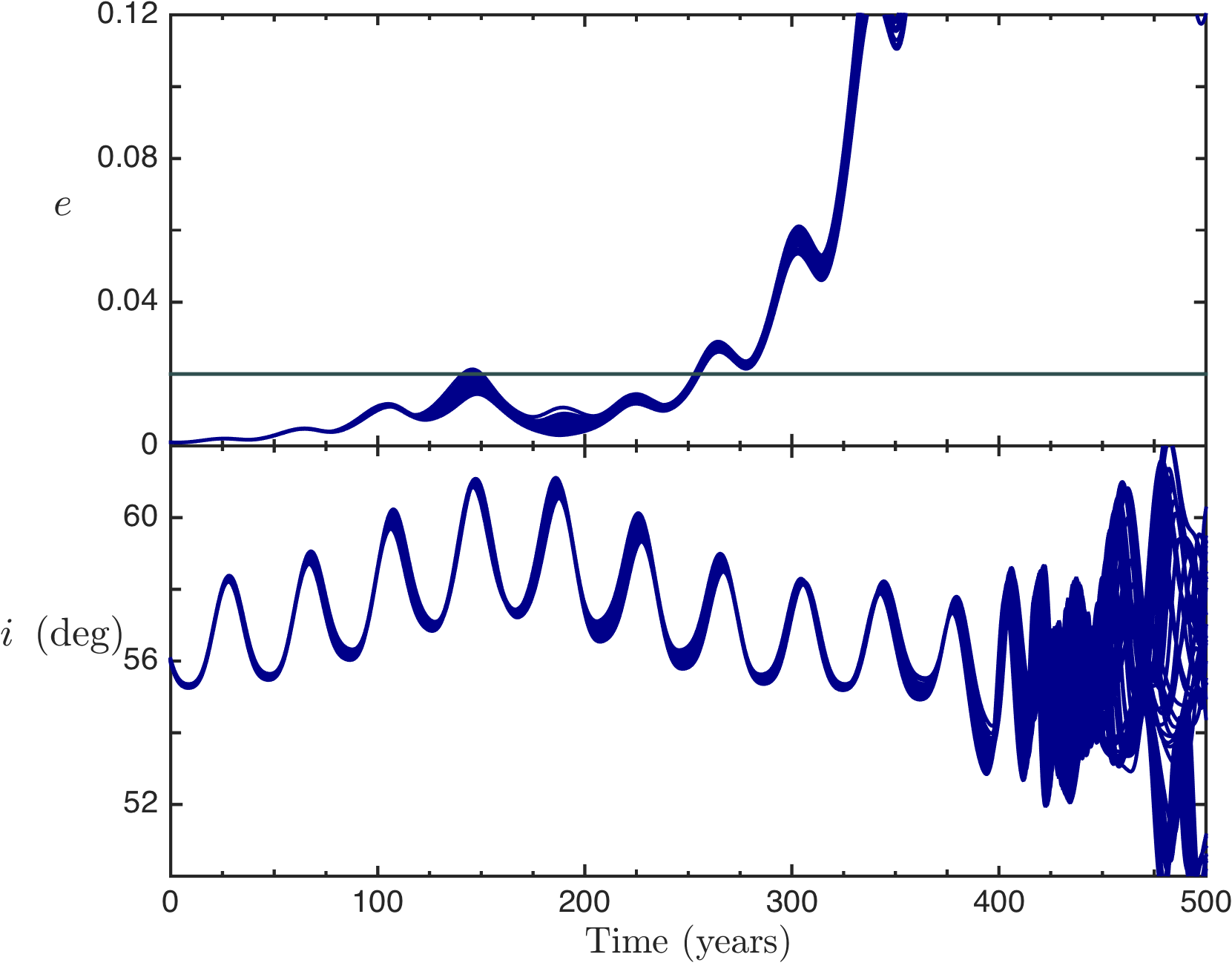} 
    \label{sfig:980226}} 
    \hspace{12pt}
    \subfigure[$\Omega_\M = 355.28^\circ$ (epoch: 12 SEPT 2006), \newline
    $\Omega = 177.64^\circ$, $\omega = 1.18^\circ$.]
    {\includegraphics[scale=0.38]{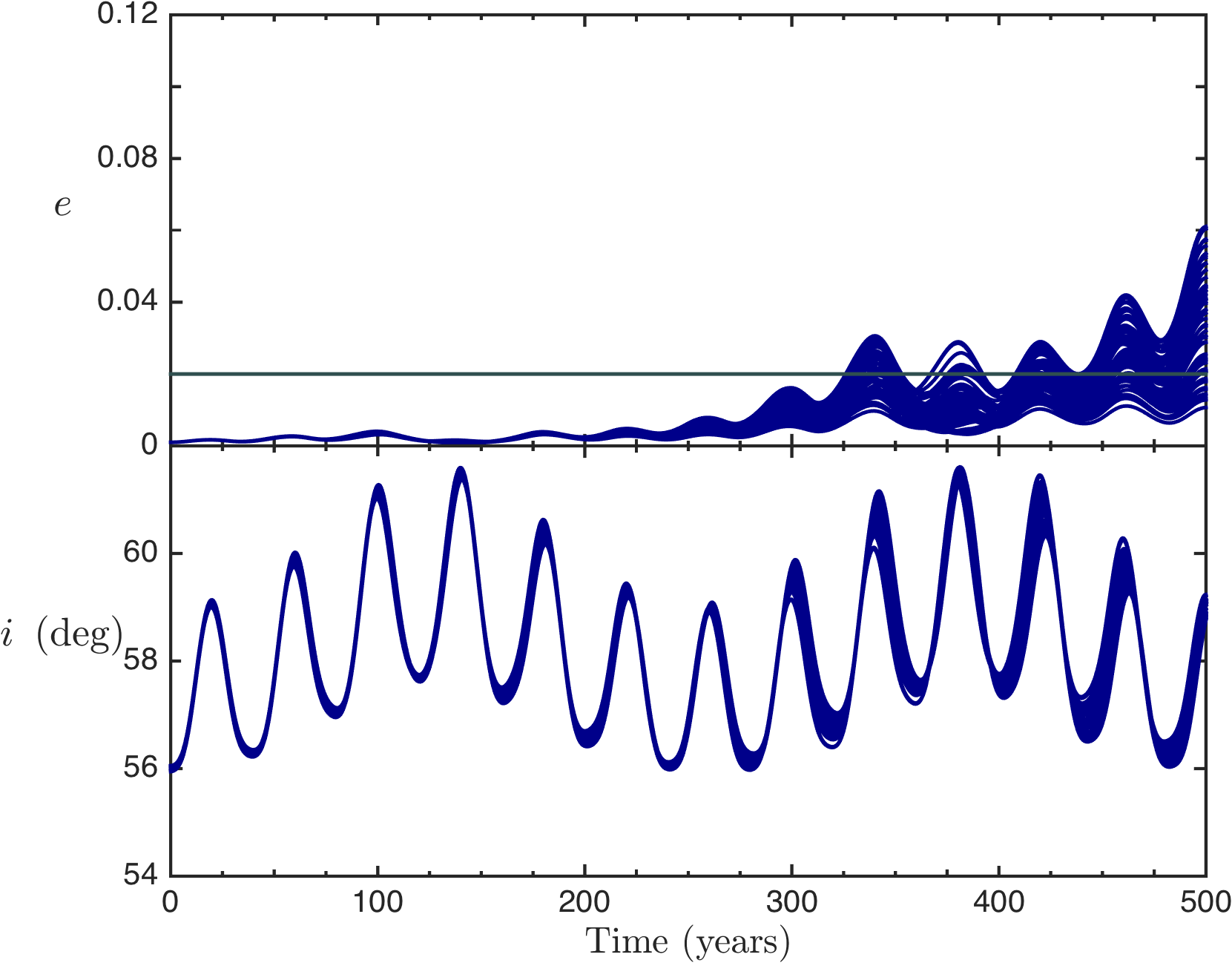}
    \label{sfig:060912}}
    \hspace{12pt}
    \subfigure[$\Omega_\M = 208.17^\circ$ (epoch: 15 APR 2014), \newline
    $\Omega = 104.08^\circ$, $\omega = 37.96^\circ$.]
    {\includegraphics[scale=0.38]{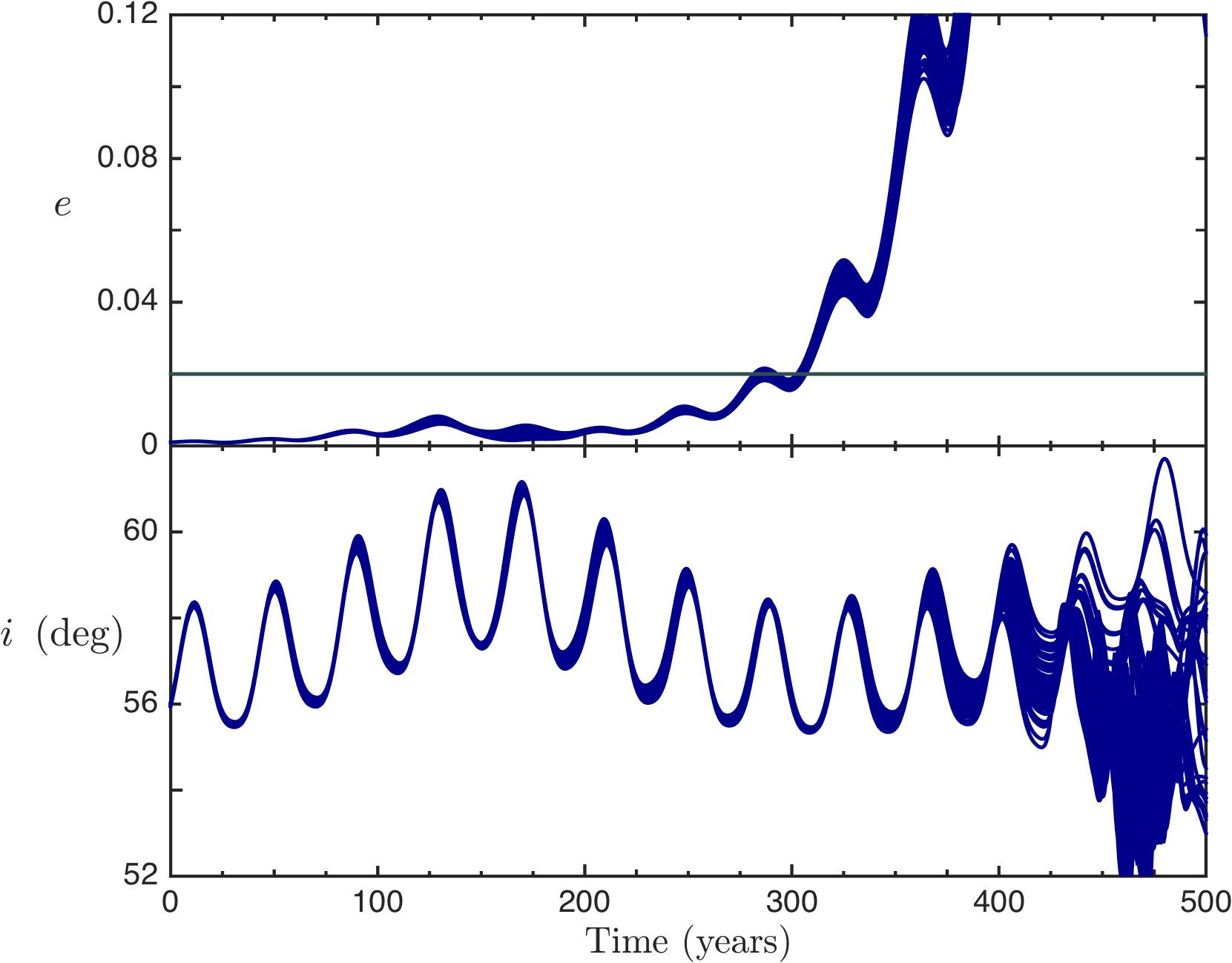}
    \label{sfig:140415}}       		            
  \caption{Fifty ensemble integrations of orbits selected according to the stable equilibria of the   
  nodal and apsidal resonances in Table~\ref{tab:fixed_points} for various disposal epochs
  ($a_0 = 30\,150$ km, $e_0 = 0.001$, $i_0 = 56^\circ$).} 
  \label{fig:GalDispStrat}
\end{figure}

It is probably unreasonable to expect space operators to have detailed FLI stability maps for each epoch or (what amounts to the same thing) initial lunar node. Accordingly, our proposed graveyard orbit criterion should be tested in the absence of such maps. There is, in this respect, an ambiguity as to which elliptic equilibria of the $2 \dot\Omega - \dot\Omega_\M \approx 0$ nodal resonance will give the wider stability band, if one even exists (cf. Fig.~\ref{fig:equilib_criteria}, where the two dashed vertical lines are located at $\Omega = 39.84^\circ$ and $\Omega = 219.84^\circ$, respectively); consequently, both should be examined to determine the more stable. Figure~\ref{fig:GalDispStrat} shows ensemble integrations of the `elliptic-elliptic' fixed-points solution (i.e., the intersection of the line $\Omega = (2 \pi + \Omega_\M)/2$ or $\Omega = \Omega_\M/2$ with $\omega = (\pi - \Omega)/2$ or $\omega = -(\pi + \Omega)/2$), for a few other epochs that were considered in \citet{eA15}. The same analysis was carried out for the remaining 34 initial epochs of that study (which incidentally turned out to sample well the various lunar nodes), from which we can loosely conclude that Fig.~\ref{sfig:060912} represents the general behaviour for initial lunar nodes in the ranges $\Omega_M \in [0^\circ,\, 125^\circ]$ and $\Omega_M \in [220^\circ,\, 360^\circ]$, while Fig.~\ref{sfig:140415} generally corresponds to the evolutions outside of these zones. Figure~\ref{sfig:980226} represents sort of an extreme behaviour found only in a few cases, where the orbits can slightly penetrate the threshold eccentricity ($e > 0.02$) within 200 years integration time and even eventually reach Earth-collision orbits. Overall, these results seem to corroborate the lunar-satellite nodal phases scheme for defining stable graveyards for the Galileo constellation. When effected, this strategy generally keeps the eccentricities below 0.02 for at least 200 years, while simultaneously locking the inclination into a long-period oscillation; this inclination behaviour should also contribute to diluting the probability of collisions within the graveyard orbits.  

The situation is not as clearcut for the proposed re-entry disposal solution, where, for the considered semi-major axis ($a_0 = 28\,100$ km), there exists only one dominant resonance (Figs.~\ref{fig:apertures} and ~\ref{fli_fxpts_GalUnstab}). However, we note from Figs.~\ref{sfig:fxpts_GalStab} and ~\ref{fig:Omega_omega_pert_param} that there exists closer (are even more circular) orbits to the nominal Galileo constellation that can lead to atmospheric reentry; and whose resonant geometry may allow for a better determination of the precise structures in the FLI maps.   

% -------------------------------------------------------------------------------------------------------------------------------------- 
%     CONCLUSION
% -------------------------------------------------------------------------------------------------------------------------------------- 
\section{Conclusion}

The earlier sections (\ref{sec:intro} and \ref{sec:parametric}) may seem to revel overmuch in our past imperfections on the MEO stability problem; yet, it is no longer possible to investigate the motion of celestial bodies without being fully conscious of the possibilities of chaos, a fact now well known to dynamical astronomers. Resonant and chaotic phenomena are ubiquitous in multi-frequency systems, and the knowledge of their long-period effects is essential for determining the stability of orbits and the lifetime of satellites. The complexity of the dynamical environment occupied by the Earth's navigation satellites is now becoming clearer \citep{aR15,jDaR15}. Resonant phenomena are widespread within the medium-Earth orbit (MEO) region as a whole, but particularly so amongst the highly inclined orbits of the navigation satellite systems, and a clear picture of the dynamics near these resonances is of considerable practical interest. We can now identify the sources of orbital instability or their absence in the MEO region and their nature and consequences in the context of long-term dynamical evolution. We examined them in terms of the detection of stability and unstable zones, with a particular view on the choice of the Galileo constellation disposal orbits. This paper links theoretical aspects of resonant and chaotic dynamics with practical applications, and lays an essential logical foundation for future developments. 

\section*{Acknowledgments}

This paper has benefited, directly or indirectly, from stimulating and useful discussions with S. Breiter, A. Celletti, T.A. Ely, C. Gale\c{s}, M. Guzzo, G. Pucacco, N. Todorovi\'{c}, and K. Tsiganis. We especially thank Fabien Gachet and Ioannis Gkolias, of the University of Rome Tor Vergata, for their technical contributions to this paper and for specific conversations about the disposal criteria. Many of the numerical stability map simulations were hosted at CNES, for which we are particularly grateful. This work is partially funded by the European CommissionÕs Framework Programme 7, through the Stardust Marie Curie Initial Training Network, FP7-PEOPLE-2012-ITN, Grant Agreement 317185. Part of this work was performed in the framework of the ESA Contract No. 4000107201/12/F/MOS ``Disposal Strategies Analysis for MEO Orbits''.

\bibliography{mnras_refs}

\bibliographystyle{mn2e}

\bsp

\label{lastpage}

\end{document}